\documentclass[12pt,preprint]{aastex}

\newcommand{\degree}{\ensuremath{^\circ}}

\begin{document}

\title {Near-Infrared Variability in the 2MASS Calibration Fields: A Search for Planetary Transit Candidates}

\author{Peter Plavchan\altaffilmark{1}, M. Jura\altaffilmark{2}, J. Davy Kirkpatrick\altaffilmark{1}, Roc M. Cutri\altaffilmark{1}, S.C. Gallagher\altaffilmark{2}} 
\altaffiltext{1}{Infrared Processing and Analysis Center, California Institute of Technology}
\altaffiltext{2}{UCLA Department of Physics and Astronomy}

\begin{abstract}
The 2MASS photometric calibration observations cover $\sim$6 square degrees on the sky in 35 ``calibration fields''  each sampled in nominal photometric conditions between 562 and 3692 times during the four years of the 2MASS mission.   We compile a catalog of variables from the calibration observations to search for M dwarfs transited by extra-solar planets.  We present our methods for measuring periodic and non-periodic flux variability.  From 7554 sources with apparent K$_s$ magnitudes between 5.6 and 16.1, we identify 247 variables, including extragalactic variables and 23 periodic variables.  We have discovered three M dwarf eclipsing systems, including two candidates for transiting extrasolar planets.
\end{abstract}


\section{INTRODUCTION}

M dwarfs comprise $\sim$70\% of the main sequence stars in the local stellar neighborhood \citep[]{mathioudakis93}.  However, the mass-radius relation and planetary companion abundance for these low mass stars are poorly constrained \citep[and references therein]{lopez07,lopez06,endl06,hebb06}.   Optical searches are hindered by the relative faintness of M dwarfs compared to solar-type stars.  M dwarfs emit much of their bolometric luminosity in the near-infrared, and consequently are the brightest and photometrically most quiescent at these wavelengths.  The near-infrared thus provides an ideal regime to look for eclipsing stellar, substellar and planetary companions to M dwarfs.

Many types of celestial objects have been studied individually to characterize photometric variability in the near-infrared.  Large-field near-infrared variability studies with 30 or fewer epochs were pioneered in several star-forming regions  \citep[]{carpenter01,carpenter02,barsony97}.   The near-infrared variability of extragalactic sources has also been explored with optical and radio-selected quasars and AGN  \citep[]{cutri85,neug89,enya02c}. A large-field search for extrasolar planets transiting M dwarfs and late-type eclipsing binaries enables an unprecedented general study of the variability of objects in the near-infrared.


\subsubsection{M Dwarf Extrasolar Planet Transits}
Among the $\sim$200 known extrasolar planets, six have been discovered orbiting four M dwarfs using the radial velocity technique -- GJ 876, GJ 436, GJ 849, \& GJ 581 \citep[and references therein]{rivera05,butler04,butler06,bonfils05}.  Additionally, \citet[]{bond04,beaulieu06} have discovered two distant M dwarfs with large separation ($\sim$3AU) Jovian companions by detecting the microlensing of a more distant star.  With the radial velocity technique, \citet[]{endl06} report on a systematic effort to constrain the frequency of close-in Jovian companions to M dwarfs.  There are initial indications that the frequency of close-in Jovian companions to M dwarfs is less than that for FGK-type stars \citep[and references therein]{endl06}.  This result implies possible differences in the formation mechanisms, planetary migration, and/or disk evolution timescales for M dwarfs relative to earlier-type stars. 

Transit searches offer another technique to search for planets around nearby M dwarfs.  More than 15 extrasolar planets have been identified and confirmed around solar-like stars using this technique \citep[and references therein]{butler06}.  The eclipse depth of a Jovian-sized planet transiting a solar-type star is $\sim$1\%.    M dwarfs have radii of $\sim$0.1-0.6R$_{\odot}$, and a transiting extrasolar planet the size of Jupiter ($\sim$0.1R$_{\odot}$) can produce an observable eclipse depth greater than 3\%.   

\subsubsection{M Dwarf Eclipsing Binaries}
There are eight known detached eclipsing binaries with M dwarf components, and only five that are well-studied \citep[and references therein]{lopez07,hebb06,creevey05,lopez05,ribas03,maceroni97}.    \citet[]{hebb06} presents a comprehensive review of known M dwarf eclipsing binaries. Theory currently underestimates M dwarf stellar radii by about 10-15\%; late-type eclipsing systems need further study to test the mass-radius relation \citep[]{burrows01,benedict00}.

\subsection{2MASS Calibration Point Source Working Database Overview}

The Two Micron All-Sky Survey \citep[2MASS]{skrutskie06} imaged the entire sky in three near infrared bands between 1997 and 2001.  Photometric calibration for 2MASS was accomplished using nightly observations of selected calibration fields around the sky.  The 35 2MASS calibration fields are distributed at approximately 2 hour intervals in right ascension near declinations of 0 and $\pm$30 deg.  Each 2MASS calibration field covers a region 8.5' wide (in right ascension) by 60' long (in declination).

Each night during 2MASS operations, the survey telescopes were directed at one of the calibration fields once per hour (before 11 Oct 1997 UT two fields were observed every two hours).  During each visit, six consecutive scans of the field were made in alternating declination directions in $\sim$10 minutes of elapsed real time (a ``scan group'').  Each scan in the set of six was offset from the preceding scan by $\sim$5'' in RA to avoid systematic pixel effects.  The calibration fields were observed using the same ``freeze-frame'' scanning strategy used for the main survey that yielded a net 7.8 sec exposure on the sky per scan.  Over the course of the 2MASS survey, between 562 and 3692 independent observations were made of each of the 35 calibration fields.  Table 1 presents a list of these fields and their aggregate properties.

The raw imaging data from each scan of a 2MASS calibration field were reduced using the same automated data processing system used to process the survey observation data \citep[${\S}$IV]{cutri06}.  The reduction process detected and extracted source positions and photometry for all objects in the images from each scan.  Measurements of the standard stars in each field were used to determine the nightly photometric zero-point solutions as a function of time, and seasonal atmospheric coefficients.  All source extractions from all scans were loaded into the 2MASS Calibration Point Source Working Database (Cal-PSWDB).  This database contains over 191 million source extractions derived from 73,230 scans of the 35 calibration fields.  Further descriptions of the Cal-PSWDB and its properties can be found in \citet[]{cutri06}.

We have carried out a new, systematic study of near-infrared variability using the 2MASS Cal-PSWDB to search for late-type eclipsing binaries and transiting extrasolar planets.  In ${\S}$2, we describe the selection of our sample of candidate variable stars from the 2MASS Cal-PSWDB. In ${\S}$3, we present our methods for quantifying and identifying variability, and finding periodicity.  In ${\S}$4, we present the results of our variability analysis, including a catalog of identified variables, periodic variables, extragalactic variables, and late-type eclipsing systems.  We present our conclusions in ${\S}$5. Constraints on the frequency of M dwarfs with extrasolar Jovian planets will be presented in a forthcoming paper. 

\section{SAMPLE SELECTION}

The sample for this study was selected in two steps.  First, candidate M-dwarf stars that fall within the 2MASS calibration field boundaries were identified in the 2MASS All-Sky PSC via their photometric color (${\S}$2.1).  All measurements of each candidate M-dwarf were then selected from the Cal-PSWDB, and the ensemble photometry for each star was analyzed for variability (${\S}$2.2).  In ${\S}$2.3, we describe photometry that are excluded from our analysis.

For nomenclature, we refer to all candidate sources by their PSC designation (2MASS Jhhmmss[.]ss$\pm$ddmmss[.]s).  We associate all Cal-PSWDB photometry with the PSC designation that was used in its selection. We use the time of the Cal-PSWDB observations to differentiate among different detections of the same PSC object.  This allows us to generate light curves of Cal-PSWDB photometry for every PSC source to investigate variability.

\subsection{PSC Selection Criteria}

To identify M dwarfs in the PSC  for our sample, we use separate selection criteria for low ($\left|b\right|<20\degree$) and high ($\left|b\right|>20\degree$) galactic latitude Cal-PSWDB fields.  For high galactic latitude fields, we use a color cut in H-K$_{s}$ (H-K$_{s} > 0.2$) to exclude earlier-type stars.  For low galactic latitude fields, we apply additional filtering to minimize contamination by faint reddened background objects, M giants, and background-confused sources.  We make two adjacent ``box'' color cuts in J-H and H-K$_{s}$ -- (0.2 $<$ H-K$_{s} \leq$ 0.35 and 0.5 $<$ J-H $<$ 0.7) OR (0.35 $<$ H-K$_{s} \leq$ 0.6 and 0.5 $<$ J-H $<$ 0.83) -- and in the apparent K$_{s}$-band magnitude (K$_{s} <$ 14.3).   The color-cuts are illustrated in Figure 1, with the PSC colors for targets in two sample fields shown in Figures 2 and 3.

Because we select our sources from the PSC, the sample matches the PSC magnitude limits, with the added color and magnitude constraints outlined above.  For high galactic latitude fields, the largest contribution of  sample `contaminants' are galaxies (see ${\S}$3.1.3), and we include these sources in our analysis.  For low galactic latitude fields, due to the relative uncertainties of PSC photometry, our selection criteria are not that successful in eliminating evolved stars and reddened earlier-type stars (See ${\S}$3.1.2 and Figures 4 and 5).

\subsection{Cal-PSWDB Selection Criteria}

From a parent sample of $\sim$10,000 PSC targets selected using the criteria outlined in ${\S}$2.1, we search the Cal-PSWDB for all detections within 2'' of the PSC coordinates.  This search radius is sufficiently large to ind all Cal-PSWDB detections associated with each PSC target given the astrometric precision of the 2MASS PSC, which is $<$0.3'' rms on each axis \citep[]{skrutskie06,zacharias05}, and the small astrometric biases that exists between the Cal-PSWDB and the PSC \citep[]{cutri06}.  We include only measurable detections in one or more bands, corresponding to non-``null'' uncertainties and \textit{[jhk]\_psfchi}$<$10. These photometry values are equivalent to the A-D photometric quality data (\textit{ph\_qual}) in the PSC; \textit{[jhk]\_psfchi} are reduced chi-squared goodness-of-fit values for each band and reflect how well each detection is fit by a single PSF (see ${\S}$2.3.3).  

We identify 7554 targets with a `sufficient' number of detections for our analysis.  We define a `sufficient'  number of detections for our analysis as follows.  We only include sources that are detected in $\geq$10\% of the repeated observations for a field at either J, H or K$_{s}$, and $\geq$50 detections at J-band.  The latter requirement eliminates sources located near the edges of fields that may not be covered by all scans.  The former constraint ensures a sufficient number of photometric measurements for our computational analysis, and corresponds to an average J-band magnitude $\sim$17.3.   These 7554 targets constitute our final sample, including 5628 sources in the 26 high galactic latitude fields and 1926 sources in the 9 low galactic latitude fields.  Basic properties of the sample are presented in Table 2.    We note that the Cal-PSWDB includes detections that would not meet the quality criteria for inclusion in the PSC \citep[]{cutri06}. 

\subsection{Excluded Photometry}

\subsubsection{Latent Image Artifacts}

We do not exclude photometry in our selection criteria in ${\S}$2.2 that are flagged with the `contamination or confusion' flag in the Cal-PSWDB, indicative of latent image artifacts or photometric confusion (\textit{cc\_flag=`c'}).  We identify by hand 7 targets with overlapping latent images of bright sources.  Overlapping latent images will appear in every other scan because the scan direction alternates between north and south.  These sources can still be of interest  -- we identify one source with an overlapping latent image that exhibits underlying periodic variability.  We note that not all sources contaminated with persistence artifacts are identified by the `contamination or confusion' flag in the Cal-PSWDB.

\subsubsection{Spurious Detections}

We eliminate from our analysis scans in which two sources are simultaneously detected within the 2'' search radius, whether the second source is real or a spurious byproduct of the Cal-PSWDB image processing.  Secondary detections are typically $\sim$0.5-1.5 mag fainter than the primary source, detected in only one band, and detected in only one scan from a group of six.  A single spurious detection included in the light curve for a source can trigger the variability flags identified in ${\S}$3.2.

\subsubsection{Poor Quality Photometry}

We do not include photometry with \textit{[jhk]psf\_chi}$\geq$10, which is equivalent to E photometric quality in the PSC.  The quoted photometric uncertainties for high chi-squared fits do not accurately reflect the photometric precision \citep[${\S}$IV.4.b]{cutri06}.  While these detections include photometric uncertainties, the PSF-fitting algorithm of the Cal-PSWDB reports a bad fit.  Possible reasons for poor point-source fits include saturation, cosmic rays, hot pixels, extended emission or partially resolved doubles \citep[${\S}$I.6]{cutri06}.  These data can be brighter by a few tenths of a magnitude, which can trigger the variability flags identified in ${\S}$3.2.  Typically, at most one of the six scans is brighter than the other five, and in only one of the three bands.

\section{ANALYSIS}

The Cal-PSWDB provides up to 3692 individual photometric measurements for each of our targets over a $\sim$4 year duration.    With this multi-band information for our large sample, we look for variability, and in particular, periodic variability.  In ${\S}$3.1, we present the characterization of our sample.  In ${\S}$3.2, we present our methods to identify variability.    A source meeting at least one of the method criteria is identified as a ``variable''.   In ${\S}$3.3, we present our methods to identify periodic variability.  In ${\S}$3.4 we discuss the identification of ``false-positive'' variability that is not intrinsic to the source.   The primary source of ``false-positive'' variability is atmospheric seeing variations.  

\subsubsection{Notation}
We adopt the following indices throughout our paper -- $m$ for band ($m\in{J,H,K_s}$), $n$ for source number; $x$ for test period; $t_6$ for time-series photometry co-added in groups of six (a ``scan-group''); and $t$ for time-series not co-added.  If $t_6$ or $t$ are not specified, the quantity has been averaged over all observations for source $n$.  We introduce other indices where appropriate.  When $m$ is not referred to as an index, we are referring to the apparent magnitude in J, H or K$_s$.  Similarly, when $t$ or $t_6$ are not indices, we are referring to the length of time since JD=2450000.0 for the particular observation.

\subsection{Sample Properties}

\subsubsection{Detection Rates}

For each target, we compute the detection rate -- the percentage of scans in each band in which a source is detected.   Sources brighter than the 99\% PSC completeness limits (in the unconfused sky), yet with detection rates $<$99\%, are typically either near the edges of a ``scan'' footprint, or in crowded fields where detections are `lost' due to confusion.   The ten faintest objects in our sample with $>$99\% detection rates have average apparent magnitudes of 16.6,15.7, and 15.1 at J, H and K$_{s}$ respectively.  Fainter than these magnitudes, the detection rates drop to zero \citep[]{skrutskie06}.  2MASS J20412236-0544317 is the faintest source in our sample with apparent magnitudes of 17.7, 16.7, and 16.1 at J, H and K$_{s}$ respectively.  The brightest object in each band with $>$99\% detection rate has an apparent magnitude of 8.5,7.7, and 7.5 at J, H, and K$_{s}$ respectively.  Brighter sources are detected, but because of image saturation the photometry are not useful for our analysis.  While we do not make any additional completion cuts based on the detection rate beyond what is described in ${\S}$2, the detection rate is useful for characterizing our sensitivity to variability (${\S}$4.4).

\subsubsection{Average Near-Infrared Magnitudes and Colors from the Cal-PSWDB}

For all targets in our sample, we determine the unweighted flux-averaged apparent magnitudes, colors, and standard deviations.  The mean apparent magnitudes, typically with milli-magnitude precision, are listed in Table 2, and near-IR colors are plotted in Figures 4 and 5 for fields with $\left|b\right|>20\degree$ and $\left|b\right|<20\degree$ respectively.  The uncertainty of colors derived from the averaged Cal-PSWDB measurements are typically 1-2 orders of magnitude smaller than the uncertainties from the single PSC measurements.  The apparent systematic difference between Cal-PSWDB and PSC colors is an artifact of the PSC color selection criteria and relative uncertaintities.  Our color selection criteria includes the distribution red ``tail'' of PSC colors (H-K$_s$$>$0.2) with Cal-PSWDB H-K$_s$$<$0.2.  Similarly, our color selection criteria excludes the distribution blue ``tail'' of PSC colors (H-K$_s$$<$0.2) with Cal-PSWDB H-K$_s$$>$0.2.  We continue to include in our analysis sources with average Cal-PSWDB colors that do not meet the PSC color criteria in ${\S}$2.1 (the PSC red ``tail'').  Due to practical limitations, we do not compute the average Cal-PSWDB colors for the entire Cal-PSWDB.   Consequently, we do not add sources to our sample with Cal-PSWDB colors that meet our PSC color criteria, but with PSC colors that do not (the PSC blue ``tail'').  This omission of sources is acceptable because we are not attempting to form a complete sample.

There is a flux-overestimation bias for sources detected with S/N$<$10 \citep[${\S}$A1.4.b.v]{cutri06}, because we do not include non-detections in computing the average flux.  In Figure 5, the population of targets with J-H$>$0.75 are consistent with the near-IR colors of reddened earlier-type stars and giants \citep[]{cutri06,allen}.  We note that this population is mostly absent in Figure 4.  We conclude that giants and reddened background stars are not a significant contribution to our sample for $\left|b\right|>20\degree$ fields.


\subsubsection{Sloan Digital Sky Survey Coverage and Optical Counterparts}

We use optical data to help categorize targets in our sample.  Eleven of 26 $\left|b\right|>20\degree$ fields have partial or full sky coverage in the  Sloan Digital Sky Survey, Data Release 5 \citep[SDSS DR5]{york00}.  We identify 1742 SDSS DR5 photometric catalog sources positionally coincident to within 5'' of our sample targets.  SDSS DR5 contains optical counterparts for all of our sample targets in nine of the eleven fields.  In the Cal-PSWDB field 90860, sample targets 2MASS J12213960-0009288 and 2MASS J12213992-0009289 are located within an arc-minute of an SDSS saturated bright star.  The SDSS DR5 catalog does not contain optical counterparts for only these two sources.  The Cal-PSWDB field 90565 is located near an edge of photometric sky coverage for SDSS DR5.  SDSS DR5 contains 33 optical counterparts to our sample.  For the remaining sources in the 90565 field, there is no SDSS DR5 catalog coverage.

For the sources without SDSS DR5 spatial coverage, we identify optical counterparts using the USNO-A2.0 and Tycho 2 catalogs with a 5'' search radius  \citep[]{monet96,hog00}.   Combined with SDSS, we identify optical counterparts for a total of 4598 and 1502 sources in $\left|b\right|>20\degree$ and $\left|b\right|<20\degree$ fields respectively.  We do not identify any optical counterparts for the remaining 1454 sources.  These sources are typically too red and faint to be included in the USNO-A2.0 and Tycho catalogs.

\subsubsection{Galaxy Identification}

In fields with SDSS DR5 coverage, we identify 392 objects classified as ``GALAXY'' or ``QSO'' by SDSS spectra and the SDSS photometric star-galaxy classification routine.    Not all extended SDSS sources are also identified as extended in the 2MASS PSC because of the larger 2MASS point-spread function and resolution.  We also query the NASA Extragalactic Database (hereafter: NED) by position for matches to within 0'.2.  NED includes the 2MASS All-Sky Extended Source Catalog (hereafter: XSC).   We identify 30 additional objects from NED identified as ``STAR'' in SDSS DR5, including 18 APM sources \citep[optical]{maddox90}, 5 extended 2MASS sources \citep[near-IR]{skrutskie06}, two NVSS sources \citep[radio]{condon98}, two quasars \citep[2dF QSO and galaxy redshift surveys, LBQS surveys]{croom04,magliocchetti02,hewett01}, one FIRST source \citep[radio]{white97}, a MGC source \citep[]{liske03}, and a ROSAT source \citep[X-ray]{bade92}.  We identify an additional 232 NED sources in fields with no SDSS DR5 coverage.  While some of the NED identified sources may be Galactic stellar objects, we classify these ``NED+SDSS''  654 objects as extragalactic.

In Figure 6, we present a color-magnitude plot of the Cal-PSWDB J-K$_{s}$ vs. SDSS-r for all 1742 sources with SDSS DR5 coverage, with 422 NED+SDSS extragalactic sources in blue and 1320 Galactic stellar objects in red.  In this color-magnitude space, the two populations are distinct and only minimally mixed  -- 9 SDSS ``STAR'' sources are plotted to the right of the green line in Figure 6, and 64 SDSS+NED objects to the left.  The 9 SDSS ``STAR'' sources are likely misidentified as stellar from their non-extended emission, and the 64 SDSS+NED objects located in the stellar locus probably include some Galactic stellar objects misidentified as extragalactic.  

\subsubsection{Observed Photometric Scatter and Modeled Photometric Scatter}

For each $n^{\mbox{th}}$ target, we compute the standard deviation of the photometry in each band $m$, $\sigma_{m,n}\pm\nu_{m,n}$.  The uncertainty, $\nu_{m,n}$, is propagated from the uncertainty in individual measurements $m_{n,t}\pm\sigma_{m,n,t}$ in computing the photometric scatter, $\sigma_{m,n}$.   Sources of scatter in the photometry such as seeing, zero-point offsets, and photon noise are discussed in more detail in \citet[]{skrutskie06}.  Figure 7 shows the distribution of the photometric standard deviations for our entire sample in each band.  Outliers with large observed photometric scatter are presumed to be variable, and we construct a model of the expected (non-variable) photometric scatter to quantitatively identify these sources.

As a function of apparent magnitude, there are four regimes to describe the observed photometric scatter, shown in Figure 7.  For region 1, corresponding to sources brighter than an apparent magnitude of $\sim$6,  the photometric scatter increases and photometric quality decreases due to detector saturation.   These sources are excluded from our sample, due to the poor photometric quality of the data (see ${\S}$2.3.3).   For sources between apparent magnitudes of 6 and 11 in J, H or K$_{s}$ (region 2), the observed photometric scatter is constant and represents a fundamental minimum in the achieved measurement accuracy due to the large 2MASS detector pixel size.   For sources fainter than an apparent magnitude of 11 (regions 3 and 4), the observed photometric scatter for individual targets increases in a fashion consistent with photon noise statistics.   There is a band-dependent ``break point''  apparent magnitude delineating regions 3 and 4, corresponding to the detection rate dropping below 100\%.  Fainter than the ``break point'' apparent magnitude (region 4), photometric scatter continues to increase due to photon noise statistics combined with flux overestimation \citep[${\S}$A1.4.b.v]{cutri06}.

For each band, we model the observed photometric scatter as a function of apparent magnitude in regions 2-4, $\sigma_{m,model}\pm\nu_{m,model}(m)$.  We identify a broken power law model for each band, shown in Figure 8 overlaid with the data, and given by the expression:
\begin{equation}
10^{\left(\sigma_{m,model}\pm\nu_{m,model}(m)\right)/2.5} = b_{m,l}\pm\sigma_{b_{m,l}}+(a_{m,l}\pm\sigma_{a_{m,l}}) 10^{m/2.5}
\end{equation}
$\{a_{m,l}\pm\sigma_{a_{m,l}},b_{m,l}\pm\sigma_{b_{m,l}}\}$ are the slope and intercept parameters and uncertainties for the fit in each band over each magnitude region $l$ for the above expression.  The first linear segment fits regions 2 and 3, while the second linear segment fits region 4.  The final derived values for $\{a_{m,l}\pm\sigma_{a_{m,l}},b_{m,l}\pm\sigma_{b_{m,l}}\}$ for each segment and each band are summarized in Table 3.  We do not analytically motivate the derived values for $\{a_{m,l},b_{m,l}\}$ in terms of the physical noise sources described above.  To arrive at these values, we take five steps.  In the first step, we fix the ``break point'' between the two linear regimes in each band to the PSC M$_{10}$ magnitude for that band.  The M$_{10}$ magnitude is defined to be the limiting magnitude for an individual PSC field in which $>$99\% of sources in the PSC have a signal-to-noise ratio (SNR) greater than 10.  For all of the PSC fields, the average M$_{10}$ apparent magnitudes are 16.422, 15.484, and 14.808 for J,H and K$_{s}$ respectively \citep[${\S}$VI.2]{cutri06}.  For the second step, we use a least squares linear regression to derive an initial set of $\{a_{m,l}\pm\sigma_{a_{m,l}},b_{m,l}\pm\sigma_{b_{m,l}}\}$ for each band on each side of the fixed ``break point''.  For the third step, we exclude sources with  $\sigma_{m,n}>  (5\nu_{m,n}+\sigma_{m,model})$ from the initial model.   This $\sigma_{m,n}$ clipping eliminates large (variable) photometric scatter sources contributing to an over-estimate of the expected (non-variable) photometric scatter.   For the fourth step, we use a least squares linear regression to derive a second and final set of $\{a_{m,l}\pm\sigma_{a_{m,l}},b_{m,l}\pm\sigma_{b_{m,l}}\}$ for each band on each side of the fixed ``break point''.  For the final step, we reset the ``break point'' magnitudes to where these two linear regimes intersect, at magnitudes of 16.53, 15.67, 15.06 for J, H and K$_{s}$ respectively.  These values, listed in Table 3 and shown in Figure 8, are fainter than the average PSC M$_{10}$ magnitudes.  This confirms that Cal-PSWDB scans are on average more sensitive relative to the PSC \citep[${\S}$VI.2]{cutri06}.  

With the values for $\{a_{m,l}\pm\sigma_{a_{m,l}},b_{m,l}\pm\sigma_{b_{m,l}}\}$ and ``break points'' listed in Table 3, we arrive at our model for the expected (nonvariable) photometric scatter, $\sigma_{m,model}\pm\nu_{m,model}$.  The uncertainties in the model $\nu_{m,model}$ are propagated from the uncertainties $\sigma_{a_{m,l}}$ and $\sigma_{b_{m,l}}$ which follow directly from the least squares linear regression.  The residuals $(\sigma_{m,n}-\sigma_{m,model})/\sqrt{\nu_{m,n}^2+\nu_{m,model}^2}$ are plotted in Figure 9.   For K$_{s}<$13,  the model slightly overestimates the observed photometric scatter ($<$0.01 mag), and this is not remedied by the above procedure.    Figure 9 demonstrates that the propagated uncertainties are potentially underestimated by a factor of $<$2.  Since we do not identify a source of additional uncertainty, we do not apply a correction factor.  Instead, we take into consideration the role of statistical fluctuations in selecting our variables (see ${\S}$3.4.2).

\subsection{Variability}

We classify objects as variable if the photometric dispersion deviates from what is expected given the uncertainties.  Sources can exhibit intrinsic variability with varying amplitudes, time-scales and frequency.   Many different methods in the literature are used to identify variability including but not limited to the Stetson index, excess photometric scatter, $\chi^2$ values, periodograms, fourier analysis, brute-force period searches, and visual inspection of light curves \citep[and references therein]{carpenter01,carpenter02,barsony97,stetson96,scargle82}.  The techniques used to identify variability have inherent strengths and disadvantages that depend on both the cadence and sensitivity of observations, and the intrinsic variability under investigation.   No single method properly identifies the variety of intrinsic variability we observe for our sample, given our sensitivities, cadence, and large number of observations.  

We adopt three complementary techniques to identify photometric variability.  Two methods are single-band ($\times$3 for all three bands) and one method is multi-band, for a total of seven measures of variability or ``flags''.  The first technique (single-band; three ``flags''), presented in ${\S}$3.2.1, is sensitive to relatively frequent small-amplitude variability ($<0.1$mag; hereafter ``flickering'').  This first technique is analogous to the identification of variability in \citet[]{carpenter01}.  The second technique (single-band; three ``flags''), presented in ${\S}$3.2.2, is sensitive to relatively infrequent, large-amplitude variability ($>0.1$mag; hereafter ``excursive'').   The third method and seventh ``flag'', presented in ${\S}$3.2.3, is the Stetson index and is a multi-band measure of correlated variability \citep[]{stetson96}.  If a source meets one or more of these seven criteria, we classify it as variable and non-variable otherwise. In ${\S}$3.2.4, we present our search for occulted sources.

\subsubsection{``Flickering'' Variability}

We apply the model $\sigma_{m,model}\pm\nu_{m,model}$ as defined in ${\S}$3.1.4 to our data for each target, $\sigma_{m,n}\pm\nu_{m,n}$.    For the first set of variability criteria, we attribute observed photometric scatter in excess of the model to underlying source variability.  We assume that this variability is gaussian, and is convolved with gaussian measurement scatter that we modeled, to produce the observed photometric scatter.  This assumption is not valid for a variety of astrophysical phenomena, but enables a simple deconvolution with photometric noise.  This assumption also enables a quantitative proxy to measure the amplitude of variability.  As in \citet[]{carpenter01}, we subtract these two standard deviations in quadrature for each band to estimate the magnitude of the source variability:

\begin{equation}
\sigma_{var,m,n}\pm\nu_{var,m,n}  = \sqrt{(\sigma_{m,n}\pm\nu_{m,n})^{2}-(\sigma_{m,model}\pm\nu_{m,model})^{2}}
\end{equation}

We plot $\sigma_{var,m,n}$ for sources with $\sigma_{m,n}>\sigma_{m,model}$ in Figure 10.  If $\sigma_{var,m,n}/\nu_{var,m,n}>5$ in  J, H or K$_{s}$, we flag that source as a ``flickering'' variable in that band with a `1'; `0' is assigned otherwise.  With the exception of  removed ``false-positive'' variables, ``flickering'' variable candidates are presented in Table 4.  Flags 1,2 or 3 correspond to sources that meet this 5-$\sigma$ variability criteria in J, H and K$_{s}$ respectively.  88 intrinsic variables meet the ``flickering'' criteria in one band, 42 in two bands, and 32 in all three bands.  In Table 10 we list the source counts for each variability flag combination (variables and ``false-positives''), and in ${\S}$3.4 we describe the removal of ``false-positive'' variables which outnumber the intrinsic variables overall.

\subsubsection{``Excursive'' Variability}

We use a single ``scan group'' observation of a field to identify short time-scale ``excursive'' variability.   Individual calibration fields were often observed two or more times during a night in order to measure the impact of atmospheric extinction.  The same field was usually not measured on consecutive calibration observations (one hour separation), but could be observed on every other observation or longer \citep[$\geq$2 hour intervals,][]{skrutskie06}.     For an individual source $n$, we compute the unweighted average apparent magnitude of the six scans for each $t_6^{th}$ scan group -- $J_{n,t_6}\pm\sigma_{J_{n,t_6}}$,$H_{n,t_6}\pm\sigma_{H_{n,t_6}}$, and $K_{s\:n,t_6}\pm\sigma_{K_{s\:n,t_6}}$ --  to improve the accuracy of the photometry and to increase the sensitivity to ``excursive'' variability.  We include groups in which the source is not detected in all six scans for a given band, but do not include upper limits from the non-detections in computing the average.  The number of scan groups for a field is one-sixth the total number of scans listed in Table 1.  

For the second set of variability criteria, we attribute group magnitudes significantly deviant from mean magnitudes ($\overline{J_n}$,$\overline{H_n}$, or $\overline{K}_{s\:n}$) to intrinsic variability.  For each group for each target in each band, we compute

\begin{equation}
\Delta_{m,n,t_6}\pm\sigma_{\Delta_{m,n,t_6}} = m_{n,t_6}\pm\sigma_{m_{n,t_6}} - \overline{m_n}\pm\sigma_{\overline{m_n}}
\end{equation}
If $\left|\Delta_{m,n,t_6}\right|/\sigma_{\Delta_{m,n,t_6}}>5$, in J, H or K$_{s}$ for any group $t_6$, we flag that source as an ``excursive'' variable in that band with a `1'; `0' is assigned otherwise.  Furthermore, we count the number of group excursions.  With the exception of removed ``false-positive'' variables described in ${\S}$3.4, ``excursive'' variable candidates are presented in Table 6.  Flags 4,5, or 6 correspond to sources that meet this 5-$\sigma$ variability detection in J, H and K$_{s}$ respectively.  85 intrinsic variables meet the ``excursive'' criteria in one band, 28 in two bands, and 31 in all three bands.  In Table 10 we list the source counts for each variability flag combination, and in ${\S}$3.4 we describe the removal of ``false-positive'' variables which outnumber the intrinsic variables overall.

\subsubsection{Stetson Index}

We adopt the Stetson Index as the third variability criteria \citep{stetson96}. This multi-band variability index was first implemented for Cepheid variables by Peter Stetson, and is used in two papers investigating variability in the 2MASS survey working database for the Orion A and Chamaeleon I molecular clouds \citep[]{carpenter01,carpenter02}.   We compute the Stetson Index for our sample, as in \citet[]{carpenter01,carpenter02,stetson96}:

\begin{equation}
S = \frac{\Sigma_{k=1}^{Z}w_k \mbox{sign}(P_k)\sqrt{|P_k|}}{\Sigma_{k=1}^{Z} w_k}
\end{equation}
where there are Z ``pairs of (simultaneous) observations to be considered, each with a weight $w_k$; $P_k=\delta_{i(k)}\delta_{j(k)}$ is the product of the normalized residuals of the two observations, $i$ and $j$, constituting the $k$th pair; and $\delta$ is the magnitude residual of a given observation from the average of all the observations in that same bandpass scaled by the standard error'':
\begin{equation}
\delta_{\alpha(k)}= \sqrt{\frac{Z}{Z-1}}\frac{m_{n,\alpha(k)}-\overline{m_{n}}}{\sigma_{m_{n,\alpha(k)}}}
\end{equation} 
where $Z>1$, $m_{n,\alpha(k)}$ is a magnitude $m_{n,t}$ from the $k^{th}$ pair, $\alpha=i$ or $j$, $\sigma_{m_{n,\alpha(k)}}$ is the uncertainty in the magnitude $m_{n,\alpha(k)}$, $w_k=\frac{2}{3}$ for observations detected in all three bands (three pairs), sign($P_k$)=$\pm$1 corresponding to whether $P_k$ is positive or negative, and $w_k=1$ for one- or two-band detections \citep[]{stetson96,carpenter01}.  We plot this index as a function of J-band apparent magnitude in Figure 11.  

In the PSC,  \citet[]{carpenter01,carpenter02} note a positive correlation for the Stetson Index, trending towards larger values for brighter apparent magnitudes.  We reproduce this trend in the Cal-PSWDB, indicating both that there exists a systematic correlation between bands in the Cal-PSWDB, and that the photometry is not entirely consistent with pure photon noise statistics.  Varying atmospheric conditions can account for the observed correlation between bands.  At the PSC 99\% completeness J-band magnitude of $\sim$16.1, the median Stetson index approaches 0 with increasing photometric noise.  The median Stetson index trends negative for sources fainter than J=16.1 due to multi-band detections biased in favor of better observing conditions.


There is no quantitative interpretation of the significance of the value of a Stetson index \citep[]{carpenter01,carpenter02}.  A Stetson index of zero would be expected for a non-variable source with non-correlated measurements.  Qualitatively, the larger the Stetson Index, the larger the multi-band variability.  We inspect Figure 11, light curves as a function of Stetson Index, and the Stetson index as a function of the variability criteria presented in ${\S}$3.2.1.  Eighteen of 23 periodic variables identified in ${\S}$4.2 have Stetson indices $>$0.2.  With a robust interpretation of periodic sources as variable, we arrive at a qualitative ``Stetson Variable'' demarcation value of 0.2.  We assign flag 7 in Table 6 a value of `1' for sources with a Stetson index in excess of 0.2.

\subsubsection{Occulted Sources}
For sources brighter than K$_{s}$=14.3, we search for naturally occurring  occultations by identifying non-detections for an entire scan-group.  Except for LHS 191  (GJ 3289, 2MASS J04261992+0336359), all non-detections in our K$_{s}$$<$14.3 sample are attributed to sources that fall outside the field footprint for that particular scan group.  For LHS 191, we attribute non-detections to a large proper motion relative to our Cal-PSWDB search radius of 2''.  No occultations are identified.

\subsection{Periodic Variability}

We examine all 7554 objects to identify periodic variability, independent of the variability criteria outlined in ${\S}$3.2.   The cadence and regularity of the Cal-PSWDB observations -- hourly and nightly cycles -- present challenges for quantitatively identifying periodicity.    We employ a ``brute force'' method to evaluate a large number of candidate periods for each individual source.  We examine a large number of light curves visually by eye to qualitatively identify periodicity.  In ${\S}$3.3.1, we present our implementation.  in ${\S}$3.3.2, we present the approximate period and photometric sensitivities of our analysis.  In ${\S}$3.3.3, we present other methods considered.

\subsubsection{Implementation}

Our period-searching method is a novel, bin-less implementation of the ``phase dispersion minimization'' approach of \citet[]{stellingwerf78}, and similar to the method of \citet[]{pilecki07} used to identify period drift in eclipsing binaries.   To identify periodic variability, we explore $\sim$15,000 periods between 0.1 and 50 days.  For each period $P_{x}$ -- not to be confused with $P_k$ in Equation 4 -- we generate a ``folded'' light curve $J(p_{x,t_6})$ from the time-series photometry $J(t_6)$:
 \begin{equation}
J(t_6) \Rightarrow J(p_{x,t_6}) \equiv J\left(\frac{t_6\%P_{x}}{P_{x}}\right)
\end{equation}
where $t_6$ is the total time elapsed since Julian Date 2450000.0 for each observation, $p_{x,t_6}$ is the phase ($p_{x,t_6} \in(0,1)$, the fraction of the period elapsed for a given time $t_6$ and period $P_x$), and \% is the decimal modulo ($t_6\%P_{x}$ is the remainder of $t_6$ divided by $P_{x}$ an integer number of times M, such that $M P_{x} < t_6 \leq (M+1)P_{x}$ and $(M + p_{x,t_6})P_{x} = t_6$).    For computational efficiency, we restrict the analysis to J-band data, and average scan groups to compute a single photometric datum for every six scans.   J-band data has both the best photometric quality and highest detection rates per source in our sample.  The H- (K$_s$-) band uncertainties are 27\% (83\%) larger than J-band uncertainties on average in our sample, and coadding these magnitudes into a ``super-magnitude'' would improve the S/N by only $\sim$19\% on average (as opposed to $\sim$43\% if the uncertainties were equal).  Since we might expect color-dependent variability that could degrade our S/N, we instead use H and K$_s$ data as a consistency check for candidate periods from the J-band data. 

The lower limit in our period search is motivated by the search for transiting planets, which are not expected to have periods much shorter than $\sim$1 day.   Additionally, for periods shorter than 0.1 days, the one-hour ``cycle'' in the Cal-PSWDB cadence of observations limits our period sensitivity.  The upper limit to our period range is motivated by the visibility of a field over the course of a year.  For periods longer than $\sim$50 days, corresponding to less than 20 observed periodic cycles, we can visually identify periodic variability from an inspection of the entire light curve.   We choose a variable period increment of $\frac{1}{2}P_{x}^2/B$ days, where  $B$ is the total time baseline of observations.   A smaller time-step produces insignificant changes in $J(p_{x,t_6})$; a larger time-step is too coarse and limits the recovery of periods from model periodic variables (see ${\S}$3.3.2).

We compute approximately 100 million ``folded'' light curves for analysis.   To reduce the number of  ``folded'' light curves to a manageable number for visual inspection, we need a measure of the significance of a candidate period.  Assuming a universal prior is ineffective because our periodic light curves can originate from different physical phenomena.   For each period, we instead generate a prior ``on the fly''.  We assume that any intrinsic periodicity will be approximately continuous and smoothly varying over the duration of one period cycle.  To evaluate the ``smoothness'' of each ``folded'' light curve, we compute a box-car smoothed light curve as our prior, with a box-car width of 0.06.  We compute the $\chi^{2}$ difference between the original light curve and the smoothed light curve, but only for the 25 worst fit data points ($\chi^{2}_{25}$).  The 25 worst-fit data points are not necessarily the same for each period.  Across all periods, we track the 5 periods with the smallest $\chi^{2}_{25}$ and identify these as candidate periods for each source.  Model periodic systems and the periodic variables we identify in ${\S}$4.2 motivate our choice of box-car width, $\chi^{2}$ metric, and the number of candidate periods to identify for each source.

At this time, we do not identify a quantitative value of $\chi^{2}_{25}$, consistent for all sources, that indicates if a candidate period is significant.   With 5 candidate periods per source, we visually inspect $\sim$35,000 ``folded'' light curves to evaluate candidate periods.  Visual inspection limits the identification of low-amplitude periodic variability (relative to the photometric scatter), and this period identification technique is inherently incomplete.  On a source-by-source basis, we re-analyze many of the variables with smaller period ranges and increments around candidate periods, integer fraction multiples of candidate periods, and other periods that look promising from the visual inspection of light curves.
  
\subsubsection{Recovery of Synthetic Eclipses}
	  
We test our implementation with model light curves of synthetic ``eclipses'', and for three earlier-type eclipsing systems in the 2MASS Cal-PSWDB discovered by \citet[]{hurt04}.    We recover all three periods for the earlier-type eclipsing systems discovered by \citet[]{hurt04}. We maximize the number of recovered periods to optimize our analysis parameters.   We take several non-variable light curves from the Cal-PSWDB, and during ``eclipse'' shift the photometry by the depth of the eclipse.   We generate 200 synthetic ``eclipses'' with a random period between 0.5 and 10 days, a random phase between 0 and 1, a random duration between 0.1\% and 4\% of the period, and a random depth between one-fifth and five times the J-band standard deviation.   We find that our algorithm can recovery periods independent of the eclipse phase and period, excluding periods near integer multiples of one-day where our sensitivity is degraded.

Our algorithm finds a minimum $\chi_{25}^2$ at the correct period -- or an integer fraction multiple thereof (1/5$\leq$P$_x$$\leq$3) when:

\begin{equation}
\mbox{transit depth(mag)}\geq \frac{8\:\sigma_{J,n}}{\sqrt{6\:(N_{s}-2.5)}}
\end{equation}
where N$_s$ is the number of scan groups in transit, the transit depth is in magnitudes, and $\sigma_{J,n}$ is the standard deviation of the 2MASS Cal-PSWDB J-band photometry in magnitudes (before coadding into scan groups).   In 7 cases out of the 200, we identify (3) or fail to identify (4) periods marginally in disagreement with the inequality in Equation 7 and hence Equation 7 is only an approximation.   From Equation 7, we see that a minimum of three scan groups in transit are required to identify a period, raising the S/N threshold to greater than 8 in the limit of a small number of observed transit events.  In the limit of a large number of observations in transit, the minimum of three to identify the period can be ignored.

\subsubsection{Comparison to Other Methods}

The ``string length'' algorithm of \citet[]{dworetsky83} is also a bin-less variation of the ``phase-dispersion minimization'' algorithm of \citet[]{stellingwerf78}.  The ``string length'' algorithm is well-suited to identify periodic variability when the number of observations are small ($\sim$20) and the amplitude of the variability is large with respect to the uncertainties in individual measurements \citep[][]{dworetsky83}.  Both the ``string length'' and our implementation are insensitive to the shape of the light curve variations, since no ``prior'' shape is assumed.  Our algorithm, however, is better suited for large number of observations (\raisebox{-4pt}{$\stackrel{>}{\sim}$}100), and our algorithm can identify periodic variability that has a  small amplitude with respect to the uncertainty for individual measurements.  The binned implementation of \citet[]{stellingwerf78} is also capable of identifying small amplitude periodic variability  with a large number of observations, but the use of bins as a function of phase $p_{x,t_6}$ can introduce period-aliasing.  For example, at periods near integer fraction multiples of one day, approximately one-half of the bins may contain all of the data, or the variable portion of a light curve might be split between two bins.  Hence, relative to the ``string length'' algorithm and the binned implementation of \citet[]{stellingwerf78}, we find that our bin-less method is more appropriate for our data and more sensitive to identifying periodic variations.

The Box Least Squares (BLS) algorithm of \citet[]{kovacs02} assumes a prior of two-state variability, and is well-suited for transit detection around solar-type stars.  Similarly, the Lomb-Scargle (L-S) periodogram of \citet[]{scargle82} numerically evaluates the discrete Fourier Transform for non-evenly sampled data.  This is equivalent to assuming a sinusoidal light curve prior, and the L-S algorithm is well-suited to identify periodic variability with sinusoidal-like light curves.  Given the variety of variability we observe in our data, we choose not to use either algorithm as a primary method to identify periods.  We avoid the assumption of a prior for our analysis, thereby decreasing the likelihood of missing a periodic signature with an uncharacteristic light curve shape.  However, both algorithms are useful to confirm periods identified with our algorithm, and both algorithms have relative computational efficiency advantages in identifying specific kinds of periodic variability.  

In Figures 12 and 13, we present the L-S periodogram, BLS periodogram, and our periodogram to compare for two periodic variables respectively -- an eclipsing system and a sinusoidal periodic variable.  Our algorithm identifies the period for both sources with greater relative significance compared to the BLS and L-S approaches.  We define the relative significance as the amplitude of the peak relative to the standard deviation of the periodogram values.  All algorithms have a periodogram peak for the eclipsing system, and the L-S algorithm and our algorithm also correctly identify the period for the sinusoidal variable.  Additionally, our algorithm identifies peaks for twice and three times the fundamental period for the sinusoidal variable.

\subsection{Systematic Sources of Variability}

The methods outlined in ${\S}$3.2 reveal systematic sources of false variability in the Cal-PSWDB.   We identify five phenomena that contribute to ``false positive'' variability in our sample -- seeing variations, statistical fluctuations, persistence artifacts, spurious detections and poor quality photometry.  We present our methods to identify ``false-positive'' variability in ${\S}$3.4.1 and screen for statistical fluctuations in ${\S}3.4.2$.   The exclusion of persistence artifacts, spurious detections and poor quality photometry are described in ${\S}2.3$.

\subsubsection{Seeing-Correlated Variability}

The Cal-PSWDB provides a wealth of information, including average seeing for every 2MASS calibration scan.  We identify photometric variations both correlated and anti-correlated with seeing variations. The correlation and amplitude of the flux variations are source-dependent and non-trivial, as shown for two sources with similar apparent magnitudes in Figure 14.    Since we cannot fully de-trend the data with a simple linear flux correction, we instead eliminate seeing correlated variables from our results without further analysis.    Seeing correlated variability is expected for an extended source in the Cal-PSWDB, since the source flux is inaccurately measured by a PSF in the Cal-PSWDB.  Seeing correlated variability due to confusion is also expected in crowded fields.  Neither the identification of a source as extended, nor the `confusion or contamination' photometric quality flag, are sufficient to identify seeing correlated variability.  A more robust identification is necessary. 

To quantitatively identify variability correlated with seeing, we employ the following process.  In ${\S}$3.4.1.1, we compute the linear Pearson $r$-correlation statistic between J-band photometry and J-band seeing.  We use this statistic to quantify how much the photometry varies in concert with seeing variations.  In ${\S}$3.4.1.2, we compute the one-dimensional Kolmogorov-Smirnov test as a function of $r$.  In the Kolmogorov-Smirnov test, we use the distribution of $r$-values for non-variables as a control distribution to compare to the distribution of $r$-values for variables.  In ${\S}$3.4.1.3, we vary the domain of $r$-values and use the resulting probability from the K-S test as an estimate of the probability that a source is not correlated with seeing.  Probabilities less than 5\% are classified as ``seeing correlated'' and variables are classified as  ``false positives'';  probabilities between 5\% and 95\% are classified as ``partially seeing correlated'' and variables are classified as  ``candidates''; probabilities of $>$95\% are classified as ``not seeing correlated''.

\noindent\textbf{3.4.1.1. Pearson r-correlation statistic}

We compute the Pearson r-correlation statistic between J-band photometry and J-band seeing for each source $n$:
  
\begin{equation}
r_n = \frac{\sum_{t=1}^{N_{J,n}} (J_{n,t}-\overline{J_n})(S_{t}-\overline{S}) }
{\sqrt{\sum_{t=1}^{N_{J,n}} (J_{n,t}-\overline{J_n})^{2}} \sqrt{\sum_{t=1}^{N_{J,n}}(S_{t}-\overline{S})^{2}}}
\end{equation}
where $S_{n}$ is the J-band seeing FWHM in arc-seconds, $\overline{S}$ is the average J-band seeing (not to be confused with the Stetson index $S$), and we sum over all J-band observations for source $n$, $N_{J,n}$.  A value of $r_n=1$ corresponds to perfectly correlated variations, and a value of $r_n=-1$ corresponds to perfectly inversely correlated variations.  Independence between the two measures yields a value approaching $r_n=0$ for large numbers of observations.

We plot the r-statistic as a function of J-band apparent magnitude for $\left|b\right|>20\degree$ and $\left|b\right|<20\degree$ fields in Figures 15 and 16, respectively.  For non-variables (0 of 7 variability criteria met), we identify a trend in the r-statistic as a function of apparent magnitude that is analogous to the trend observed for the Stetson Index.  A small positive correlation with seeing exists, on average and with large scatter, for sources brighter than J=16.1.  The average r-statistic drops to zero at J=16.1, and then an inverse correlation is observed for fainter sources.  The similarity between the trends in the r-statistic for non-variables and the Stetson Index as a function of apparent magnitude is expected.  If photometry is correlated with seeing in one band, then it is likely to also be correlated with seeing in the other two bands.  Since the Stetson Index measures correlations between bands, an overall correlation with seeing in all three bands would also be apparent as a correlation between bands.  As noted in ${\S}$3.2.3, varying atmospheric conditions can account for the observed correlations \citep[]{cutri06}.

\noindent\textbf{3.4.1.2. One-Dimensional Kolmogorov-Smirnov Test}

We use the one-dimensional Kolmogorov-Smirnov (1-D K-S) Test to determine the domain of $r_n$-values that contains variables that are not seeing correlated, and conversely, to exclude ``false-positives''.  We use non-variables as the comparison population.  We treat $\left|b\right|<20\degree$ and $\left|b\right|>20\degree$ separately, given the difference in range of apparent magnitudes, spatial source density, and the resulting $r_n$-values.  We compute the K-S test statistic, $D$, as in \citet[]{numrec}.

For illustrative purposes, in Figures 17 and 18 we plot ``smoothed'' distribution functions for variables (blue line) and non-variables (red line) for $\left|b\right|>20\degree$ and $\left|b\right|<20\degree$ fields respectively.  These ``smoothed'' distribution functions are derived from a numerical derivative of the cumulative distribution function or CDF, where the distribution function $DF(r_i)$, before normalization, is given by:
\begin{equation}
DF(r_{n}) = \frac{CDF(r_{n}+0.025) -  CDF(r_{n}-0.025)}{0.05}
\end{equation}
The separate populations of seeing-correlated and intrinsic variables are apparent in Figure 17, whereas seeing-correlated variables dominate the intrinsic variables in Figure 18.  

\noindent\textbf{3.4.1.3. Estimating the probability that a variable is not correlated with seeing}

``False-positive'' variables exhibit both large positive and large negative values of $r_n$ relative to non-variables.  Intrinsic variables exhibit values closer to $r_n=0$ than non-variables, since increasing variations due to intrinsic variability decreases the overall correlation with seeing.  It follows that there exists two domains of ``moderate'' $r_n$-values where non-variables and variables (that we assume are not correlated with seeing) have similar distributions as a function of $r_n$.  We denote these domains as $\{r_{a},r_{b}\}$ and $\{r_{c},r_{d}\}$, such that $-1 < r_{a} < r_{b} < r_{c} < r_{d} < 1$.  To identify these two domains, we maximize the K-S test probability that the variable and the non-variable populations within these two domains have identical distributions.   We identify the maximum K-S test probability at 99.7\% for $(r_{a},r_{b},r_{c},r_{d}) = (-0.276, -0.153, 0.188, 0.327)$ for $\left|b\right|>20\degree$ fields and at 99.9\% for (-0.245, -0.182, 0.031, 0.039) for $\left|b\right|<20\degree$ fields.  These domains are shown in orange in Figures 17 and 18, respectively.  

We next equate the K-S test probability of 99.7\% (99.9\%) to the probability that a source is not correlated with seeing for $r_n$-values contained in $\{-0.276,0.327\}$ ($\{-0.245,0.039\}$) for $\left|b\right|>20\degree$ ($\left|b\right|<20\degree$) fields.  By holding $r_{a}$,$r_{b}$, and $r_{c}$ fixed while increasing $r_{d}$, or alternatively holding $r_{b}$, $r_{c}$, and $r_{d}$ fixed while decreasing $r_{a}$, we use the K-S test probability to estimate the probability that a source is not correlated with seeing for values outside of these domains.  These probabilities are plotted in green in Figures 17 and 18.  The probability drops rapidly for small decreases to $r_{a}$ for all fields, and for small increases to $r_{d}$ for $\left|b\right|<20\degree$ fields.  There are no sources with $r_n>$0.66 in $\left|b\right|>20\degree$ fields in our sample due to a lack of source crowding (and only two $>$0.5).  Consequently, we are unable to estimate probabilities accurately for $r_n>$0.5 with our approach.  We consider these sources on an individual basis, as any observed variability is likely a false-positive due to spatial confusion.  We classify as a ``false-positive'' the variables that have a $<$5\% probability of being uncorrelated with seeing (white regions in Figures 17 and 18).  We classify as a ``candidate'' the variables that have a probability of being correlated with seeing between 5\% and 95\% (brown regions).   If the probability of being uncorrelated with seeing is $>$95\%, we classify a variable source as intrinsically variable (grey regions).

Our methods do fail to identify some intrinsic variables.  For example, the QSO B31456+375 (2MASS 14584479+3720216) exhibits large amplitude variability ($\sim$0.25 magnitude ``flickering'') that is only partially accounted for by seeing variations due to a flux overestimation bias (J band Pearson $r_n=-0.378$). Additionally, 4 of the 23 periodic variables identified in ${\S}$4.3 --  2MASS J18510526-0437311, 2MASS J18510882-0436123, 2MASS J18512929-0412407, and 2MASS J18513115-0424324 -- have probabilities of being uncorrelated with seeing of $<$95\%.   These four periodic variables are in $\left|b\right|<20\degree$ fields, and visual inspection of the photometry supports the conclusion that they are correlated with seeing variations in addition to exhibiting periodic variability.  We ``promote'' these five sources to intrinsic variables in our results.

\subsubsection{Statistical Fluctuations}

Given our sample size and number of scan groups, two or more ``excursions'' of a scan group in a single band at 5-$\sigma$ significance are sufficient for a statistically robust identification of variability.  Likewise, two or more flagged variability criteria are also sufficient.  To improve the reliability of the identification of variability, we place additional constraints on flagged variables that fall into two categories.  The first category are flagged variables that possess only a single ``excursion'' in a single band at 5-$\sigma$, meeting no other variability criteria.  For these sources, we require that one of the other two bands meet $\left|\Delta_{m,n,t_6}\right|/\sigma_{\Delta_{m,n,t_6}}>3$ for the same ``excursive'' scan group.   

The second category are variables that are flagged in only one band for  ``flickering'', meeting no other variability criteria.  The residual scatter shown in Figure 9 implies $\nu_{var,m,n}$ could be underestimated by a factor of $<$2. Consequently, a ``flickering'' variable  with a significance of 5-$\sigma$ in a single band is not necessarily statistically robust for a 7554-object sample.  Since we do not identify an additional source of error, we do not apply a correction to $\nu_{var}$.   For these sources, we instead require that $\sigma_{var,m,n}/\nu_{var,m,n}>3$ in one of the other two bands.  Sources that do not meet these additional criteria are identified as not variable in our results.  

110 variables  and 34 candidate variables with only one flag meet these additional criteria.  Approximately 60 sources for each variability flag and 365 sources total do not.   We list the numbers by flag combination in Table 10. While some of these sources may be intrinsically variable, we cannot distinguish them from false positives due to statistical fluctuations.

\section{RESULTS}

In ${\S}$4.1, we present our catalog of variables.   In ${\S}$4.2, we present our catalog and properties of periodic variables.   In ${\S}$4.3, we present a classification of all the identified variables.  In ${\S}$4.4, we present sensitivity estimates for our sample.

\subsection{Variables}

We identify 247 variables and 58 candidate variables that meet one or more of the criteria defined in ${\S}$3.2.  Variable sources do not exhibit ``false-positive'' or systematic variability defined in ${\S}$3.4, whereas candidate variables exhibit photometric variations that are partially correlated with seeing variations.  For candidate variables, we estimate a probability that the observed variability is intrinsic to the source (see ${\S}$3.4.1).  The properties of the variability for every variable and candidate variable are presented in Tables 4 and 5 respectively.   Each table contains the variable source PSC designation; the number of variability criteria met out of 7; the specific combination of variability criteria met; the magnitudes of the ``flickering'' if any and 3-$\sigma$ upper-limits otherwise; the number of  ``excursions'' and a sign to indicate whether a source is dimming (+) or brightening (-); and finally the Stetson Index.  Table 5 additionally contains an estimated probability for each candidate that the observed variability is not due to seeing variations; probabilities range from 5 to 95\%. 

The fraction of sample sources that are variable and are not correlated with seeing is 3.5\% (174/4952) for $\left|b\right|>20\degree$ fields, and 16.9\% (73/431) for $\left|b\right|<20\degree$ fields.  Including partially correlated sources inflates these statistics because some of the candidate variables are not intrinsically variable.   With more reddened earlier type stars and giants in the $\left|b\right|<20\degree$ fields, we expect a larger variety of astrophysical variability phenomenon to explain the higher fraction of variables.  In Table 6 we present counts for variables, candidate variables, and non-variables broken down by field.  We additionally break down the counts of non-variables to indicate the degree to which these sources are correlated with seeing.  In Table 7, we present the same counts for sources identified as extragalactic only.  

In Figures 19 and 20, we present color-color plots of the variables in $\left|b\right|>20\degree$ and $\left|b\right|<20\degree$ fields respectively, analogous to Figures 4 and 5.   In Figure 21, we plot the number of sources and number of variables for each field as a function of galactic latitude.  We note that the number of sources, and with larger scatter the number of variables, increases for decreasing latitudes.  This is expected as lower galactic latitude fields look through more of the galactic disk and the majority of identified variables are Galactic.  The larger scatter for the number of variables per field is partially due to the differing number of observations for each field which is a limiting factor in the identification of variability (see ${\S}$4.4).  

\subsection{Periodic Variables}

We identify 23 periodic variables with periods between 0.12 and 8.09 days. In Table 8, we present periods, ephemerides, variability flag combinations, and inferred object types.   Light curves and color curves in order of right ascension are presented in Figures 22-44 and 45-67 respectively.  For the plotted light curves, each point corresponds to an averaged ``scan group'', where the number of scan groups is equal to the number of observations of a field divided by 6.   In ${\S}$4.2.1 we present three newly discovered detached M-type eclipsing systems.  In ${\S}$4.2.2, we present two newly discovered periodically variable YSOs in Rho Ophiuchi.  In ${\S}$4.2.3 and ${\S}$4.2.4, we present the remaining periodic variables in $\left|b\right|>20\degree$ and $\left|b\right|<20\degree$ fields respectively.

\subsubsection{M-Type Eclipsing Systems}

2MASS J01542930+0053266, 2MASS J04261603+0323578, and 2MASS J04261900+0314008 possess near-IR colors consistent with M dwarfs.  The light-curves, shown in Figures 22,24 and 25, indicate that these three systems are likely to be newly discovered detached eclipsing binaries with M dwarf components.  2MASS J01542930+0053266 has a period of 2.6390 days, and the components are probably M0 type dwarfs with slightly unequal sizes.  In Figure 22, the primary and secondary eclipses have different depths.

For  2MASS J04261603+0323578 (and 2MASS J04261900+0314008), we are currently unable to distinguish between two eclipsing mid-M dwarfs with a period of 1.7644 (2.15262) days, and a mid-M dwarf primary with an unseen, possibly Jovian-sized secondary with a period of 0.8822 (1.07631) days. Due to uncertainty in the periods  of $\pm$0.00003 days and photometric uncertainties in the available data, we are not able to distinguish between these two scenarios without radial velocity follow-up \citep[]{plavchan07}.

\subsubsection{Rho Ophiuchi Periodic Variables}

The two periodic variables reddest in H-K$_{s}$, 2MASS J16271273-2504017 and 2MASS J16272658-2425543, lie in the 90009 field centered on the star-forming region Rho Ophiuchi.   One of the two Rho Ophiuchi periodic variables, 2MASS J16272658-2425543, is confirmed as a member of the Rho Ophiuchi complex.   These two sources are likely to be young stellar objects (YSOs), and exhibit large-amplitude dimming ($>$0.3 mag) on time-scales of 0.88 days and $\sim$3 days.  Possible origins of the variability include rotationally modulated spots, and veiling due to the circumstellar accretion driven by a companion.  Spectroscopic and photometric monitoring of these two systems is warranted to explain the observed variability.  From the near-infrared color light curve in Figures 52, dimming of up to 0.3 magnitudes produces no detectable color change for 2MASS J16271273-2504017. This source reddens in H-K$_{s}$ and J-H  by $<$0.1 magnitudes across nearly a factor of 2 change in apparent brightness.

\subsubsection{$\left|b\right|>20\degree$  Periodic Variables}

Eight of the 23 periodic variables are located in $\left|b\right|>20\degree$ fields, including the three late-type eclipsing binaries presented in ${\S}$4.2.1.  Of the remaining 5, the two that are bluest in J-H, 2MASS J18391777+4854001 and 2MASS J20310630-4914562, have the shortest periods in our sample at $\sim$3 hours.  These short periods are inconsistent with main-sequence rotation periods for M dwarfs.  Along with 2MASS J01545296+0110529 with a period of $\sim$4.5 hours, we suspect that these three sources are cataclysmic variables (CVs).  We infer this from the light curves, short periods and blue B-K$_s$ colors relative to the H-K$_s$ colors indicative of flux from a white dwarf at B band in addition to a red dwarf.  Additionally, three CVs in our sample are consistent with the expected number from the number density of CVs \citep[$\sim$2$\times$10$^{-5}$pc$^{-3}$]{townsley05}, the number density of M dwarfs \citep[$\sim$0.07 pc$^{-3}$]{reid02}, and the number of M dwarfs in our sample \citep[]{plavchan07}.  We do not, however, identify any ROSAT X-ray sources coincident with these three short-period variables.

The final two periodic variables, 2MASS J08512729+1211484 and 2MASS J15001192-0103090, have periods of 1.24 and 3.26 days respectively.  These two objects possess sinusoidal light curves and near-infrared colors that imply spectral types earlier than M.  We note that 2MASS J08512729+1211484 is spatially coincident with a latent image from a bright star in the Cal-PSWDB.  This latent image produces photometric variability not intrinsic to the source, but only during every other scan (see ${\S}$2.3.1).  We presume that the periodic variability is intrinsic to this source because the periodic variability is still observed for the scans not affected by the latent image.

\subsubsection{$\left|b\right|<20\degree$  Periodic Variables}

Fifteen of the 23 periodic variables are located in $\left|b\right|<20\degree$ fields, including the two YSOs presented in ${\S}$4.2.2.  Seven have light curves and periods that are consistent with earlier-type eclipsing binaries -- 2MASS J08255405-3908441, 2MASS J18510479-0442005, 2MASS J18510526-0437311, 2MASS J18512034-0426311, 2MASS J18512261-0409084, 2MASS J18512929-0412407, and 2MASS J19020989-0439440.   While the near-IR colors for these eclipsing binaries are consistent with an M spectral type, the eclipse durations indicate that these are reddened earlier-type stars (see figures).  Two of the seven earlier-type eclipsing binaries, 2MASS J18510526-0437311 and 2MASS J18512929-0412407, have a factor of two ambiguity in their period.  This ambiguity arises since the observed eclipses all have the same shape, depth and duration.  If the components are equal-sized, then the primary and secondary eclipses would be same.  Under such a scenario, the longer of the two periods listed in Table 8 is the correct one.  We do not rule out a secondary component smaller than the primary that produces an unseen secondary eclipse.  With this scenario, the shorter period would be the correct period.

 Four periodic variables are suspected cataclysmic variables with periods of less than one day and sinusoidal light curves -- 2MASS J18513076-0432148, 2MASS J18513115-0424324, 2MASS J19014393-0447412, and 2MASS J19014985-0432493.  The final two periodic variables, 2MASS J18511786-0355311 and 2MASS J18510882-0436123, also have periods shorter than a day.  The former has a light curve that indicates it is possibly a distorted eclipsing binary with unequal sized components since the primary and secondary ``eclipses'' have different depths (see Figure 35).  The latter source, 2MASS J18510882-0436123, has the most peculiar light curve out of all the periodic variables at low galactic latitudes (see Figure 34).  It is uncertain whether the actual period for this source is $\sim$0.44 or $\sim$0.88 days.

\subsection{Classification of Variables}

We classify variables by their variability properties, identified counterparts, and colors in ${\S}$4.3.1 through ${\S}$4.3.5 to provide a starting point for understanding the variety of the observed variability.  This classification is summarized in Tables 12 and 13.  In ${\S}$4.3.5, we cross-correlated the variables in the 90067 field, with the optical variability search of \citet[]{stassun02}.

\subsubsection{Variability Properties}

For periodic variables, the period and shape of the light curve provides information about the type of object and the origins of the observed variability.   Stetson-index identified variables exhibit multi-band, correlated variability.  For ``excursive'' variables, the sign of the excursions listed in Tables 6 and 7 are indicators of whether a source is dimming (+) or flaring (-).   For ``flickering'' variables, however, the magnitude of the variability does not provide a similar indicator.  Instead, we measure skews in the photometric scatter for ``flickering'' variables of 0.76, 0.53, and 0.63 with standard deviations of 1.57, 0.86, and 1.39 in J, H and K$_{s}$ respectively.  In Table 9, we present variables with skews more than one standard deviation outside the mean.   A large positive skew corresponds to a source that exhibits variability by getting fainter, and a large negative skew corresponds to a source that exhibits variability by getting brighter.


We observe a variety of astrophysical sources of variability that are both color and time-scale dependent, indicated by the different variability flag combinations we observe.  In general, the variables with more variability criteria met have larger amplitudes of variability.  We present in Table 10 source counts by variability flag combination.  By breaking down the number of variables in this fashion, we note that we detect the most variables at J-band.  For  ``flickering'' variables, we detect more at H than K$_{s}$.  This is expected, and corresponds to the order of bands with the best overall photometric quality -- we are generally more sensitive to variability at J, then H and finally K$_{s}$, although this is source color dependent.  For ``excursive'' variables, we detect more variables at K$_{s}$ than H.  This result implies short-term and relatively infrequent variability observed in our sample, such as would be detected by the ``excursive'' variability criteria, tends to have a larger amplitude at K$_{s}$ than at H.

We observe flickering and excursive variability for sources in our sample with optical and near-infrared colors consistent with M dwarfs.  We do not identify periodic variability for these sources given the relatively small amplitudes of variability.  The observed variability is likely due to rotationally modulated spots or flares that are more readily detected at shorter wavelengths for active M dwarfs \citep[]{plavchan07}.

 \subsubsection{Galactic Variables With Identified Counterparts}

Most of the variables are stellar, but lack optical identification or known spectral types.      2MASS J03320092+3727391 is identified as a member of the stellar open cluster NGC 1342.   2MASS J11214924-1313084 is identified as LHS 2397a, an M8 plus late-L brown dwarf close binary with a $\sim$3AU separation \citep[]{freed03}.   While LHS 2397a is a 2MASS standard calibrator star, LHS 2397a flares only three times (detected once at J-band and twice at K$_{s}$) with amplitudes of $<$0.1 mag, and has a J-band flickering variability of $<$0.02mag.  For the candidate variables, 2MASS J14405094-0023368 is identified as BD+00 3222, a high proper motion K0III star.  2MASS J16264814+0615056 is also identified as the K0 star SAO 121605.  Finally, 2MASS J18513086-0433412 is positionally coincident with IRAS 18488-0437, but the infrared flux might be produced by another source in the large IRAS beam.

 \subsubsection{Extragalactic Variables With Identified Counterparts}

In Table 11, we present 5 variables and 3 candidate variables that are identified as extragalactic, and include SDSS DR5 colors and redshifts when available.  76\% (500/654) of identified extragalactic sources exhibit photometry that are correlated or partially correlated with seeing (including the variable 2MASS J14584479+3720216).  Of the remaining 24\%, 2.6\% (4/154) are variable.  2MASS J14584479+3720216 is identified as the z=0.333 quasar B31456+375.  The variables 2MASS J00333270-3922457 and 2MASS J14411477-0057254, are known quasars with z$>$2 -- [ICS96] 003106.6-393917 and 2QZ J144115.5-005726 respectively.   The other two extragalactic variables -- 2MASS J16312442+2953016 (RX J1631.3+2953) and 2MASS J23182671+0030561	(APMUKS(BJ) B231552.98+001431.1) -- exhibit J-K$_{s}$ colors that are consistent with other identified galaxies.  2QZ J144115.5-005726, APMUKS(BJ) B231552.98+001431.1, and B31456+375 are unresolved with SDSS DR5. B31456+375 and RX J1631.3+2953 are spectroscopically identified as a QSO by SDSS DR5, and are previously known quasars.  [ICS96] 003106.6-393917 is not located in a field with SDSS DR5 spatial coverage.  None of the five variable extragalactic sources are in the 2MASS Extended Source Catalog.

\subsubsection{Colors}

For the  $\left|b\right|>20\degree$ variables, we identify 153 B-magnitudes from the PSC optical counterparts taken from USNO-A2.0 and Tycho 2 \citep[Johnson B and photographic B respectively]{monet96,hog00}; SIMBAD with a 5'' search radius; and SDSS DR5.  In the latter case, we use the transformation of \citet[]{lupton05},

\begin{equation}
B=u - 0.8116\times(u - g) + 0.1313
\end{equation}
to derive approximate Johnson B-magnitudes from SDSS colors.    We classify variables into seven categories, `A',`B',`C',`D',`E',`F', and `G', based on galactic latitudes, B-K$_{s}$,J-H, and H-K$_{s}$ colors.  Table 12 summarizes the criteria for each category, and the likely type of Galactic object corresponding to each category (e.g., category `D' variables are most likely M dwarfs).  Due to interstellar reddening in the $\left|b\right|<20\degree$ fields, we do not attempt a similar color classification (see Figure 21).  In Figures 68 and 69, we present B-K$_{s}$ vs. J-H and H-K$_{s}$ vs J-H plots for the $\left|b\right|>20\degree$ variables, with different colors and symbols for each of the seven categories outlined in Table 12.  We do not identify any clear trends in the observed variability within each category (e.g. flaring, dimming, flickering, etc.).  In Tables 13 and 14, we summarize the colors, categories, and notes for each variable and candidate variable respectively.

\subsubsection{M67 Variables}
We identify 24 variables in the 90067 field containing the $\sim$4 Gyr stellar open cluster M67 \citep[]{pols98}.  The number of variables identified in 90067 is large relative to the number of variables found in other high galactic latitude fields, but it is consistent with the number of sources in this field, the galactic latitude of $\sim$32$\degree$, and the increased sensitivity (this field has the highest number of  calibration observations -- 3692).  16 of the 24 variables identified in the 90067 field are identified as cluster members of M67 (Table 13).   2MASS J08512729+1211484 is also periodic with a period of 1.237 days.  

We identify 83 sources in our sample that are coincident to within 5'' with sources in the sample of M67 cluster members monitored for variability using B,V and I filters by \citet[]{stassun02}.  An additional 43 sources in our sample are located within the spatial coverage of \citet[]{stassun02}, but do not have optical counterparts that meet the selection criteria of \citet[]{stassun02}.  An additional 440 sources in the sample of \citet[]{stassun02} are located within the 90067 field, but either do not have 2MASS counterparts or do not meet the color criteria for inclusion in our sample.

Among the 83 sources common to our sample and the sample of \citet[]{stassun02}, \citet[]{stassun02} identify zero variables, and we identify three variables -- 2MASS J08510192+1149532, 2MASS J08511496+1203597, and 2MASS J08512916+1207014.  The corresponding source identification numbers used by \citet[]{stassun02} and defined in \citet[]{fan96} are 2899, 3271, and 3728 respectively.  2MASS J08510192+1149532 is spatially coincident to within 0.2' with X-ray source CX 128.  All three near-infrared variables common to our sample and the sample of \citet[]{stassun02} exhibit similar variability properties.  Each dims once by $\sim$0.1 magnitudes in J and H band, and less than 0.05 magnitudes in K$_{s}$.  The single `excursive' instances of variability for all three sources would not necessarily be identified by \citet[]{stassun02}, who had fewer epochs of observations than our survey.

\subsection{Sensitivity Estimates}

In ${\S}$4.4.1 and ${\S}$4.4.2, we present ``bulk'' approximations of the sensitivities of our variability criteria detailed in ${\S}$3.2.1 and ${\S}$3.2.2 respectively.  These ``bulk'' approximations are applicable across our entire sample, excluding sources that exhibit systematic sources of variability as outlined in ${\S}$3.4.   We present expressions based on a limited number of parameters -- $a_{m,l},b_{m,l},m_n,c_{m,n}$, and $N_{m,n}$-- where $a_{m,l}$ and $b_{m,l}$ are the photometric scatter model parameters listed in Table 3, $m_n\in(J_n,H_n,K_{s\:n})$ are apparent magnitudes for a particular source $n$, $c_{m,n}\equiv N_{m,n}/N_{scans}$ are the detection rates in each band $m$ for a particular source $n$, $N_{m,n}$ are the number of detections in each band $m$ for a particular source $n$, and $N_{scans}$ are the number of scans for an entire field (see Table 1, column 2).

\subsubsection{``Flickering'' Variables}

We identify ``flickering'' variables when $\sigma_{var,m,n}>5\nu_{var,m,n}$ in a particular band $m$ (see Equation 3 and ${\S}$3.2.1).  We note this can be rewritten as $\sigma_{var,m,n}>\sqrt{5\nu_{var,m,n}\sigma_{var,m,n}}>5\nu_{var,m,n}$.  We write the preceding since $\nu_{var,m,n}$ is inversely correlated with $\sigma_{var,m,n}$ and is inherently noisy. However,  $\sqrt{5\nu_{var,m,n}\sigma_{var,m,n}}$ can be approximated as:

\begin{equation}
\frac{\sqrt{5}}{\sqrt[4]{N_{m,n}}}  \sigma_{m,model}
\end{equation}
It follows that we identify ``flickering'' variability at 5-$\sigma$ significance when:
\begin{equation}
\sigma_{var,m,n} > \frac{2.5\sqrt{5}}{\sqrt[4]{N_{m,n}}} \log\left[b_{m,l}+a_{m,l} 10^{m_{n}/2.5}\right]
\end{equation}
For sources brighter than J$\sim$16.1 and not located near the edge of a field, $N_{m,n}\sim N_{scans}$.  For sources that meet no other variability criteria, we require a 3-$\sigma$ significance detection of variability in one of the other two bands (see ${\S}$3.4.2).

\subsubsection{``Excursive'' Variables}

We flag ``excursive'' variables when $\left|\Delta_{m,n,t_6}\right|>5\sigma_{\Delta_{m,n,t_6}}$ in a particular band (see Equation 3 and ${\S}$3.2.2).  On average, $\sigma_{\Delta_{m,n,t_6}}$ can be approximated by:

\begin{equation}
\sigma_{\Delta_{m,n,t_6}}(m_n)\sim\frac{1}{\sqrt{6c_{m,n}} } \sigma_{m,model}(m_n)
=\frac{2.5} {\sqrt{6c_{m,n}}}   \log\left[b_{m,l}+a_{m,l} 10^{m_n/2.5}\right]
\end{equation}
For sources brighter than the completeness limit, $c_{m,n}\sim1$.  For sources that have only a single ``excursion'' in a single band, and that meet no other variability criteria, we require a 3-$\sigma$ significance detection of variability in one of the other two bands for the same ``excursive'' group (see ${\S}$3.4.2).

\section{CONCLUSIONS}

The 2MASS Cal-PSWDB provides a wealth of near-infrared photometry covering $\sim$6 square degrees on the sky, a $\sim$4 year baseline and a few thousand repeated observations.   We present techniques to identify variability, periodicity, and to screen for ``false-positive'' variability due to a variety of effects.  From a subset of $\sim$7500 candidate sources in this database, we identify $\sim$250 variables.  We  present properties of the variables and the observed variability.  We identify 23 periodic variables, including three M dwarf eclipsing systems.  We characterize the sensitivity of our techniques to identify both long-term (``flickering'') and short-term (``excursive'') variability.

\subsection{Future Work}

The ability of our sample selection to recover M dwarfs in the 2MASS Cal-PSWDB and our resulting sample completeness will be discussed in \citet[]{plavchan07}.  With a color-selected sample of 7554 objects, much of the 2MASS Cal-PSWDB remains unexplored.  The full database can be analyzed with the techniques outlined in this work.  For the variables identified in this work, follow-up research is needed to identify objects types, and to identify the physical mechanisms generating the observed variability.


In our analysis, we identify objects that exhibit photometry correlated with seeing variations producing ``false positive'' detections of variability.  In some instances, the number of ``false-positive'' variables detected can outnumber intrinsic variables by a factor of $\sim$8.  These objects are primarily extragalactic or located in crowded fields close to the Galactic plane.  We are not able to identify variability for these objects with the same sensitivity we obtain for the rest of our sample.  We plan to de-trend the photometry for these sources with a linear correction to the flux as a function of the seeing FWHM, excluding photometry in the worst seeing conditions (FWHM$>$3'').  While this will result in a fewer number of observations per source and increased photometric uncertainties, such an analysis will open for exploration the variability properties of these sources.

\acknowledgements
This publication makes extensive use of data products from the Two Micron All Sky Survey, which is a joint project of the University of Massachusetts and the Infrared Processing and Analysis Center/California Institute of Technology, funded by the National Aeronautics and Space Administration and the National Science Foundation.  

This publication makes use of the NASA/IPAC Extragalactic Database (NED) which is operated by the Jet Propulsion Laboratory, California Institute of Technology, under contract with the National Aeronautics and Space Administration.

This research has made use of the NASA/ IPAC Infrared Science Archive, which is operated by the Jet Propulsion Laboratory, California Institute of Technology, under contract with the National Aeronautics and Space Administration.

This publication makes use of the Sloan Digital Sky Survey, Data Release 5 (http://www.sdss.org/).
 
Thanks to Karen Peterson, Thayne Currie, Robert Hurt, Eric Feigelson, and our anonymous referee for their conversations and comments.  

Parts of the research described in this publication was carried out at the Jet Propulsion Laboratory, California Institute of Technology, under a contract with the National Aeronautics and Space Administration.

\clearpage

\begin{deluxetable}{lrrrrr p{4cm}}
\tablecolumns{7}
\tabletypesize{\scriptsize}
\tablewidth{0pc}
\tablecaption{2MASS Calibration Working Database Fields and Aggregate Properties\tablenotemark{a}}
\tablehead{
\colhead{2MASS}	&	\colhead{Scans\tablenotemark{b}}	&	\colhead{RA\tablenotemark{c}}	&	\colhead{Dec\tablenotemark{c}}	&	\multicolumn{2}{c}{Galactic}	&	\colhead{Notes, Objects}\\
\colhead{Field}	&	\colhead{}	&	\colhead{(J2000)}	&	\colhead{(J2000)}	&	\colhead{l}	&	\colhead{b}	&	\colhead{of Interest}

}
\startdata
90021	&	2025	&	00 24 25.5	&	-01 58 22.6	&	107.36651	&	-64.02523	&	HD2023 - K1III, BRI0021-0214 - nearby, high proper motion M9.5V	\\
90294	&	2621	&	00 33 15.9	&	-39 24 05.5	&	318.91629	&	-77.15726	&	HD98784 - K0/K1III	\\
90004	&	2977	&	01 54 38.6	&	+00 43 00.9	&	154.11677	&	-58.27527	&	\nodata	\\
90301	&	3430	&	03 26 54.4	&	-39 50 33.6	&	244.63466	&	-55.50638	&	\nodata	\\
90247	&	1962	&	03 32 03.9	&	+37 20 37.9	&	155.04346	&	-15.31752	&	HD18552 - B8Vne	\\
90533	&	1839	&	03 41 03.3	&	+06 56 11.3	&	179.51313	&	-36.67609	&	\nodata	\\
90191	&	2086	&	04 26 21.4	&	+03 37 24.3	&	190.90892	&	-29.7781	&	LHS191 - nearby, high proper motion M6.5V	\\
90013	&	3515	&	05 57 08.3	&	+00 01 08.0	&	206.63167	&	-12.04885	&	\nodata	\\
\enddata
 \tablenotetext{a}{Adapted from \citet[Appendix 4.1]{cutri06}}
\tablenotetext{b}{Approximately 7.8 second integrations per scan; identical observing strategy as employed for the main 2MASS survey \citep[${\S}$III.2]{cutri06}.}
\tablenotetext{c}{Averaged coordinates for the center of each scan footprint.}
\end{deluxetable}

\begin{deluxetable}{l r r r r r }
\tablecolumns{6}
\tablewidth{0pc}
\tablecaption{Sample Properties}
\tabletypesize{\scriptsize}
\tablehead{
\colhead{Name}	&	\colhead{Field}	&	\colhead{N$_{obs}$\tablenotemark{b}} 	&	\colhead{$J$\tablenotemark{c}}&	\colhead{$H$\tablenotemark{c}} & \colhead{$K_{s}$\tablenotemark{c}} \\
\colhead{``2MASS J''\tablenotemark{a}}	&	\colhead{}	&	\colhead{}	&	\colhead{(mag)}&	\colhead{(mag)} & \colhead{(mag)}
}
\startdata
00240748-0140366	&	90021	&	152	&	16.4349$\pm$0.0287	&	15.7866$\pm$0.0444	&	15.3409$\pm$0.0395	\\
00240770-0218052	&	90021	&	271	&	16.1539$\pm$0.0133	&	15.7748$\pm$0.0228	&	15.5703$\pm$0.0371	\\
00240808-0155312	&	90021	&	609	&	16.9092$\pm$0.016	&	16.1137$\pm$0.0201	&	15.6066$\pm$0.0237	\\
00240811-0207138	&	90021	&	647	&	16.7314$\pm$0.0122	&	15.9586$\pm$0.015	&	15.3307$\pm$0.0133	\\
00240927-0146231	&	90021	&	1722	&	15.5633$\pm$0.0029	&	14.8828$\pm$0.0036	&	14.3658$\pm$0.0031	\\
\multicolumn{6}{c}{Abbreviated...}\\
\enddata
 \tablenotetext{a}{PSC catalog designation.  The number corresponds to the sexagesimal right ascension and declination in J2000 coordinates formatted as hhmmss[.]ss$\pm$ddmmss[.]s.  We note that all Cal-PSWDB detections have unique identifiers, and we spatially associate each with the PSC catalogue source and designation in our sample (${\S}$2.2)}
\tablenotetext{b}{Number of detections, either in J, H or K$_{s}$ (equivalent to $N_{m,n}$ in text).   See Table 1 and ${\S}$3.1 for discussion.}
\tablenotetext{c}{Unweighted mean apparent magnitude of Cal-PSWDB photometry (equivalent to $m_n$ in text).}
\end{deluxetable} 

\begin{deluxetable}{llrr}
\tablecolumns{4}
\tablewidth{0pc}
\tablecaption{Model Fit Parameters for Observed Photometric Scatter\tablenotemark{a}}
\tablehead{
\colhead{Band}	& \colhead{Range\tablenotemark{b}}&	\colhead{$a_{m,l}\pm\sigma_{a_{m,l}}$}	&	\colhead{$b_{m,l}\pm\sigma_{b_{m,l}}$}	}
\startdata
J & $<$16.53 & (3.1868$\pm$0.0098)$\times$10$^{-8}$ & 1.01780$\pm$0.00019 \\
J & $>$16.53 & (2.092$\pm$0.038)$\times$10$^{-8}$ & 1.0627$\pm$0.0019 \\
H & $<$15.67 & (7.2100$\pm$0.0038)$\times$10$^{-8}$ & 1.01862$\pm$0.00030 \\
H & $>$15.67 & (4.319$\pm$0.056)$\times$10$^{-8}$ & 1.0721$\pm$0.0013 \\
K$_{s}$ & $<$15.06 & (1.284$\pm$0.009)$\times$10$^{-7}$ & 1.01820$\pm$0.00038 \\
K$_{s}$ & $>$15.06 & (6.66$\pm$0.12)$\times$10$^{-8}$ & 1.0836$\pm$0.0019 \\
\enddata
 \tablenotetext{a}{See ${\S}$3.1.4 for explanation of parameters.}
 \tablenotetext{b}{Range in apparent magnitude; apparent magnitudes of $<$6 are excluded from this model.}
 \end{deluxetable}

\begin{deluxetable}{l r r r r r  r r  r r}
\rotate
\tablecolumns{10}
\tablewidth{0pc}
\tablecaption{Variables}
\tabletypesize{\scriptsize}
\tablehead{
\colhead{Name}	&	\colhead{Number}	&	\colhead{Flags\tablenotemark{b}}	&	\colhead{J Flicker\tablenotemark{c}}	&	\colhead{H Flicker\tablenotemark{c}} 	&	\colhead{K$_{s}$ Flicker\tablenotemark{c}}&	\colhead{J Excursion} & \colhead{H Excursion} & \colhead{K$_{s}$ Excursion} &  \colhead{Stetson}  \\
\colhead{``2MASS J''\tablenotemark{a}}	&	\colhead{of Flags}	&	\colhead{}	&	\colhead{(mag)}	&	\colhead{(mag)}	&	\colhead{(mag)}&	\colhead{Number\tablenotemark{d}} & \colhead{Number\tablenotemark{d}} &  \colhead{Number\tablenotemark{d}}  & \colhead{Index\tablenotemark{e}} 
}
\startdata
00240949-0140038	&	1	&	1000000	&	0.0217$\pm$0.0043	&	0.0231$\pm$0.0065	&	$<$ 0.0305	&	\nodata	&	\nodata	&	\nodata	&	0.047	\\
00241293-0133302	&	3	&	1101000	&	0.0195$\pm$0.0032	&	0.0306$\pm$0.0036	&	$<$ 0.0324	&	+1	&	\nodata	&	\nodata	&	0.112	\\
00242118-0219071	&	2	&	0001100	&	$<$ 0.0237	&	$<$ 0.0191	&	$<$ 0.0461	&	+1	&	+1	&	\nodata	&	0.128	\\
00243221-0216483	&	5	&	1111001	&	0.076$\pm$0.0068	&	0.0747$\pm$0.0091	&	0.0473$\pm$0.0094	&	-1	&	\nodata	&	\nodata	&	0.214	\\
00243850-0129122	&	1	&	0100000	&	0.0138$\pm$0.0033	&	0.0255$\pm$0.0027	&	$<$ 0.0163	&	\nodata	&	\nodata	&	\nodata	&	0.169	\\
00243956-0220288	&	2	&	1000100	&	0.0205$\pm$0.0028	&	0.0193$\pm$0.0042	&	0.0212$\pm$0.0053	&	\nodata	&	+1	&	\nodata	&	0.054	\\
00325745-3931347	&	1	&	1000000	&	0.0148$\pm$0.0015	&	0.0114$\pm$0.0023	&	0.0092$\pm$0.0026	&	\nodata	&	\nodata	&	\nodata	&	0.004	\\
00333270-3922457	&	1	&	1000000	&	0.0102$\pm$0.0018	&	0.0086$\pm$0.0027	&	$<$ 0.0084	&	\nodata	&	\nodata	&	\nodata	&	-0.043	\\
\enddata
 \tablenotetext{a}{PSC catalog designation.  The number corresponds to the sexagesimal right ascension and declination in J2000 coordinates formatted as hhmmss[.]ss$\pm$ddmmss[.]s.}
\tablenotetext{b}{The seven variability ``flags''  defined in ${\S}$3.2.   A `1'  is an identification of variability for the particular band and criteria for that flag; a `0' is listed otherwise.  Flags 1-3 correspond to $>5\sigma$ flickering in J, H and K$_{s}$ -- the standard deviation of the source variability that reproduces the observed photometric scatter in excess of the expected photometric scatter (equivalent to $\sigma_{var,m,n}\pm\nu_{var,m,n}$ in the text).  Flags 4-6 identify the number of excursions in J, H and K$_{s}$ -- individual scan groups that are $>5\sigma$ deviant from the mean magnitude.  Flag 7 corresponds to variables with Stetson indices in excess of 0.2.}
\tablenotetext{c}{Magnitudes of ``flickering'' variability for flags 1-3.  3$\sigma$ upper-limits and detections are listed, but 5$\sigma$ is required to trigger the variability flags.  See ${\S}$3.2.1.}
\tablenotetext{d}{The number of excursions for flags 4-6.  The ``excursive'' scan groups need not be the same between bands.  A positive number corresponds to the largest excursion being fainter than the mean magnitude, and vice-versa for a negative number.  See ${\S}$3.2.2}
\tablenotetext{e}{A Stetson Index in excess of 0.2 triggers the Stetson variability flag.  See ${\S}$3.2.3.}
\end{deluxetable}

\begin{deluxetable}{l r r r r r r  r r  r r}
\rotate
\tablecolumns{11}
\tablewidth{0pc}
\tablecaption{Candidate Variables}
\tabletypesize{\scriptsize}
\tablehead{
\colhead{Name}	&	\colhead{Probability\tablenotemark{b}}	&	\colhead{Number}	&	\colhead{Flags\tablenotemark{c}}	&	\colhead{J Flicker\tablenotemark{d}}	&	\colhead{H Flicker\tablenotemark{d}} 	&	\colhead{K$_{s}$ Flicker\tablenotemark{d}}&	\colhead{J Excursion} & \colhead{H Excursion} & \colhead{K$_{s}$ Excursion} &  \colhead{Stetson}  \\
\colhead{``2MASS J''\tablenotemark{a}}	& \colhead{}  &	\colhead{of Flags}	&	\colhead{}	&	\colhead{(mag)}	&	\colhead{(mag)}	&	\colhead{(mag)}&	\colhead{Number\tablenotemark{e}} & \colhead{Number\tablenotemark{e}} &  \colhead{Number\tablenotemark{e}}  & \colhead{Index\tablenotemark{f}} 
}
\startdata
01542788+0027106	&	50\%	&	1	&	1000000	&	0.0487$\pm$0.0094	&	0.0526$\pm$0.0128	&	$<$ 0.0562	&	\nodata	&	\nodata	&	\nodata	&	0.019	\\
03320092+3727391	&	53\%	&	2	&	0100010	&	$<$ 0.0043	&	0.0175$\pm$0.002	&	$<$ 0.0027	&	\nodata	&	\nodata	&	+1	&	0.095	\\
04263108+0354319	&	38\%	&	4	&	0101101	&	0.007$\pm$0.0015	&	0.0165$\pm$0.0014	&	$<$ 0.0056	&	-2	&	-3	&	\nodata	&	0.295	\\
05570658+0024590	&	5\%	&	1	&	0001000	&	$<$ 0.0031	&	$<$ 0.0037	&	$<$ 0.0038	&	+1	&	\nodata	&	\nodata	&	0.112	\\
05571081+0002357	&	18\%	&	1	&	0000010	&	$<$ 0.0037	&	$<$ 0.0052	&	$<$ 0.0044	&	\nodata	&	\nodata	&	+1	&	0.143	\\
06295229-5953131	&	9\%	&	4	&	1100101	&	0.1285$\pm$0.0167	&	0.1083$\pm$0.0081	&	0.1115$\pm$0.0242	&	\nodata	&	-2	&	\nodata	&	0.401	\\
08254051-3913250	&	6\%	&	1	&	0001000	&	$<$ 0.0039	&	$<$ 0.0043	&	$<$ 0.0034	&	+1	&	\nodata	&	\nodata	&	0.113	\\
11215635-1321534	&	87\%	&	2	&	1000001	&	0.1725$\pm$0.0122	&	$<$ 0.3859	&	$<$ 0.6598	&	\nodata	&	\nodata	&	\nodata	&	0.207	\\
\enddata
 \tablenotetext{a}{PSC catalog designation.  The number corresponds to the sexagesimal right ascension and declination in J2000 coordinates formatted as hhmmss[.]ss$\pm$ddmmss[.]s.}
 \tablenotetext{b}{The estimated probability that the observed variability is intrinsic to the source, rather than due to seeing variations.  See ${\S}$3.3.1 for discussion.}
\tablenotetext{c}{The seven variability ``flags''  defined in ${\S}$3.2.   A `1'  is an identification of variability for the particular band and criteria for that flag; a `0' is listed otherwise.  Flags 1-3 correspond to $>5\sigma$ flickering in J, H and K$_{s}$ -- the standard deviation of the source variability that reproduces the observed photometric scatter in excess of the expected photometric scatter (equivalent to $\sigma_{var,m,n}\pm\nu_{var,m,n}$ in the text).  Flags 4-6 identify the number of excursions in J, H and K$_{s}$ -- individual scan groups that are $>5\sigma$ deviant from the mean magnitude.  Flag 7 corresponds to variables with Stetson indices in excess of 0.2.}
\tablenotetext{d}{Magnitudes of ``flickering'' variability for flags 1-3.  3$\sigma$ upper-limits and detections are listed, but 5$\sigma$ is required to trigger the variability flags.  See ${\S}$3.2.1.}
\tablenotetext{e}{The number of excursions for flags 4-6.  The ``excursive'' scan groups need not be the same between bands.  A positive number corresponds to the largest excursion being fainter than the mean magnitude, and vice-versa for a negative number.  See ${\S}$3.2.2}
\tablenotetext{f}{A Stetson Index in excess of 0.2 triggers the Stetson variability flag.  See ${\S}$3.2.3.}\end{deluxetable}

\begin{deluxetable}{lr|rr|rrrrrr|rr}
\tabletypesize{\scriptsize}
\tablecolumns{12}
\tablewidth{0pc}
\tablecaption{2MASS Calibration Database Sample Counts - All Sources}
\tablehead{
\colhead{} & \colhead{} & \colhead{} & \colhead{Candidate} & \multicolumn{6}{c}{Not Variable - Category\tablenotemark{c}}  & \colhead{SDSS} &	\colhead{SDSS} \\
\colhead{2MASS}	&	\colhead{Total}	&	\colhead{Variable\tablenotemark{a}}	&	\colhead{Variable\tablenotemark{b}} & \colhead{Total}	&	\colhead{i}	&	\colhead{ii}	&	\colhead{iii}	&	\colhead{iv}	 & \colhead{v}	 &\colhead{Imaging\tablenotemark{d}} & \colhead{Spectra\tablenotemark{d}}
}
\startdata
ALL\tablenotemark{e}	&	7554	&	247	&	58	&	7248 	&	4961 &	345	&	857	&	1079 &	6	&	1742 	&	119	\\
$\left|b\right|<20\degree$\tablenotemark{e}	&	1926	&	73	&	42	&	1811	&	431	&	191	&	557	&	630	&	2	&	0	&	0	\\
$\left|b\right|>20\degree$\tablenotemark{e}	&	5628	&	174	&	16	&	5437	&	4531	&	154	&	300	&	449	&	4	&	1742  	&	119	\\
90021	&	141	&	6	&	0	&	135	&	75	&	3	&	17	&	40	&	0	&	\nodata	&	\nodata	\\
90294	&	108	&	3	&	0	&	105	&	83	&	9	&	6	&	7	&	0	&	\nodata	&	\nodata	\\
90004	&	115	&	7	&	1	&	107	&	65	&	4	&	11	&	27	&	0	&	115	&	14	\\
90301	&	114	&	0	&	0	&	114	&	106	&	1	&	5	&	2	&	0	&	\nodata	&	\nodata	\\
90247	&	56	&	1	&	1	&	54	&	21	&	10	&	13	&	10	&	0	&	\nodata	&	\nodata	\\
\enddata
 \tablenotetext{a}{These sources meet a minimum of 1 (of 7) variability criteria as outlined in ${\S}$3.2, and the observed variability is not correlated with seeing  (see ${\S}$3.4).}
\tablenotetext{b}{These sources meet a minimum of 1 (of 7) variability criteria as outlined in ${\S}$3.2, and the observed variability is partially correlated with seeing (see ${\S}$3.4)}
\tablenotetext{c}{We classify these sources as not variables, but sub-classify them into the following five categories -- `i' sources are not correlated with seeing and meet 0 (of 7) variability criteria; `ii' sources are partially correlated with seeing and meet 0 variability criteria; `iii' sources are correlated with seeing variations and meet 0 variability criteria; `iv' sources are correlated with seeing and meet 1 or more variability criteria; `v' sources meet 1 or more variability criteria but the observed variability is due to latent images from bright stars overlaid on the source every other scan (see ${\S}$2.3.1).}
\tablenotetext{d}{Cal-PSWDB sample targets coincident to within 5" with a SDSS DR5 object.}
\tablenotetext{e}{Aggregate counts for all, $\left|b\right|<20\degree$, and $\left|b\right|>20\degree$ fields.}
\end{deluxetable}

\begin{deluxetable}{lr|rr|rrrrrr|rr}
\tabletypesize{\scriptsize}
\tablecolumns{12}
\tablewidth{0pc}
\tablecaption{2MASS Calibration Database Sample Counts - Identified Extragalactic Sources}
\tablehead{
\colhead{} & \colhead{} & \colhead{} & \colhead{Candidate} & \multicolumn{6}{c}{Not Variable - Category\tablenotemark{c}}  & \colhead{SDSS} &	\colhead{SDSS} \\
\colhead{2MASS}	&	\colhead{Total}	&	\colhead{Variable\tablenotemark{a}}	&	\colhead{Variable\tablenotemark{b}} & \colhead{Total}	&	\colhead{i}	&	\colhead{ii}	&	\colhead{iii}	&	\colhead{iv}	 & \colhead{v}	 &\colhead{Imaging\tablenotemark{d}} & \colhead{Spectra\tablenotemark{d}}
}
\startdata
ALL\tablenotemark{e}	&	654	&	5	&	3	&	646	&	149	&	60	&	136	&	300	&	1	&	422	&	113	\\
$\left|b\right|<20\degree$\tablenotemark{e}	&	9	&	0	&	0	&	9	&	0	&	0	&	2	&	7	&	0	&	0	&	0	\\
$\left|b\right|>20\degree$\tablenotemark{e}	&	645	&	5	&	3	&	637	&	149	&	60	&	134	&	293	&	1	&	422	&	113	\\
90021	&	50	&	0	&	0	&	50	&	8	&	1	&	5	&	36	&	0	&	\nodata	&	\nodata	\\
90294	&	30	&	1	&	0	&	29	&	12	&	6	&	4	&	7	&	0	&	\nodata	&	\nodata	\\
90004	&	49	&	0	&	1	&	48	&	7	&	4	&	11	&	26	&	0	&	49	&	14	\\
90301	&	20	&	0	&	0	&	20	&	13	&	1	&	4	&	2	&	0	&	\nodata	&	\nodata	\\
90247	&	3	&	0	&	0	&	3	&	0	&	0	&	0	&	3	&	0	&	\nodata	&	\nodata	\\
\enddata
 \tablenotetext{a}{These sources meet a minimum of 1 (of 7) variability criteria as outlined in ${\S}$3.2, and the observed variability is not correlated with seeing  (see ${\S}$3.4).}
\tablenotetext{b}{These sources meet a minimum of 1 (of 7) variability criteria as outlined in ${\S}$3.2, and the observed variability is partially correlated with seeing (see ${\S}$3.4)}
\tablenotetext{c}{We classify these sources as not variables, but sub-classify them into the following five categories -- `i' sources are not correlated with seeing and meet 0 (of 7) variability criteria; `ii' sources are partially correlated with seeing and meet 0 variability criteria; `iii' sources are correlated with seeing variations and meet 0 variability criteria; `iv' sources are correlated with seeing and meet 1 or more variability criteria; `v' sources meet 1 or more variability criteria but the observed variability is due to latent images from bright stars overlaid on the source every other scan (see ${\S}$2.3.1).}
\tablenotetext{d}{Cal-PSWDB sample targets coincident to within 5" with a SDSS DR5 object.}
\tablenotetext{e}{Aggregate counts for all, $\left|b\right|<20\degree$, and $\left|b\right|>20\degree$ fields.}
\end{deluxetable}

\clearpage
\newpage
\begin{deluxetable}{lrrlr p{3.1cm} r}
\tablecolumns{7}
\tablewidth{0pc}
\tablecaption{Periodic Variables}
\tabletypesize{\scriptsize}
\tablehead{
\colhead{Name}	& \colhead{Flags}&	\colhead{Period}	&	\colhead{Phase\tablenotemark{c}}&\colhead{Field}  & \colhead{Object} & \colhead{Figure} \\
\colhead{``2MASS J''\tablenotemark{a}}	& \colhead{} &	\colhead{(Days)\tablenotemark{b}}	&	\colhead{} &	\colhead{} & \colhead{Type\tablenotemark{d}} & \colhead{Numbers}
}
\startdata
01542930+0053266	&	1111110	&	2.639	&	0.07	&	90004	&	M-type eclipsing binary\tablenotemark{f}	&	22, 45	\\
01545296+0110529	&	1100000	&	0.18603\tablenotemark{e}	&	0.95	&	90004	&	suspected CV	&	23, 46	\\
04261603+0323578	&	1001110	&	0.8832, 1.7664	&	0.487	&	90191	&	M-type eclipsing binary\tablenotemark{f}	&	24, 47	\\
04261900+0314008 &	0001000 & 1.07631, 2.15262 & 0.425 & 90191 & M-type eclipsing binary\tablenotemark{f} & 25, 48\\ 
08255405-3908441	&	1111111	&	8.08986	&	0.825	&	90312	&	eclipsing binary	&	26, 49	\\
08512729+1211484	&	0001110	&	1.23725	&	0.7	&	90067	&	sinusoidal	&	27, 50	\\
15001192-0103090	&	1111111	&	3.262	&	0.11	&	90868	&	sinusoidal	&	28, 51	\\
16271273-2504017	&	1111111	&	0.831445	&	0.26	&	90009	&	YSO	&	29, 52	\\
\enddata
 \tablenotetext{a}{PSC catalog designation.  The number corresponds to the sexagesimal right ascension and declination in J2000 coordinates formatted as hhmmss[.]ss$\pm$ddmmss[.]s.}
 \tablenotetext{b}{Uncertainty in period is approximately $\pm1$ in the least significant digit.  When two periods are quoted, the system is an eclipsing binary with indistinguishable eclipses, or an inconclusive Object Type.  See Figures 22-44 for light curves and 45-67 for color curves of these objects folded to these periods.}
 \tablenotetext{c}{Corresponding phase in Figures 22-67 for the Julian Date = 2450000.0.  When two periods are listed, this phase is for the larger period, which is also the one plotted in Figures 33,34, and 38.}
 \tablenotetext{d}{Object Type inferred from period, colors, light curve appearance, and Galactic latitude. Sinusoidal objects and objects without any listed Object Type have inconclusive identifications from available data.}
 \tablenotetext{e}{Period from Andrew Becker.}
 \tablenotetext{f}{See ${\S}$4.2.1 for discussion of these objects.}
 \tablenotetext{g}{The identification of this period is uncertain.  The object exhibits variability on a 3-day timescale, but the observed variability is not entirely consistent with this period derived from our analysis.}
 \end{deluxetable}

\newpage\clearpage

\begin{deluxetable}{lrrrrl}
\tablecolumns{6}
\tablewidth{0pc}
\tablecaption{Variables With Large Skew In Observed Photometry}
\tabletypesize{\scriptsize}
\tablehead{
\colhead{Name}	& \colhead{Flags}&	\colhead{J Skew\tablenotemark{b}}	&	\colhead{H Skew\tablenotemark{b}}	&	\colhead{K$_{s}$ Skew\tablenotemark{b}} & \colhead{Period (days)/}	\\
\colhead{``2MASS J''\tablenotemark{a}}	& \colhead{} &	\colhead{}	&	\colhead{}	&	\colhead{} & \colhead{Probability\tablenotemark{c}}
}
\startdata
00242118-0219071	&	0001100	&	2.6	&	1.97	&	0.55	&	\nodata	\\
01542930+0053266	&	1111110	&	1.8	&	1.46	&	1.2	&	2.64 days	\\
08255405-3908441	&	1111111	&	4.28	&	3.63	&	3.47	&	8.09 days	\\
08512916+1207014	&	0001000	&	3.07	&	1.47	&	0.49	&	\nodata	\\
11214462-1341508	&	0001000	&	2.39	&	2.01	&	0.87	&	\nodata	\\
11215454-1340473	&	0001010	&	2.99	&	1.71	&	0.99	&	\nodata	\\
14410068-0011348	&	0100000	&	-0.02	&	-0.34	&	0.45	&	\nodata	\\
16271273-2504017	&	1111111	&	2.32	&	2	&	1.35	&	0.831 days	\\
\enddata
 \tablenotetext{a}{PSC catalog designation.  The number corresponds to the sexagesimal right ascension and declination in J2000 coordinates formatted as hhmmss[.]ss$\pm$ddmmss[.]s.}
 \tablenotetext{b}{For each source, we compute the skew of the photometric distribution in each band.  In this table we list sources with skews in J, H or K$_{s}$ in excess of 1 standard deviation from the means for all variables.  See ${\S}$4.3.1 for discussion.}
\tablenotetext{c}{Percentages are listed for candidate variables in the rightmost column, corresponding to the estimated probability that the observed variability is intrinsic to the source as opposed to correlated with seeing variations.  Periods, in days, are listed for periodic variables.} 
 \end{deluxetable}
 
\newpage\clearpage

 \begin{deluxetable}{lrrrrrrr}
\tabletypesize{\scriptsize}

\tablecolumns{8}
\tablewidth{0pc}
\tablecaption{Variability Flag Combination Source Counts}
\tablehead{
\colhead{}	& \colhead{} & \colhead{} & \multicolumn{3}{c}{False Positives\tablenotemark{b}} & \colhead{Partially}&\colhead{} \\
\cline{4-6}	
\colhead{}	&	\colhead{} & \colhead{Candidate} & \colhead{} &  \colhead{Statistical} & \colhead{Seeing} & \colhead{Seeing} & \colhead{}\\
\colhead{Flags\tablenotemark{a}}	&	\colhead{Variable} & \colhead{Variable} & \colhead{Persistence} &  \colhead{Fluctuation} & \colhead{Correlated} & \colhead{Correlated\tablenotemark{c}} & \colhead{Total}
}
\startdata
1******	&	113	&	23	&	6	&	73	&	824	&	28	&	1010	\\
**1*****	&	96	&	16	&	6	&	71	&	796	&	19	&	965	\\
***1****	&	59	&	16	&	4	&	41	&	555	&	19	&	660	\\
****1***	&	100	&	10	&	4	&	66	&	592	&	15	&	753	\\
*****1**	&	63	&	11	&	2	&	47	&	582	&	15	&	689	\\
******1*	&	71	&	17	&	2	&	67	&	381	&	27	&	517	\\
*******1	&	83	&	16	&	6	&	0	&	563	&	16	&	668	\\
\hline															
all 7	&	19	&	1	&	1	&	0	&	189	&	1	&	210	\\
\enddata
 \tablenotetext{a}{The seven variability ``flags''  defined in ${\S}$3.2.   A `1'  is an identification of variability for the particular band and criteria for that flag; a `0' is listed otherwise.  A `*' can either be a `0' or a `1'.}
\tablenotetext{b}{We include these columns as a note of caution to illustrate the number of sources with apparent variability that is not intrinsic to the source.  These sources meet one or more of the seven variability criteria, but are not included as variable since the detected variations are correlated with seeing.} 
\tablenotetext{c}{If a partially correlated seeing source is not a candidate variable, it is a false positive due to statistical fluctuations, or does not meet any variability criteria.}
 \end{deluxetable}

\newpage\clearpage

\begin{deluxetable}{lrrrrrrrrr}
\rotate
\tablecolumns{10}
\tablewidth{0pc}
\tablecaption{Extragalactic Variables}
\tabletypesize{\scriptsize}
\tablehead{
\colhead{Name\tablenotemark{a}}	& \colhead{Flags}&	\colhead{SDSS-u\tablenotemark{b}}	&	\colhead{SDSS-g}	&	\colhead{SDSS-r}	&	\colhead{SDSS-i} &	\colhead{SDSS-z} & \colhead{Redshift} & \colhead{Object} & \colhead{SDSS}\\
\colhead{(2MASS PSC Name)}	& \colhead{} &	\colhead{}	&	\colhead{}	&	\colhead{} 	&	\colhead{}  	&	\colhead{}  	&	\colhead{z} 	& \colhead{Type} &	\colhead{Spectra?}  
}
\startdata
[ICS96] 003106.6-393917	&	1000000	&	\nodata	&	\nodata	&	\nodata	&	\nodata	&	\nodata	&	2.2	& QSO &	N	\\
(2MASS J00333270-3922457) & & & & & & & & &\\
2QZ J144115.5-005726	&	1100000	&	17.801	&	16.362	&	15.873	&	15.707	&	15.653	&	2.258$\pm$0.005	& QSO &	N	\\
(2MASS J14411477-0057254) & & & & & & & & & \\
B3 1456+375                    & 1111111 & 20.666	& 19.977	& 19.363	& 18.938	& 18.531 & 0.3330$\pm$0.0002 & QSO & Y \\
(2MASS J14584479+3720216) & & & & & & & & & \\
RX J1631.3+2953	&	1011010	&	16.954	&	16.922	&	16.841	&	16.711	&	16.478	&	0.254	& QSO &	Y	\\
(2MASS J16312442+2953016) & & & & & & & & & \\
\enddata
 \tablenotetext{a}{Name is derived either from SDSS DR5 or NED.}
  \tablenotetext{b}{SDSS DR5 magnitudes (modelMag\_*)}
  \tablenotetext{c}{It is unclear for these sources whether the observed variability is due to intrinsic source variability or variability correlated with seeing variations.  The photometry for these sources exhibit moderate correlation with seeing variations (-0.235$<$r$<$-0.185).  Detrending the photometry will be needed to ascertain the source of the observed variability.  See ${\S}$5.1 for discussion.}
 
 \end{deluxetable}

\begin{deluxetable}{lllllp{4cm}}
\tablecolumns{6}
\tablewidth{0pc}
\tablecaption{Variable Classification Categories}
\tablehead{
\colhead{Category}	& \colhead{$\left|b\right|$} & \colhead{B-K$_{s}$\tablenotemark{a}}&	\colhead{H-K$_{s}$}	&	\colhead{J-H} & \colhead{Probable Object Type\tablenotemark{b}}
	}
\startdata
A & $>$20$\degree$ &$<$4.97 & $<$0.2 & \nodata & K-type star or earlier\\
B & $>$20$\degree$ &$<$4.97 & 0.2-0.7 & \nodata & Extragalactic, cataclysmic variable\\
C & $>$20$\degree$ &$<$4.97 &  $>$0.7 & \nodata & Extragalactic\\
D & $>$20$\degree$ &$>$4.97 & \nodata & $<$0.79 & M dwarf\\
E & $>$20$\degree$ & $>$4.97\tablenotemark{c} & \nodata & $<$0.79 & M dwarf, extragalactic\\
F & $>$20$\degree$ & $>$4.97 & \nodata & $>$0.79 & red giant\\
G & $<$20$\degree$ & \nodata & \nodata & \nodata & \nodata  \\
\enddata
\tablenotetext{a}{We use a B-K$_s$=4.97 for an M0 dwarf, from average colors of B-V=1.41 \citep[]{hartigan94} and a V-K=3.56 \citep[]{kirkpatrick94,white98}.  The H-K$_s$ color cuts follow from \citet[]{kirkpatrick94,white98}, and we determine the J-H cut by visual inspection of the mean Cal-PSWDB colors for giants.}
 \tablenotetext{b}{Assuming a source is galactic; we do not exclude the possibility that a variable in each category is an unidentified extragalactic source.}
 \tablenotetext{c}{There are no identified B magnitudes for these sources.  Instead we assume that B$>$21, and infer that B-K$_{s}>$4.97.  While this inference is consistent with the H-K$_{s}$ and J-H colors for these sources, the assumption might not be valid in all cases.}
 \end{deluxetable}

\begin{deluxetable}{l rrrrr p{4cm}}
\tablecolumns{7}
\tablewidth{0pc}
\tablecaption{Classification of Variables}
\tabletypesize{\scriptsize}
\tablehead{
\colhead{Name}	&	\colhead{K$_{s}$}	&	\colhead{B-K$_{s}$}	&	\colhead{H-K$_{s}$}	&	\colhead{J-H} 	&	\colhead{Color}&	\colhead{Object}   \\
\colhead{``2MASS J''\tablenotemark{a}} & \colhead{} &\colhead{} &\colhead{} &\colhead{} &\colhead{Category\tablenotemark{b}} & \colhead{Notes\tablenotemark{c}}
}
\startdata
00240949-0140038	&	14.289	&	4.71	&	0.209	&	0.624	&	B	&	\nodata	\\
00241293-0133302	&	14.091	&	\nodata	&	0.191	&	0.536	&	E	&	\nodata	\\
00242118-0219071	&	13.175	&	4.62	&	0.191	&	0.642	&	A	&	\nodata	\\
00243221-0216483	&	14.774	&	2.33	&	0.834	&	0.725	&	C	&	\nodata	\\
00243850-0129122	&	13.660	&	4.54	&	0.192	&	0.662	&	A	&	\nodata	\\
00243956-0220288	&	13.898	&	\nodata	&	0.228	&	0.622	&	E	&	\nodata	\\
00325745-3931347	&	11.824	&	5.78	&	0.256	&	0.592	&	D	&	\nodata	\\
00333270-3922457	&	11.403	&	6.20	&	0.268	&	0.605	&	D	&	QSO [ICS96] 003106.6-393917       	\\
\enddata
 \tablenotetext{a}{PSC catalog designation.  The number corresponds to the sexagesimal right ascension and declination in J2000 coordinates formatted as hhmmss[.]ss$\pm$ddmmss[.]s.}
\tablenotetext{b}{Categories `A',`B',`C',`D',`E' and `F' corresponds to sources in $\left|b\right|>20$ fields.  Category `G' corresponds to sources in $\left|b\right|<20$ fields.  Category `A' corresponds to sources with B-K$_{s}<$4.97 and H-K$_{s}<$0.2; `B' corresponds to sources with B-K$_{s}<$4.97 and 0.2$<$H-K$_{s}<$0.7; `C' corresponds to B-K$_{s}<$4.97 and H-K$_{s}>$0.8; `D' corresponds to B-K$_{s}>$4.97, excluding category  `F' sources; `E' sources have no identified B counterpart and typically H-K$_{s}>$0.2; `F' sources have colors consistent with evolved giants.  See Figures 68 and 69, and ${\S}$4.3.}
\tablenotetext{c}{This column lists such information as any SIMBAD object identification, any SDSS DR5 object identification (e.g. `STAR'), any coincident extragalactic source name, a period if the source is periodic, and the type of object if a reasonable determination can be made (e.g. `eclipsing binary').  This list is not complete.}
\end{deluxetable}

\begin{deluxetable}{l rrrrr p{4cm}}
\tablecolumns{7}
\tablewidth{0pc}
\tablecaption{Classification of Candidate Variables}
\tabletypesize{\scriptsize}
\tablehead{
\colhead{Name}	&	\colhead{K$_{s}$}	&	\colhead{B-K$_{s}$}	&	\colhead{H-K$_{s}$}	&	\colhead{J-H} 	&	\colhead{Color}&	\colhead{Object}   \\
\colhead{``2MASS J''\tablenotemark{a}} & \colhead{} &\colhead{} &\colhead{} &\colhead{} &\colhead{Category\tablenotemark{b}} & \colhead{Notes\tablenotemark{c}}
}
\startdata
01542788+0027106	&	15.242	&	4.16	&	0.517	&	0.691	&	B	&	APMUKS(BJ) B015153.87+001227.7	\\
03320092+3727391	&	6.625	&	5.78	&	0.222	&	0.670	&	G	&	NGC 1342 38 -- Star in Open Cluster	\\
04263108+0354319	&	7.496	&	6.00	&	0.216	&	0.825	&	F	&	red giant	\\
05570658+0024590	&	13.114	&	6.39	&	0.280	&	0.678	&	G	&	\nodata	\\
05571081+0002357	&	13.027	&	5.47	&	0.248	&	0.631	&	G	&	\nodata	\\
06295229-5953131	&	15.121	&	\nodata	&	-0.173	&	1.550	&	F	&	\nodata	\\
08254051-3913250	&	12.348	&	5.35	&	0.224	&	0.661	&	G	&	\nodata	\\
11215635-1321534	&	15.852	&	\nodata	&	0.611	&	0.483	&	E	&	\nodata	\\
\enddata
\tablenotetext{a}{PSC catalog designation.  The number corresponds to the sexagesimal right ascension and declination in J2000 coordinates formatted as hhmmss[.]ss$\pm$ddmmss[.]s.}\tablenotetext{b}{Categories `A',`B',`C',`D',`E' and `F' corresponds to sources in $\left|b\right|>20$ fields.  Category `G' corresponds to sources in $\left|b\right|<20$ fields.  Category `A' corresponds to sources with B-K$_{s}<$4.97 and H-K$_{s}<$0.2; `B' corresponds to sources with B-K$_{s}<$4.97 and 0.2$<$H-K$_{s}<$0.7; `C' corresponds to B-K$_{s}<$4.97 and H-K$_{s}>$0.8; `D' corresponds to B-K$_{s}>$4.97, excluding category  `F' sources; `E' sources have no identified B counterpart and typically H-K$_{s}>$0.2; `F' sources have colors consistent with evolved giants.  See Figures 68 and 69, and ${\S}$4.3.}
\tablenotetext{c}{This column lists such information as any SIMBAD object identification, any SDSS DR5 object identification (e.g. `STAR'), any coincident extragalactic source name, a period if the source is periodic, and the type of object if a reasonable determination can be made (e.g. `eclipsing binary').  This list is not complete.}
\end{deluxetable} 

\clearpage

\begin{figure}
\plotone{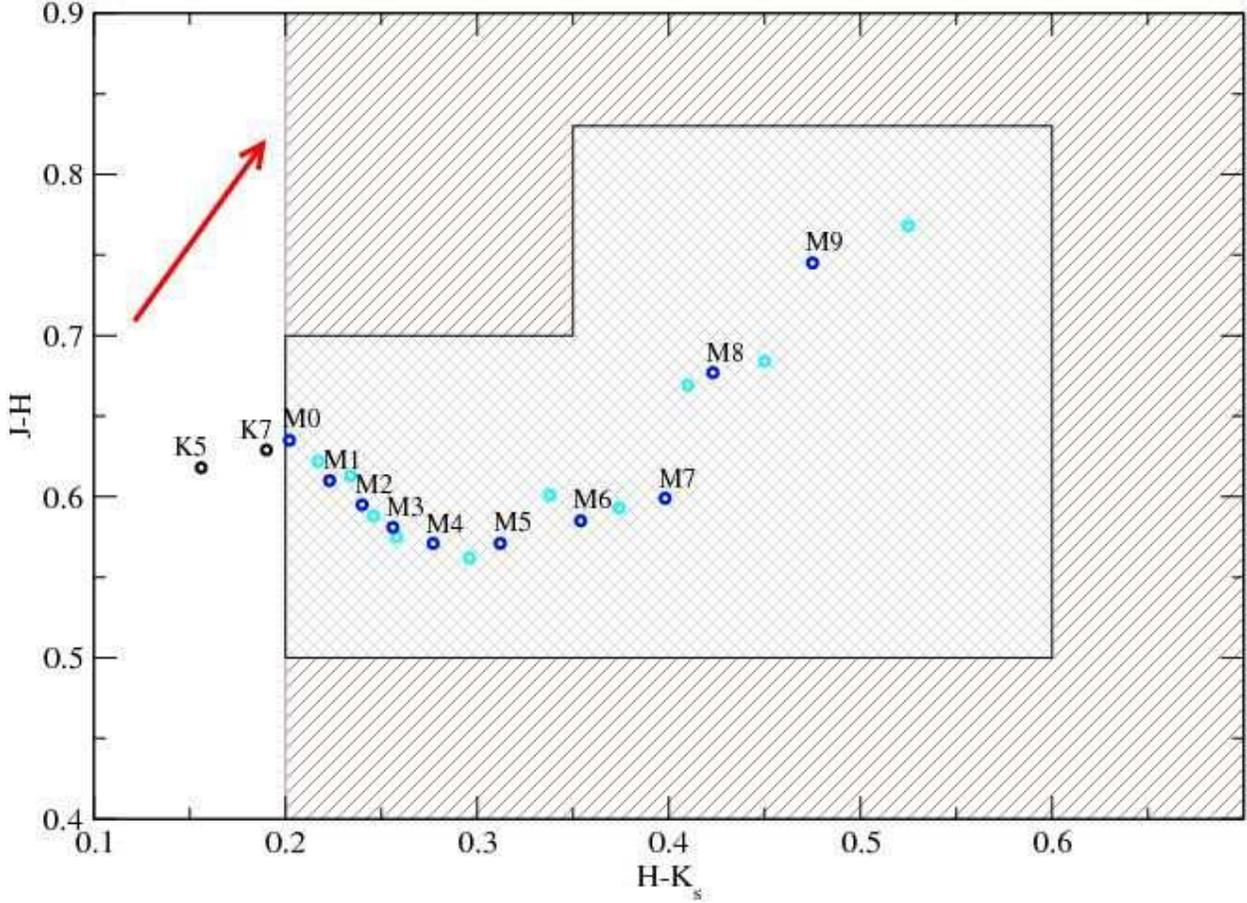}
\caption {Color cuts to identify M dwarfs in the 2MASS PSC.  For the 26 $\left|b\right|>20\degree$ fields, we select PSC sources with H-K$_{s} >$  0.2 to exclude bluer earlier-type stars.  For the remaining 8 calibration fields, we use the grey ``two-box'' cut, and an apparent magnitude cut ($K_{s}<14.3$).  These cuts are chosen to select nearby M dwarfs, and to exclude reddened background objects and red giants \citep[J-H$>$0.75,][]{allen}.  An A$_{v}$=1 interstellar reddening vector derived from \citet[]{cohen81,carpenter01b} is show in red.  Mean PSC colors for K5 and K7 dwarfs are shown with black circles, and for M dwarfs with blue (M0,M1,M2, etc.) and cyan (M0.5,M1.5,M2.5, etc.) circles. See ${\S}$2.1 for discussion \citep[]{cohen81,cutri06,kirkpatrick94}.}
\end{figure}
\clearpage
\begin{figure}
\plotone{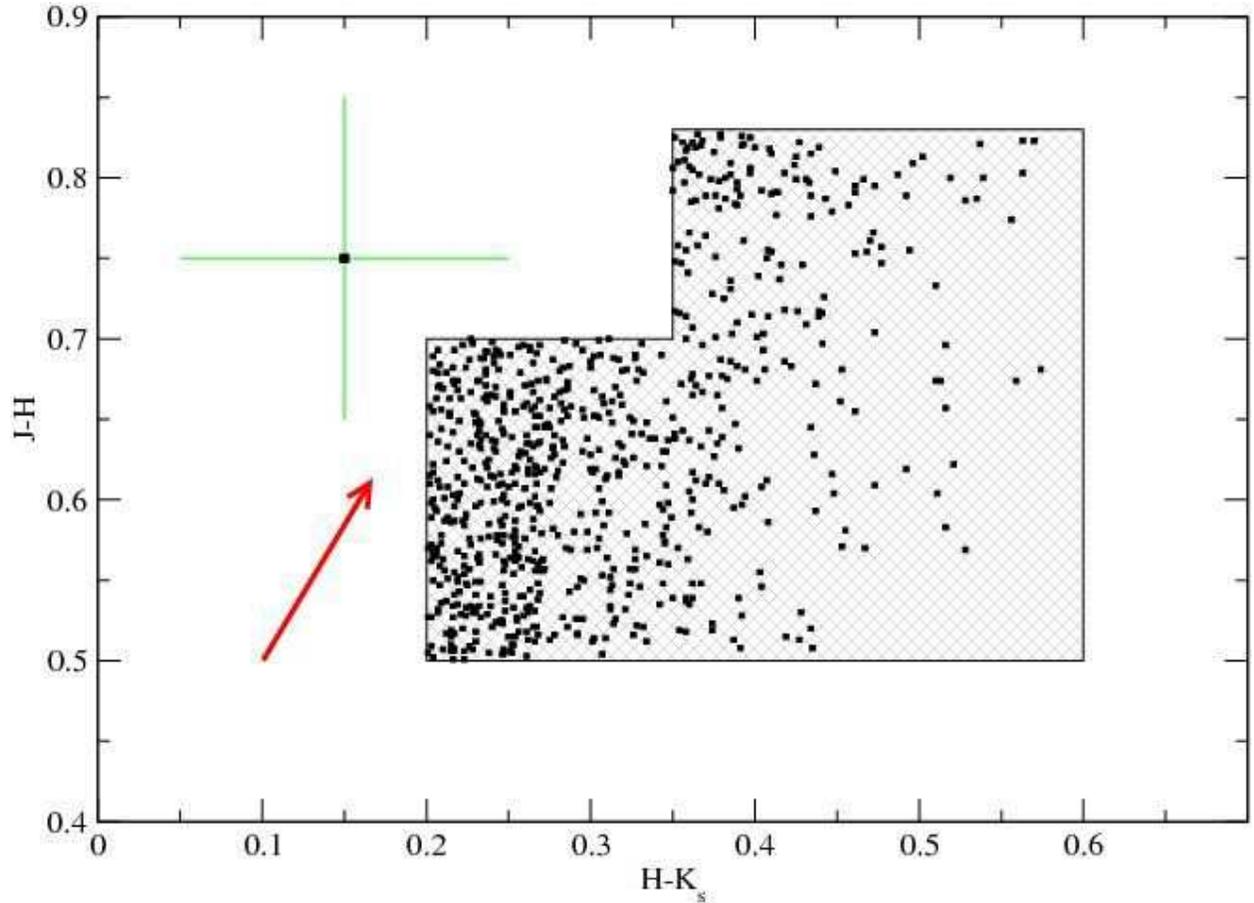}
\caption {PSC near-IR color-color diagram for targets in the 90547 calibration field.  This is a $\left|b\right|<20\degree$ field, so all targets are brighter than 14.3 at K$_{s}$ and lie within the ``two-box'' color cut.  Data are shown in black, with typical PSC error bars shown in green. This figure, along with Figure 5, illustrates the large uncertainty in the PSC colors relative to the Cal-PSWDB colors, and the relative difficulty in removing background contaminants from $\left|b\right|<20\degree$ fields. An A$_{v}$=1 interstellar reddening vector is shown in red.}
\end{figure}
\clearpage
\begin{figure}
\plotone{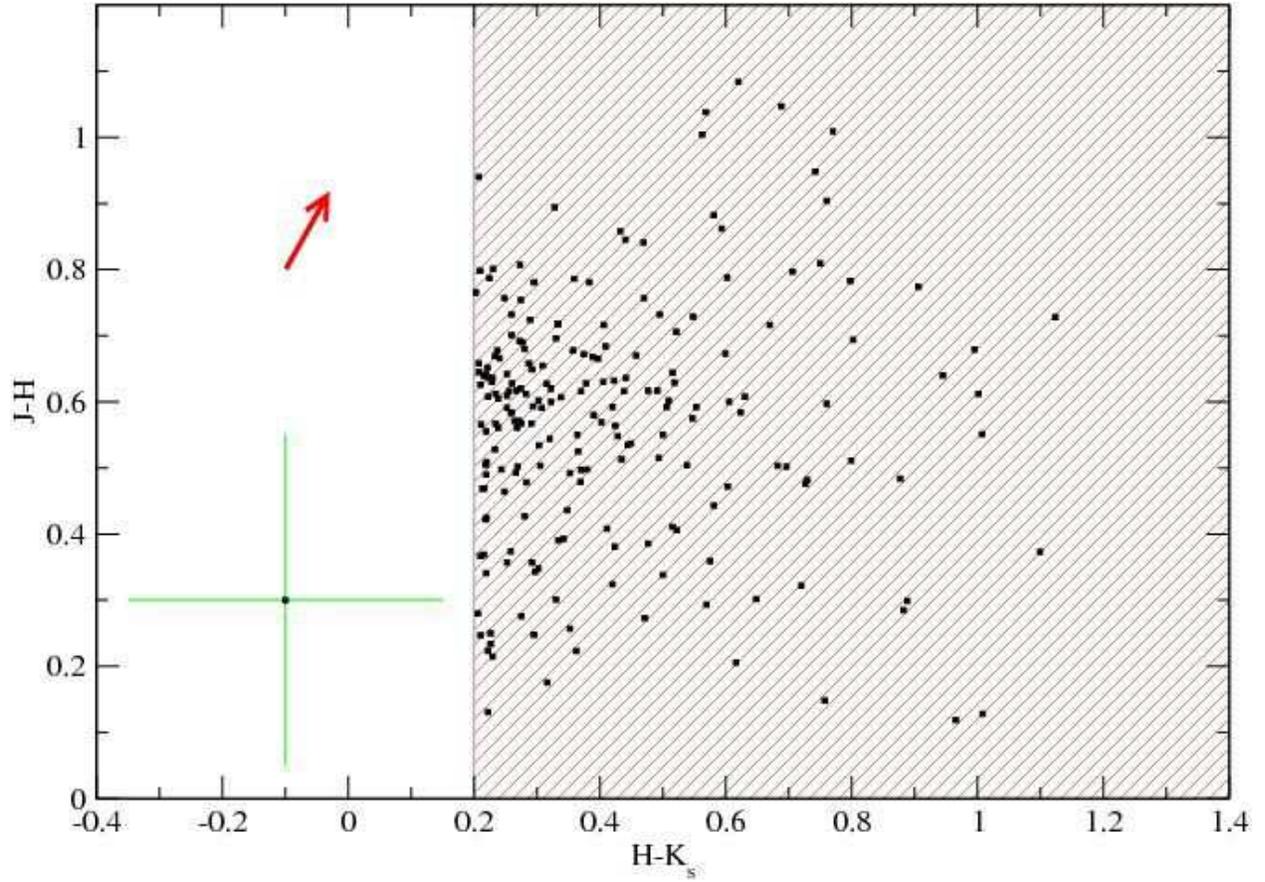}
\caption {PSC near-IR color-color diagram for targets in the 90182 calibration field, showing all PSC sources with H-K$_{s}>$0.2 in this $\left|b\right|>20\degree$ field.  Data are shown in black, with typical PSC error bars displayed in green.  An A$_{v}$=1 interstellar reddening vector is shown in red.}
\end{figure}
\clearpage
\begin{figure}
\plotone{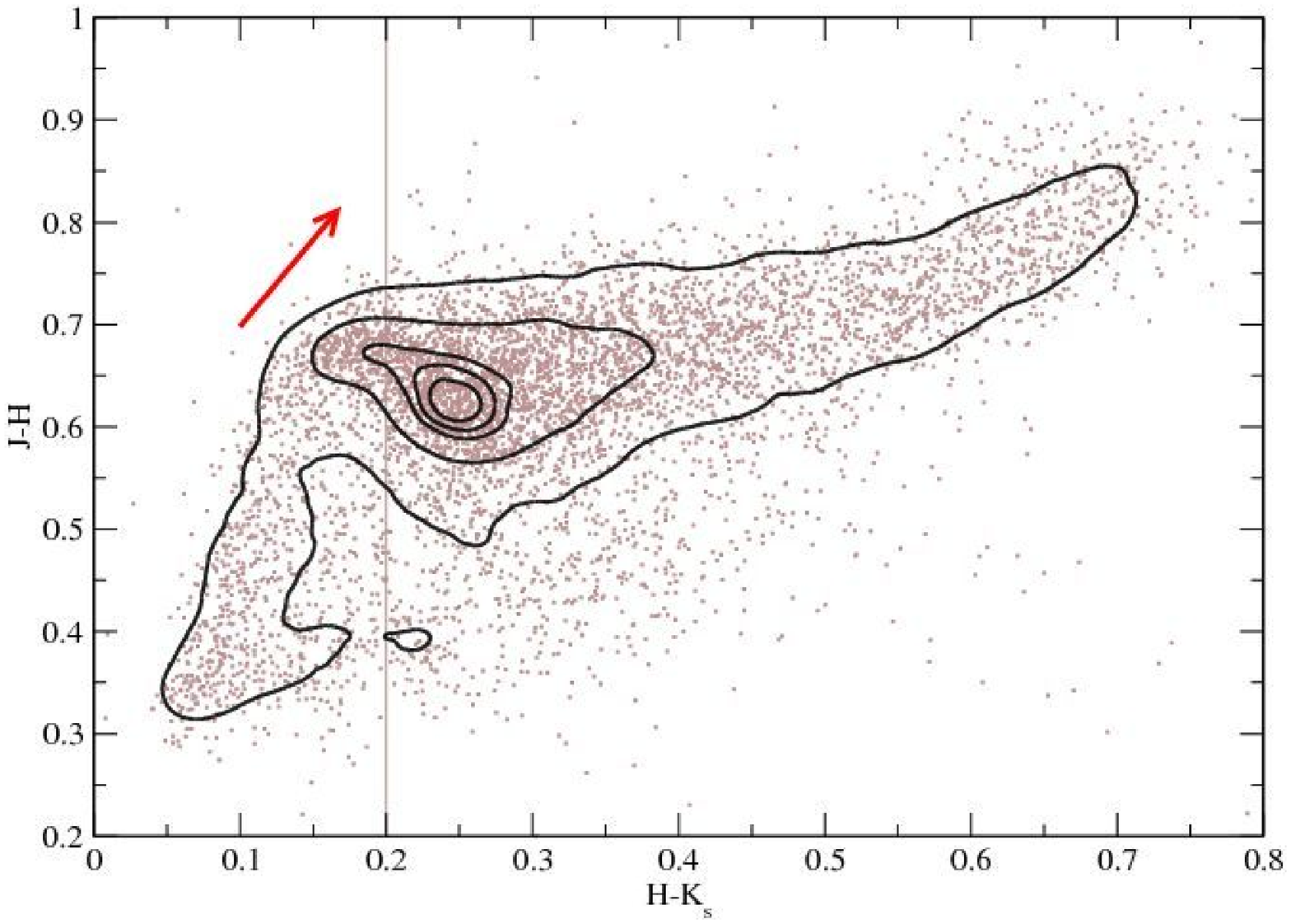}
\caption {Mean Cal-PSWDB colors for sources in sample with $\left|b\right|>20\degree$.  Data are shown in brown, and error bars are suppressed.  An A$_{v}$=1 interstellar reddening vector is shown in red.  Source density contours of 12\%,40\%,70\%,80\% and 90\% of the maximum source density are shown in black. The line at H-K$_{s}$=0.2 illustrates the PSC color selection criteria.  The Cal-PSWDB colors for many sources fall outside the PSC color selection criteria because of the larger errors in PSC photometry.}
\end{figure}
\clearpage
\begin{figure}
\plotone{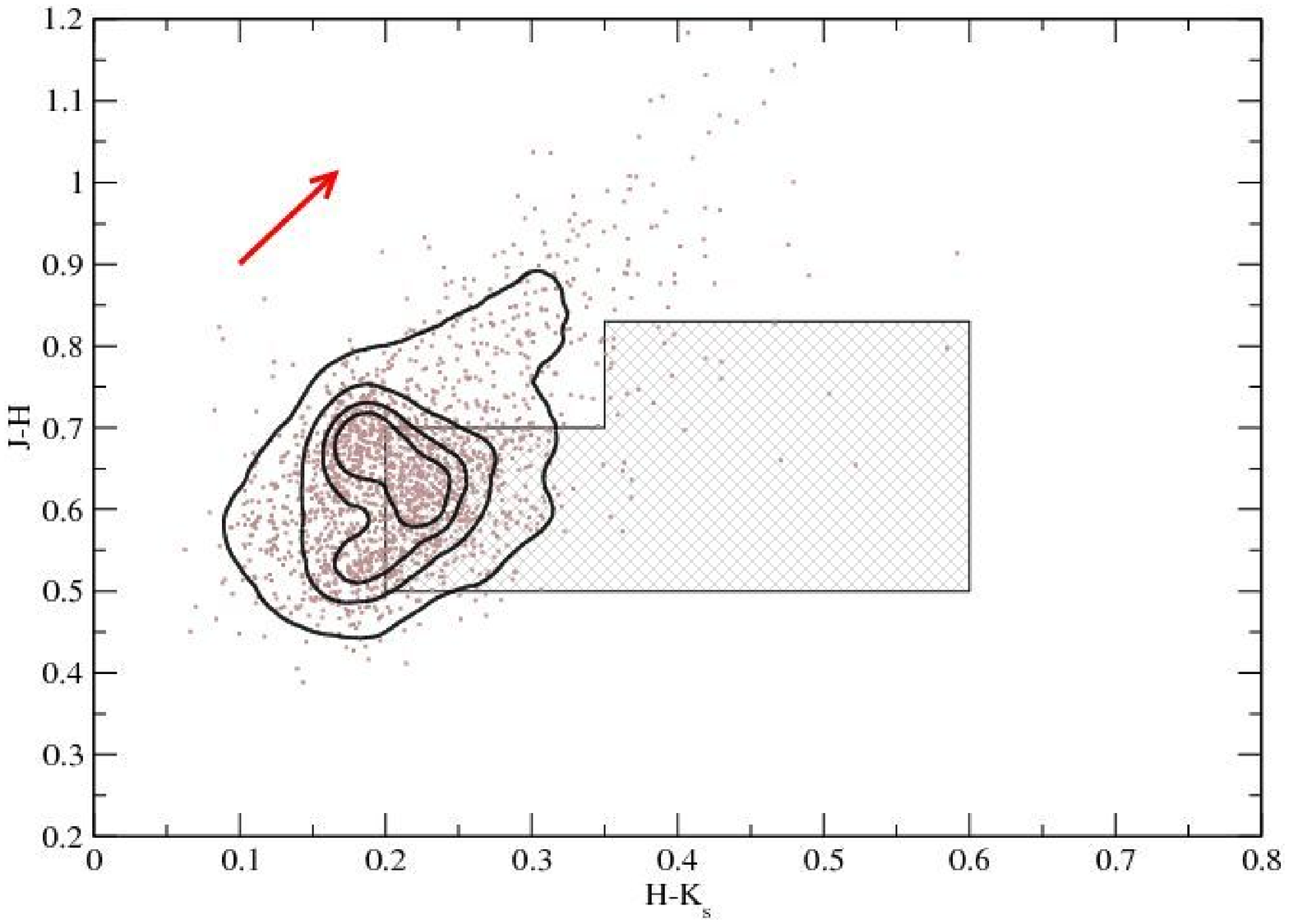}
\caption {Mean Cal-PSWDB colors for sources in sample with $\left|b\right|<20\degree$. Data are shown in brown, and error bars are suppressed. An A$_{v}$=1 interstellar reddening vector is shown in red.  Source density contours of 10\%,40\%,65\%, and 80\% of the maximum source density are shown in black. The box illustrates the PSC color selection criteria. The Cal-PSWDB colors for many sources fall outside the PSC color selection criteria because of the larger errors in PSC photometry.}
\end{figure}
\clearpage
\begin{figure}
\plotone{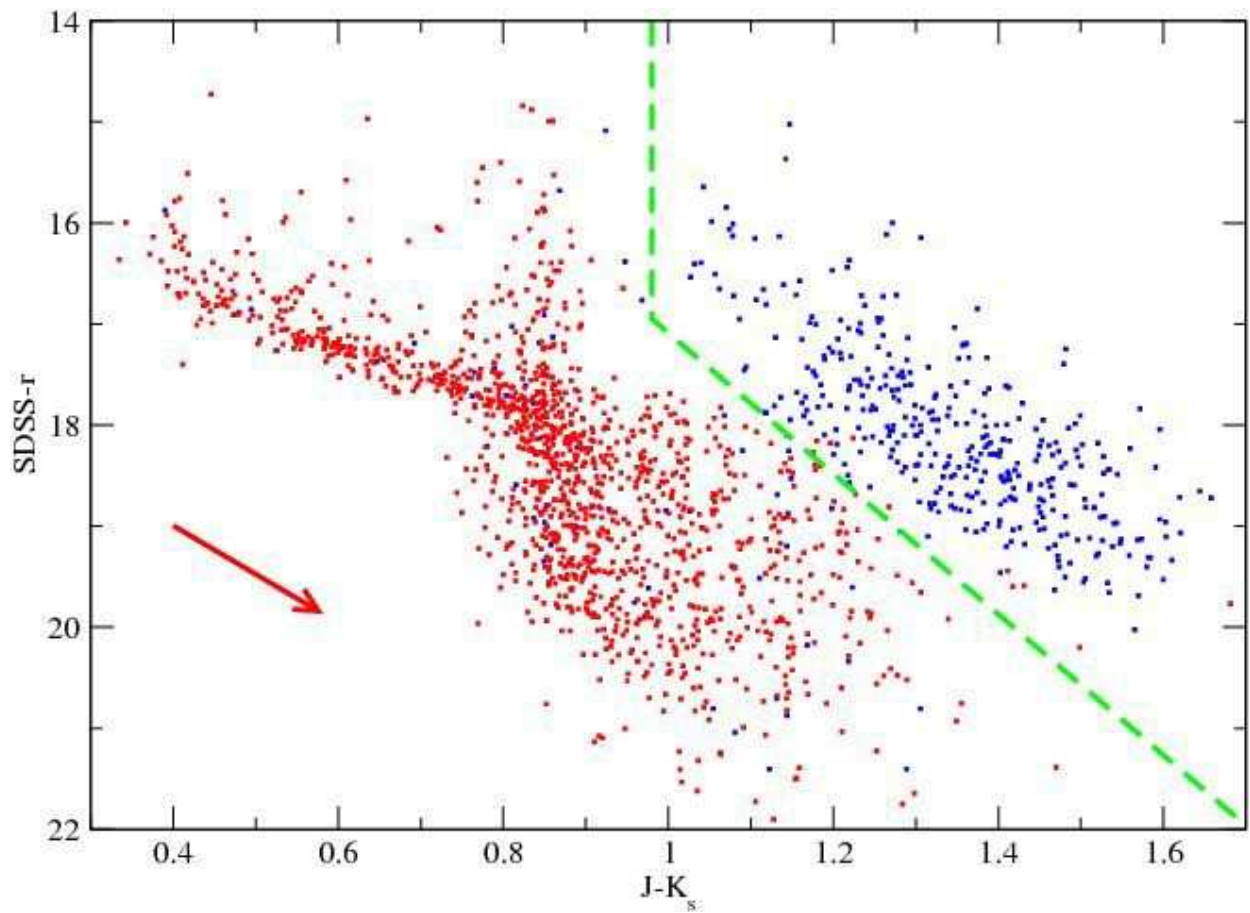}
\caption {For fields with SDSS DR5 coverage, we plot the average Cal-PSWDB J-K$_{s}$ color vs. SDSS r$^\prime$ apparent magnitude.  Photometric uncertainties are on average 1-3\% for both axes.  In blue are targets identified as extragalactic by any one of the following -- NED, the 2MASS XSC, and the SDSS star-galaxy classification routine.  The rest of the targets are displayed in red, and are identified as ``STAR'' by the SDSS star-galaxy classification routine.  Most galaxies  -- not including unresolved quasars, AGN, etc. -- separate from galactic sources with this choice of plotted parameters, qualitatively represented by the green dashed segmented line.  An A$_{V}$=1 interstellar reddening vector derived from \citet[]{carpenter01b,cohen81,schlegel98} is shown in red.  See ${\S}$3.1.3 for discussion.}
\end{figure}
\clearpage
\begin{figure}
\plotone{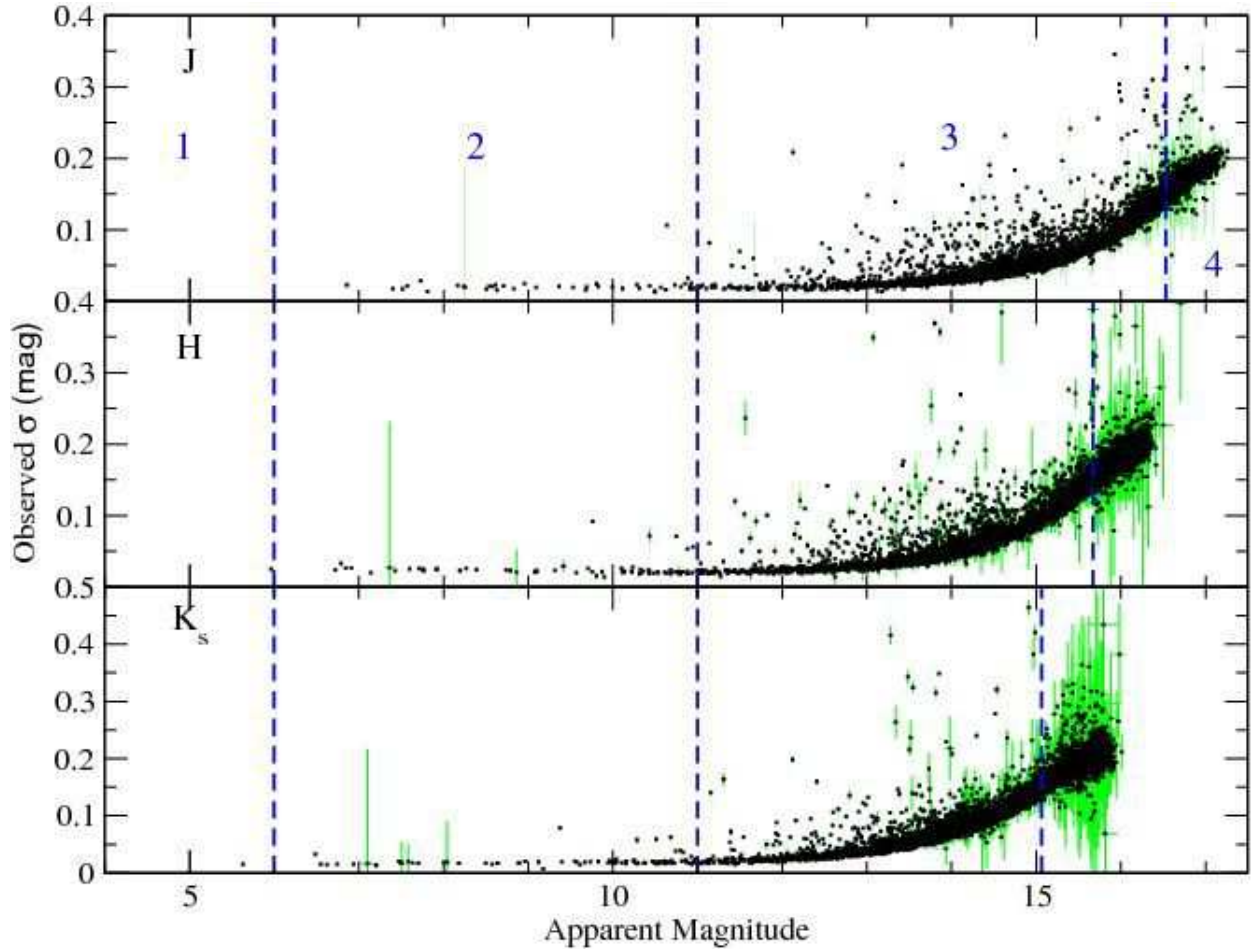}
\caption {Standard deviation of photometry, as a function of apparent magnitude, for each target in sample.  One-sigma uncertainties are shown in green.  Regions 1 through 4 are delineated in each band with blue dashed vertical lines, and numbered for J-band.  From this data we compute an expected standard deviation of photometry to measure variability of individual sources.  See ${\S}$3.1.4 for discussion.}
\end{figure}
\clearpage
\begin{figure}
\plotone{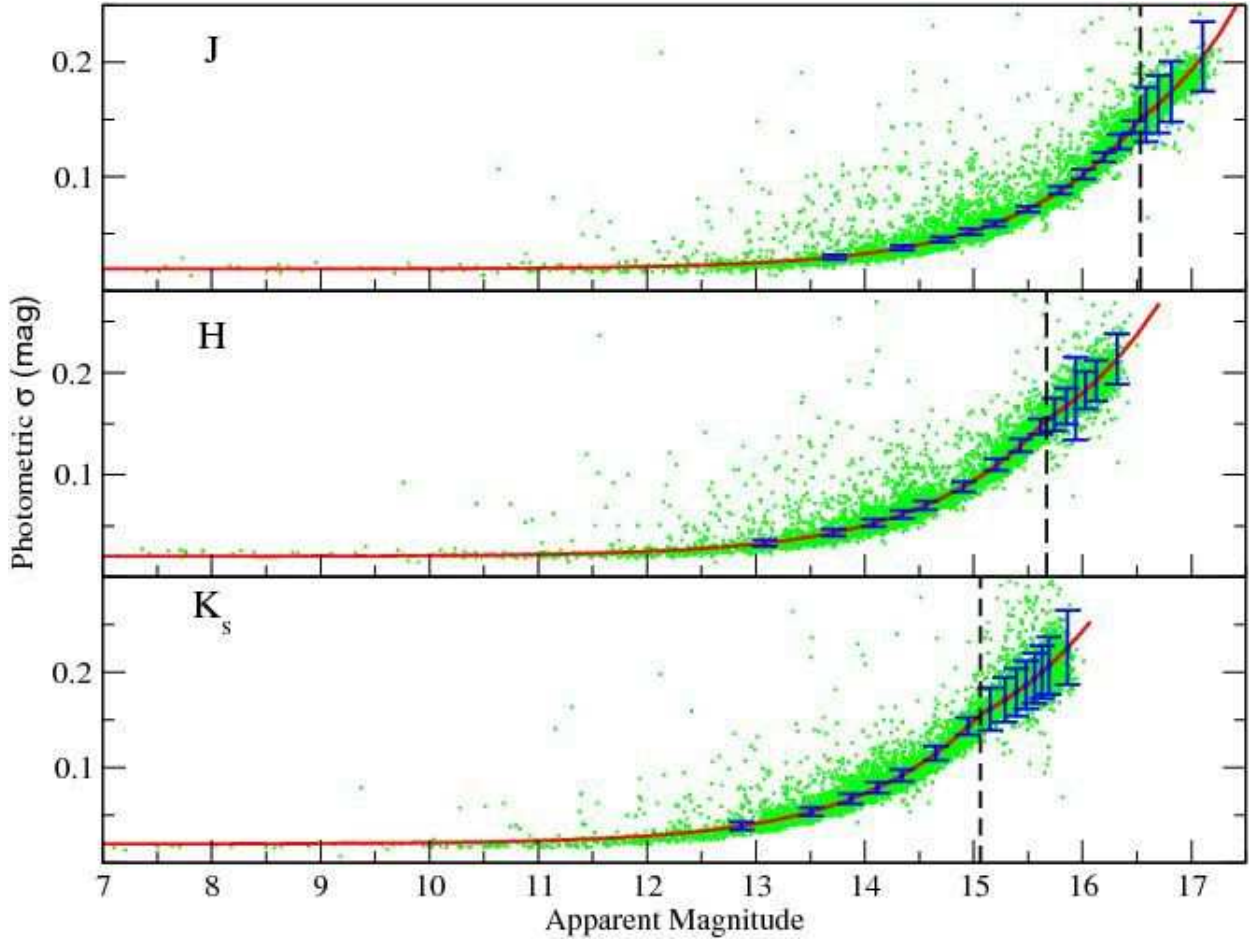}
\caption {Model of the standard deviation of photometry, as a function of apparent magnitude, shown in red.  At every 500th target apparent magnitude, 10-$\sigma$ model uncertainties (10$\nu_{m,model}$) are shown in blue.  Shown in green are the observed photometric standard deviations for the sample, with error bars suppressed for clarity. The black dashed vertical lines correspond to the ``break point'' magnitudes, where the linear relationship changes between photon statistics and flux overestimation from incomplete detections.  These occur at $J=16.53$, $H=15.67$, and $K_{s}=15.06$ and are within 1-$\sigma$ of the mean M$_{10}$ magnitudes, corresponding to a $>$$99\%$ completeness for SNR$>$10 in the PSC \citet[${\S}$VI.2]{cutri06}.  Propagated uncertainties are dominated by the uncertainty in the observed photometric scatter.  See ${\S}$3.1.4 for discussion.}
\end{figure}
\clearpage
\begin{figure}
\plotone{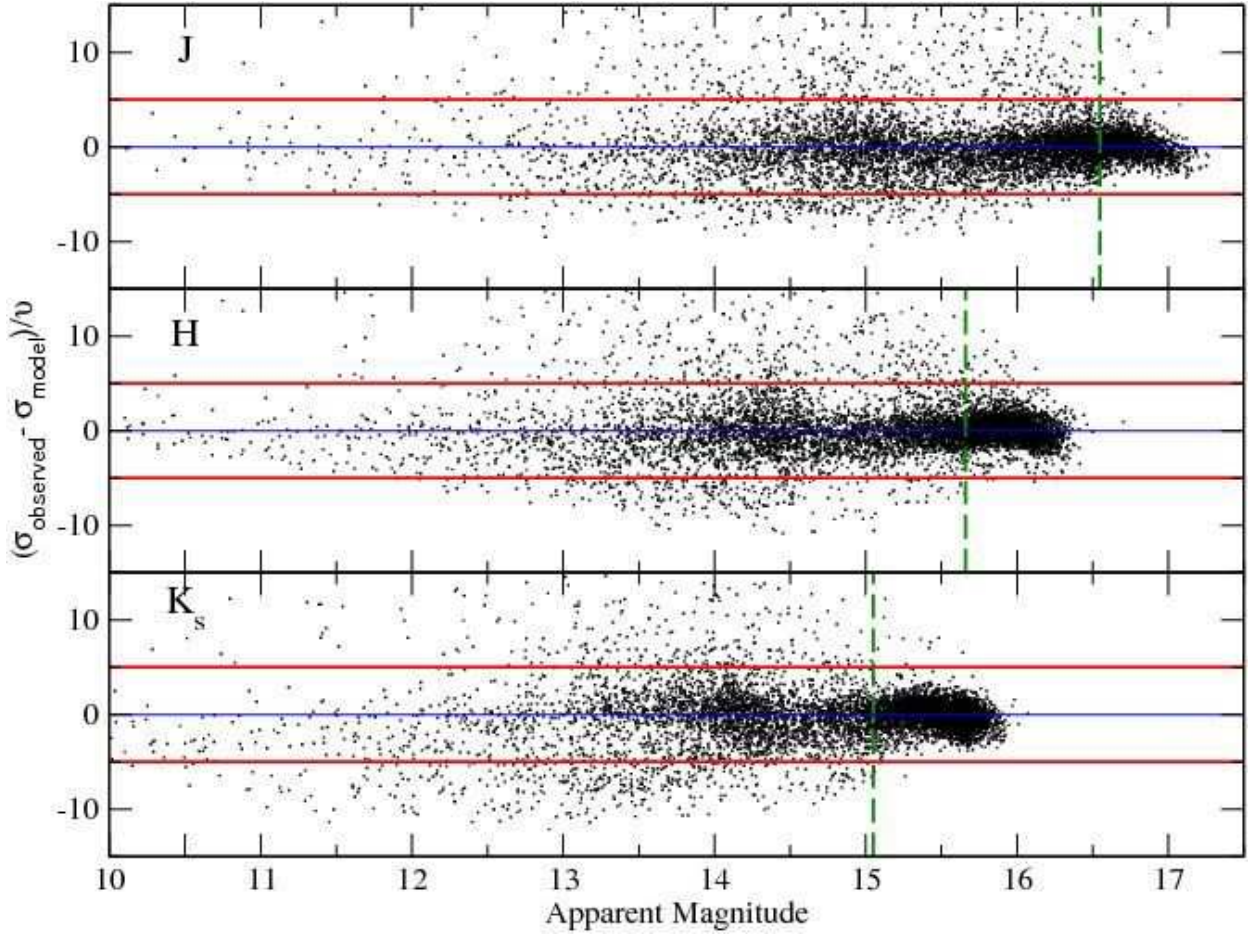}
\caption {As a function of apparent magnitude, we plot for each source the observed photometric scatter subtracted from the model, and divided by the propagated uncertainty ($\nu$ in the vertical axis title is equal to $\sqrt{\nu_{m,n}^2+\nu_{m,model}^2}$).  Note that this is not equivalent to the subtraction in quadrature for the identification of source variability.  Data are shown in black.  The green vertical dashed lines correspond to the ``break point'' magnitudes.  5-$\sigma$ horizontal lines are shown in red.  Sources on this plot with y-axis values of $<$-5 demonstrate that propagated uncertainties are underestimated by a factor of $<$2.  Since we do not identify a source of additional uncertainty, we do not apply a correction factor. Instead, in ${\S}$3.4.2, we ensure that the identified variables are statistically robust.  See ${\S}$3.1.4.}
\end{figure}
\clearpage
\begin{figure}
\plotone{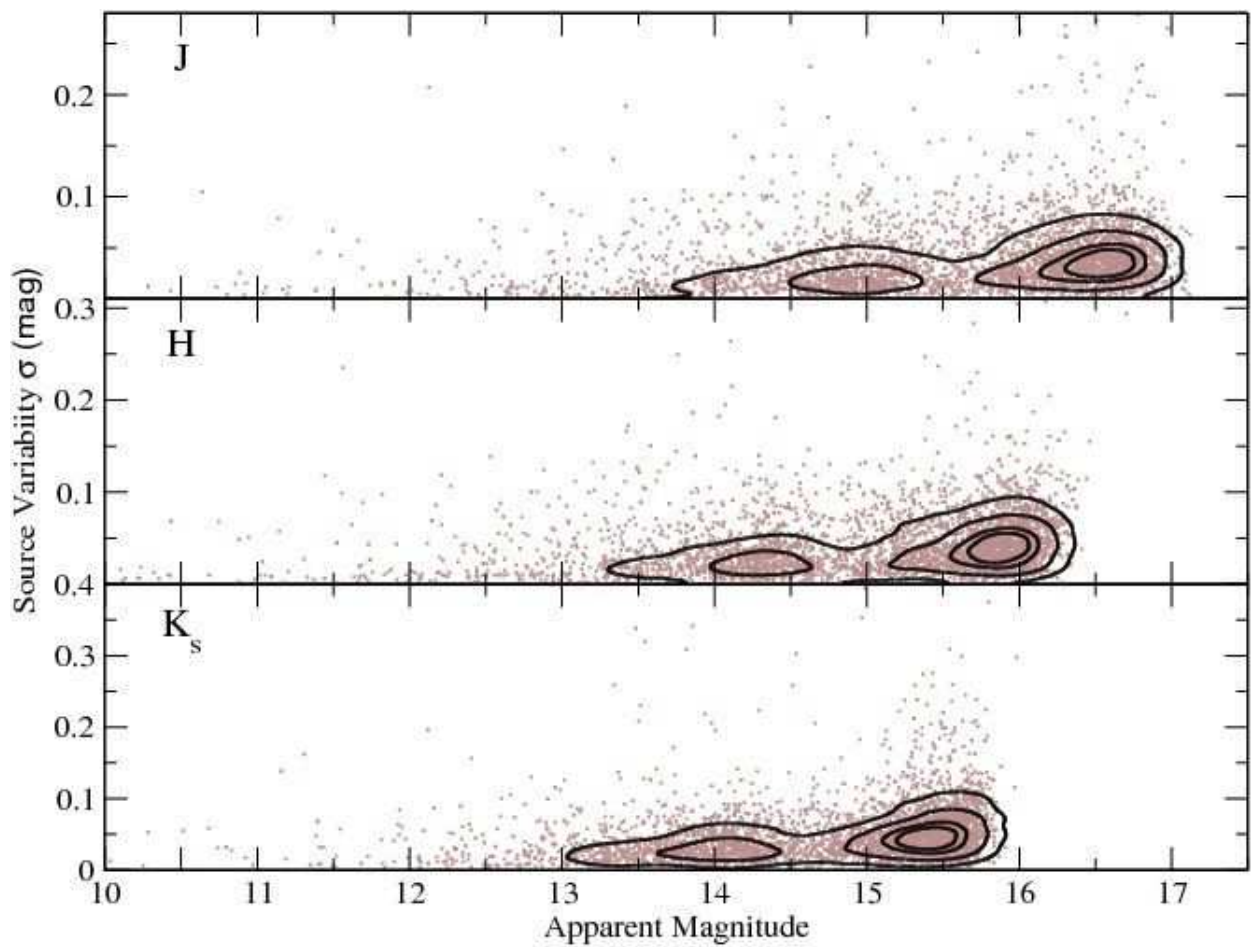}
\caption {For sources with $\sigma_{m,n}$$>$$\sigma_{m,model}$, we plot the standard deviation of photometric variability, $\sigma_{var,m,n}$, without ``false-positive'' variables removed, shown as a function of apparent magnitude.  We subtract the observed photometric scatter from the modeled measurement scatter in quadrature to arrive at a measure of the intrinsic source variability.  Data points are shown in brown, with 15\%,35\%,65\% and 80\% source density contours shown in black.  See ${\S}$3.2.1 for discussion.}
\end{figure}
\clearpage
\begin{figure}
\plotone{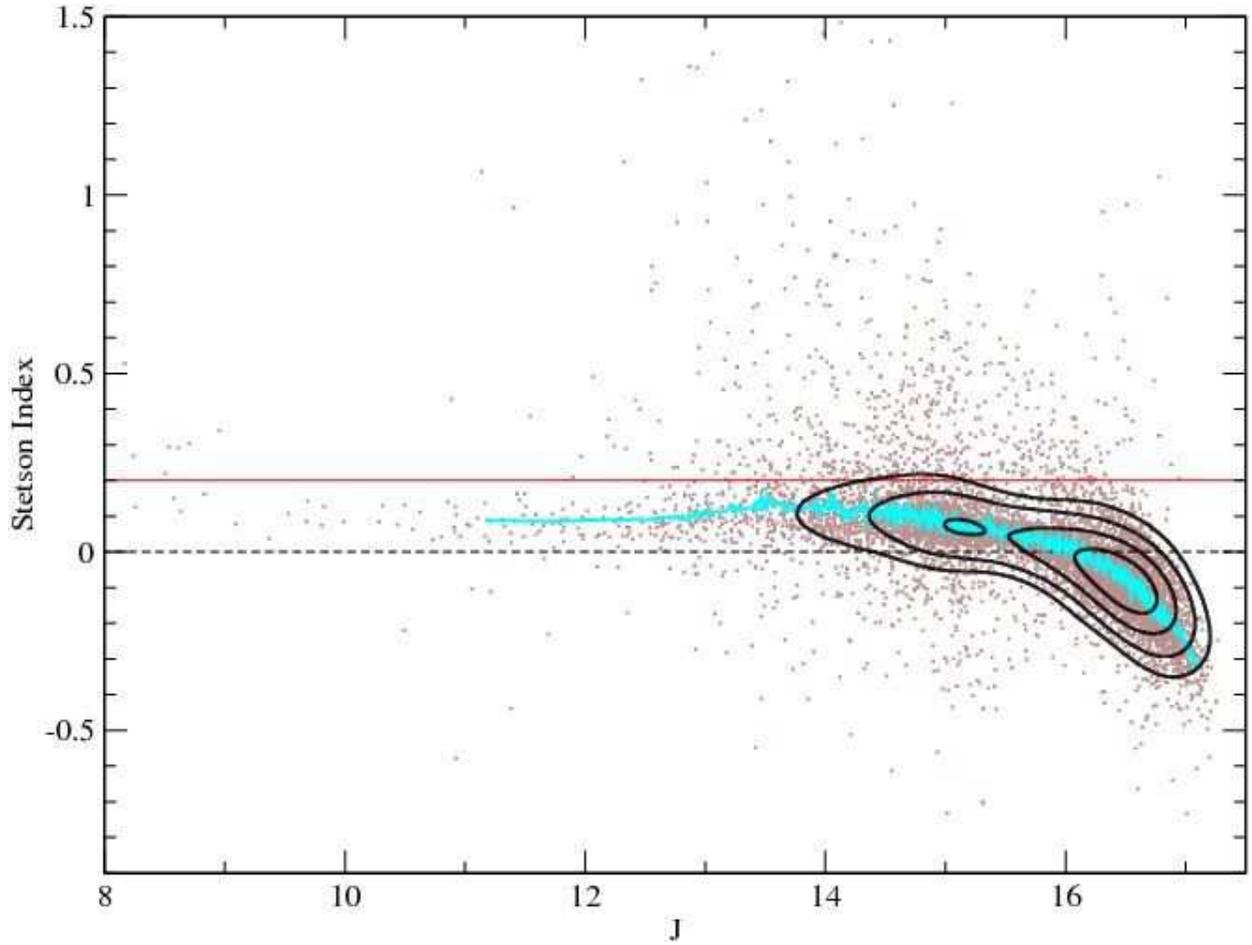}
\caption {Stetson index plotted for all 7554 sample targets as a function of apparent J-band magnitude, with data points in brown.  15\%,30\%,55\%, and 80\% source density contours are shown in black.  A horizontal red line is shown for a Stetson Index of 0.2, corresponding to the adopted line demarcating Stetson variables from non-variables. A running, 100-point median filter is overlaid in cyan. See ${\S}$3.2.3 for discussion.}
\end{figure}
\clearpage
\begin{figure}
\plotone{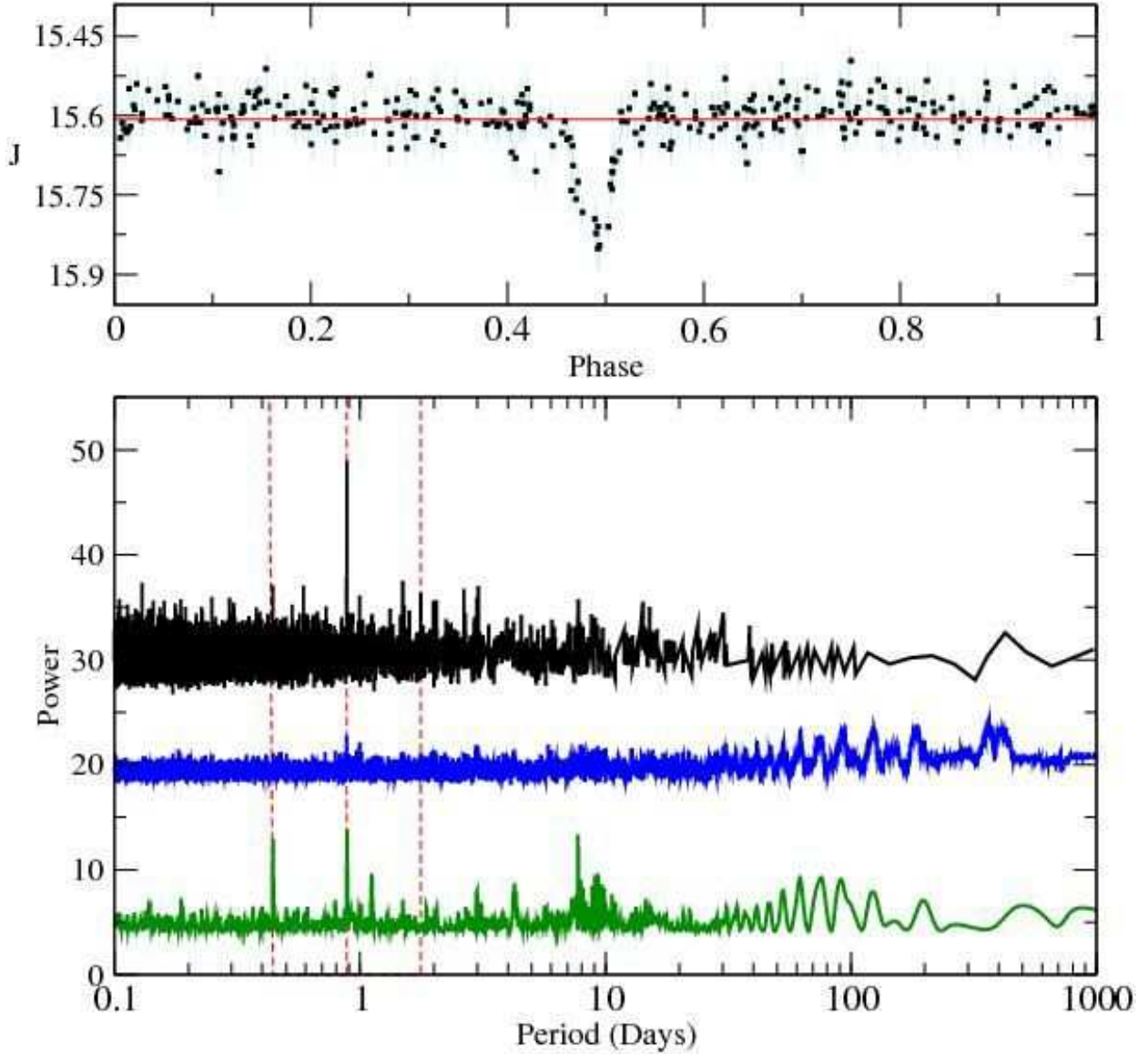}
\caption {Top Panel:  J-band Cal-PSWDB data for 2MASS J04261603+0323578 folded to a period of 0.8832 days.   Data are shown in black, with 1-$\sigma$ error bars in teal.  Each data point corresponds to the unweighted average of one group of six scans ($m_{n,t_6}$).  The mean apparent magnitude is shown with a red horizontal line.   Bottom Panel: Periodograms for 2MASS J04261603+0323578, calculated from J-band Cal-PSWDB data.  Before applying a relative offset for clarity, all periodograms have been shifted to a mean of 0 and normalized to a standard deviation of 1.  Black: our algorithm (1/$\chi_{25}^{2}$).  Blue: Box Least Squares.  Green: Lomb-Scargle.  Periods are on the horizontal axis in days, and the corresponding periodogram power is on the vertical axis.  Overlaid are dashed vertical red lines at 0.4416, 0.8832, and 1.7664 days -- half the period, the period, and twice the period discovered for this periodic variable. See ${\S}$3.3.3 for discussion.}
\end{figure}
\clearpage
\begin{figure}%

\plotone{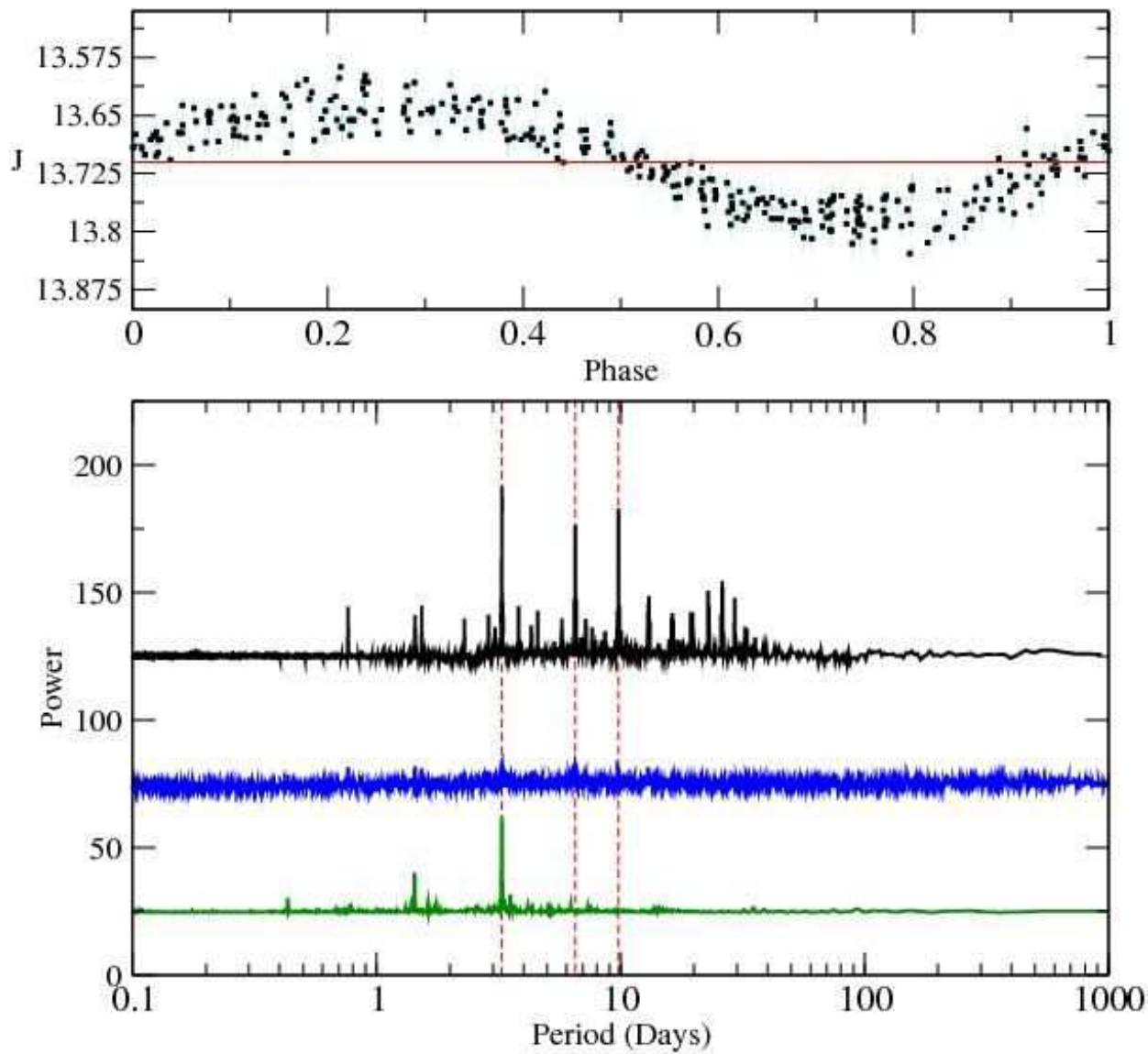}
\caption {Top Panel:  J-band Cal-PSWDB data for 2MASS J15001192-0103090 folded to a period of 3.262 days.   Data are shown in black, with 1-$\sigma$ error bars in teal.  Each data point corresponds to the unweighted average of one group of six scans ($m_{n,t_6}$).  The mean apparent magnitude is shown with a red horizontal line.   Bottom Panel: Periodograms for 2MASS J15001192-0103090, calculated from J-band Cal-PSWDB data.  Before applying a relative offset for clarity, all periodograms have been shifted to a mean of 0 and normalized to a standard deviation of 1.  Black: our algorithm (1/$\chi_{25}^{2}$).  Blue: Box Least Squares.  Green: Lomb-Scargle.  Periods are on the horizontal axis in days, and the corresponding periodogram power is on the vertical axis.  Overlaid are dashed vertical red lines at 3.262, 6.524, and 9.786 days -- the period, twice the period, and three times the period discovered for this periodic variable. See ${\S}$3.3.3 for discussion.}
\end{figure}
\clearpage
\begin{figure}
\plotone{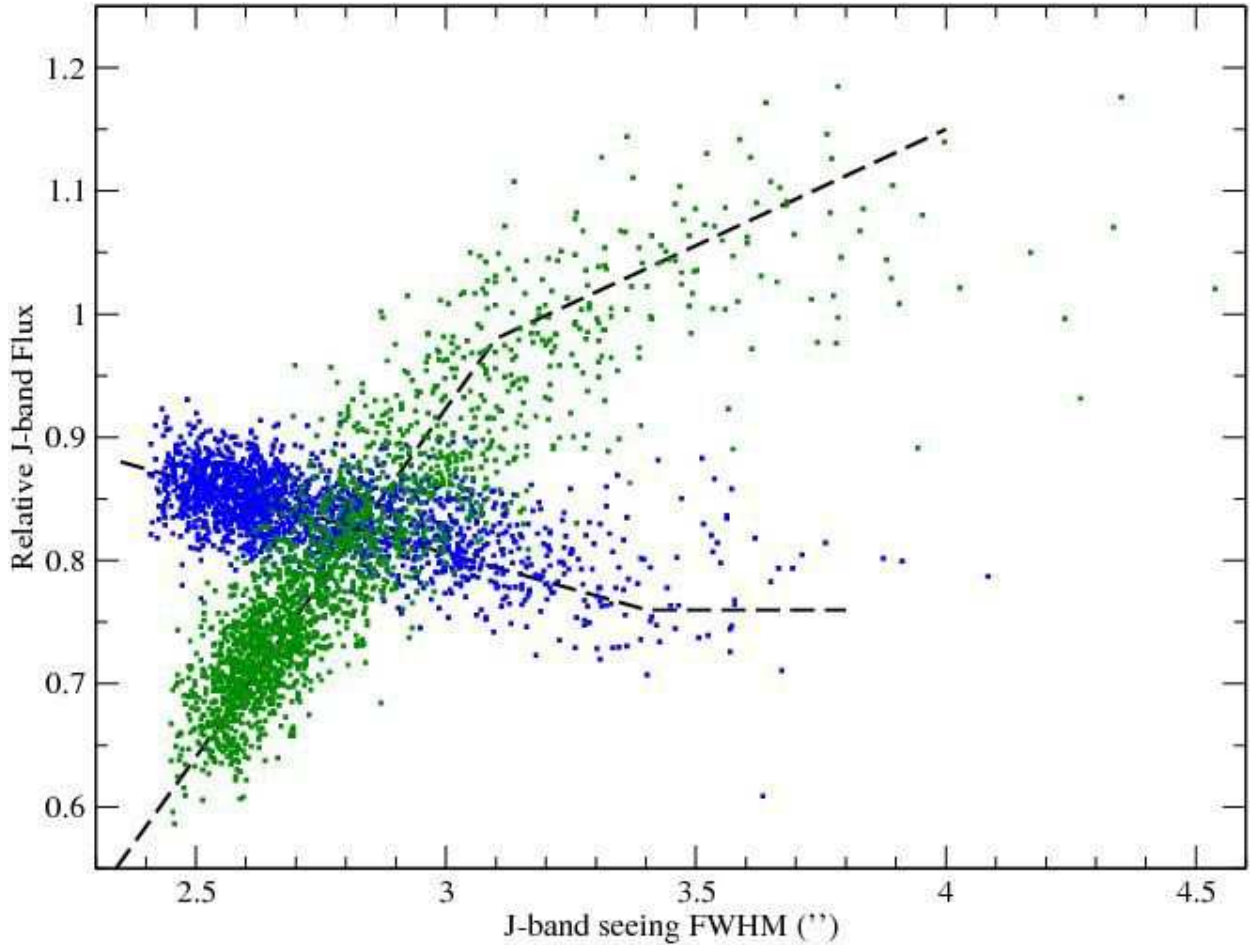}
\caption {Relative J-band flux (arbitrary units), plotted as a function of J-band seeing FWHM (arc-seconds), for  2MASS J19020417-0454110 (blue) and 2MASX J03320335+3658101(green) Cal-PSWDB data.  Error bars are suppressed.  Black dashed segmented lines illustrate the dependence of apparent magnitude with seeing.  These sources are extreme examples of (anti-)correlated variations observed in our sample, with J-band r-correlation statistics of 0.63 and -0.82 respectively.   See ${\S}$3.4 for discussion.}
\end{figure}
\clearpage
\begin{figure}
\plotone{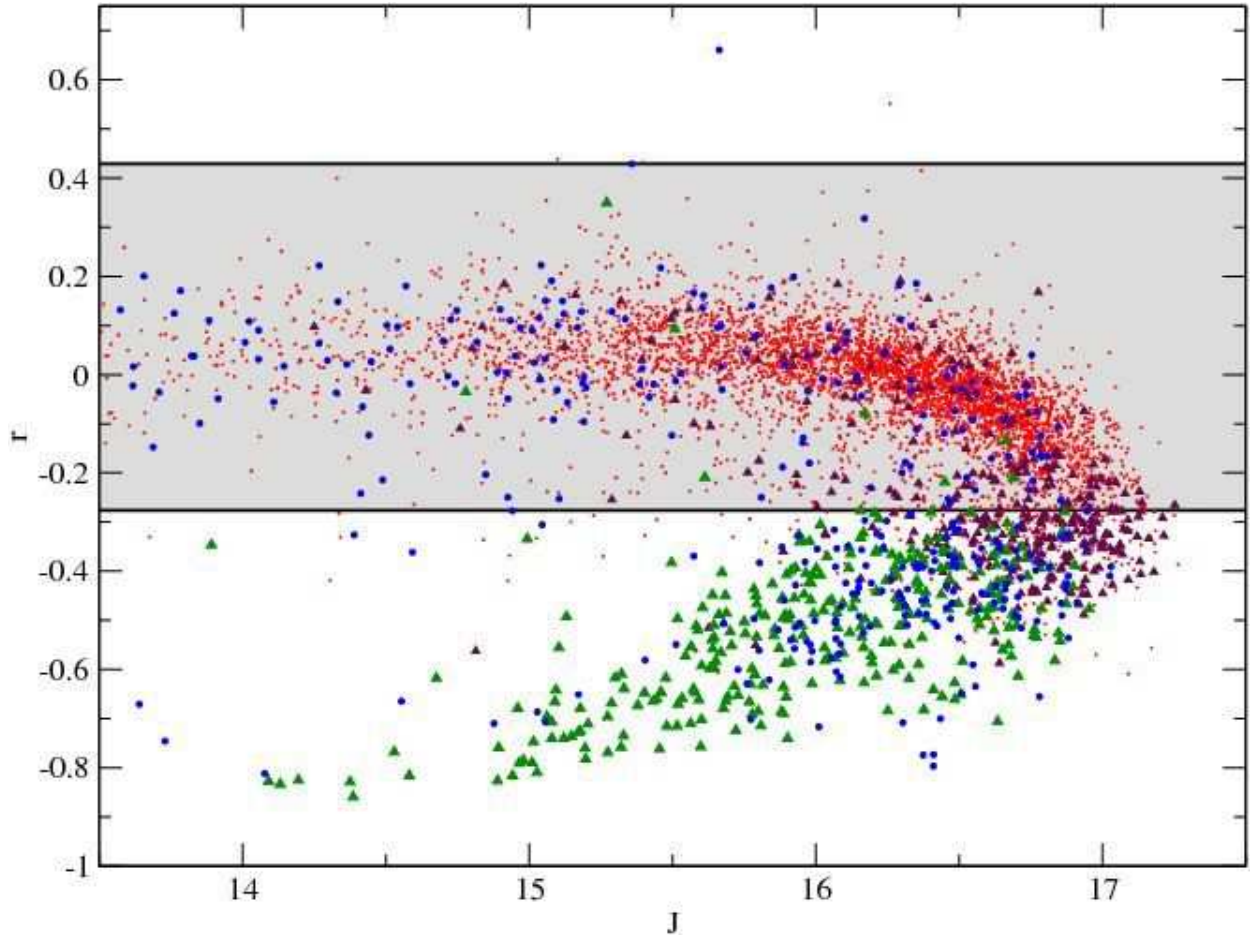}
\caption {As a function of apparent magnitude for our $\left|b\right|>20\degree$ sample, we plot the r-correlation statistic between J-band photometry and J-band seeing.   Non-variables are shown as red squares and black triangles, the latter are identified as extragalactic by 2MASS, NED, or SDSS.  Variables are shown as blue circles and green triangles, the latter are identified as extragalactic.  Almost all of the identified extragalactic variables are extended and exhibit variability that is anti-correlated with seeing (see Figure 14).  This variability is not intrinsic to the source.  The grey region corresponds to sources with photometry that is neither correlated nor anti-correlated with seeing variations.  See ${\S}$3.4 for discussion.}
\end{figure}
\clearpage
\begin{figure}
\plotone{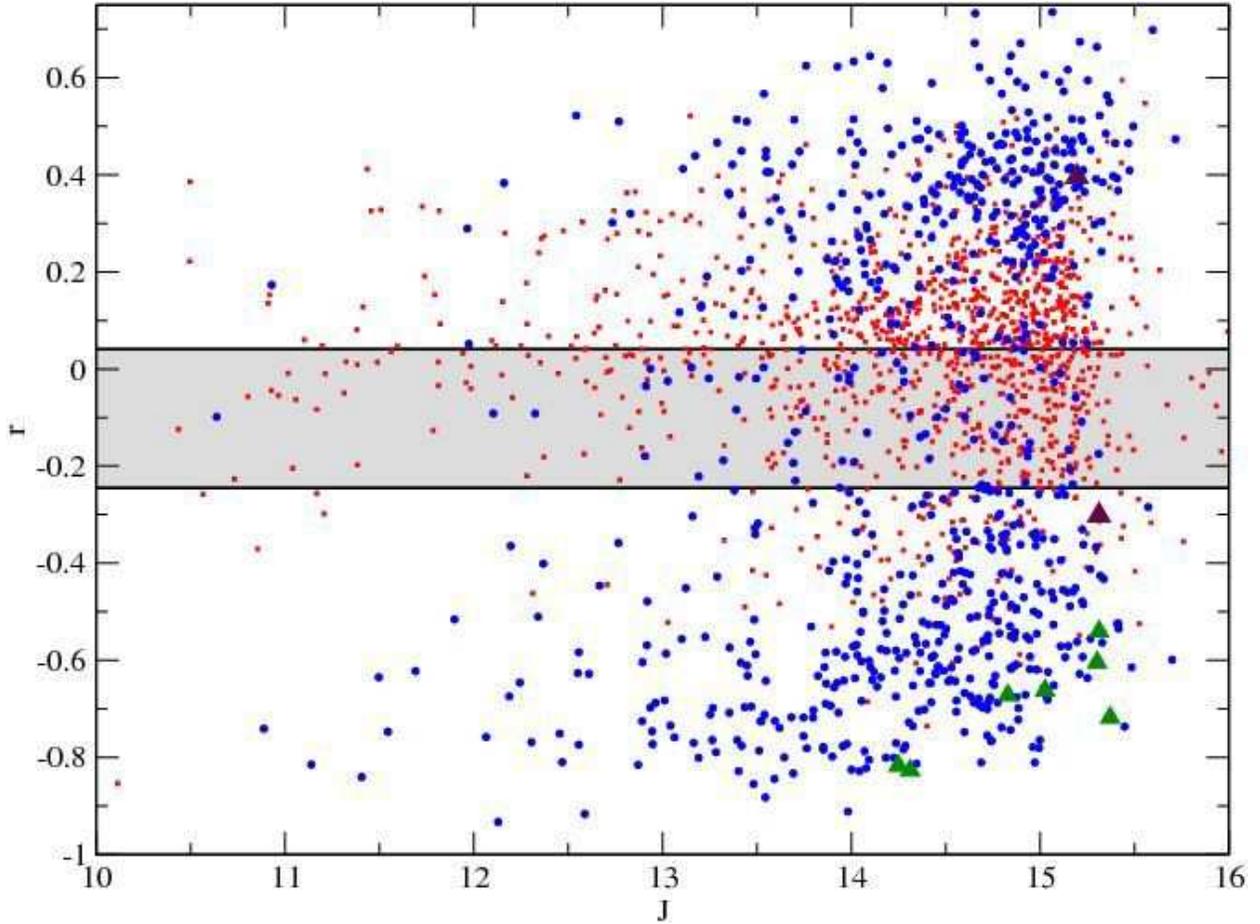}
\caption {Identical plot to Figure 15 for our $\left|b\right|<20\degree$ sample.   We observe variable sources that are correlated with seeing (see Figure 14) in addition to anti-correlated sources.  
See ${\S}$3.4 for discussion.}
\end{figure}
\clearpage
\begin{figure}
\plotone{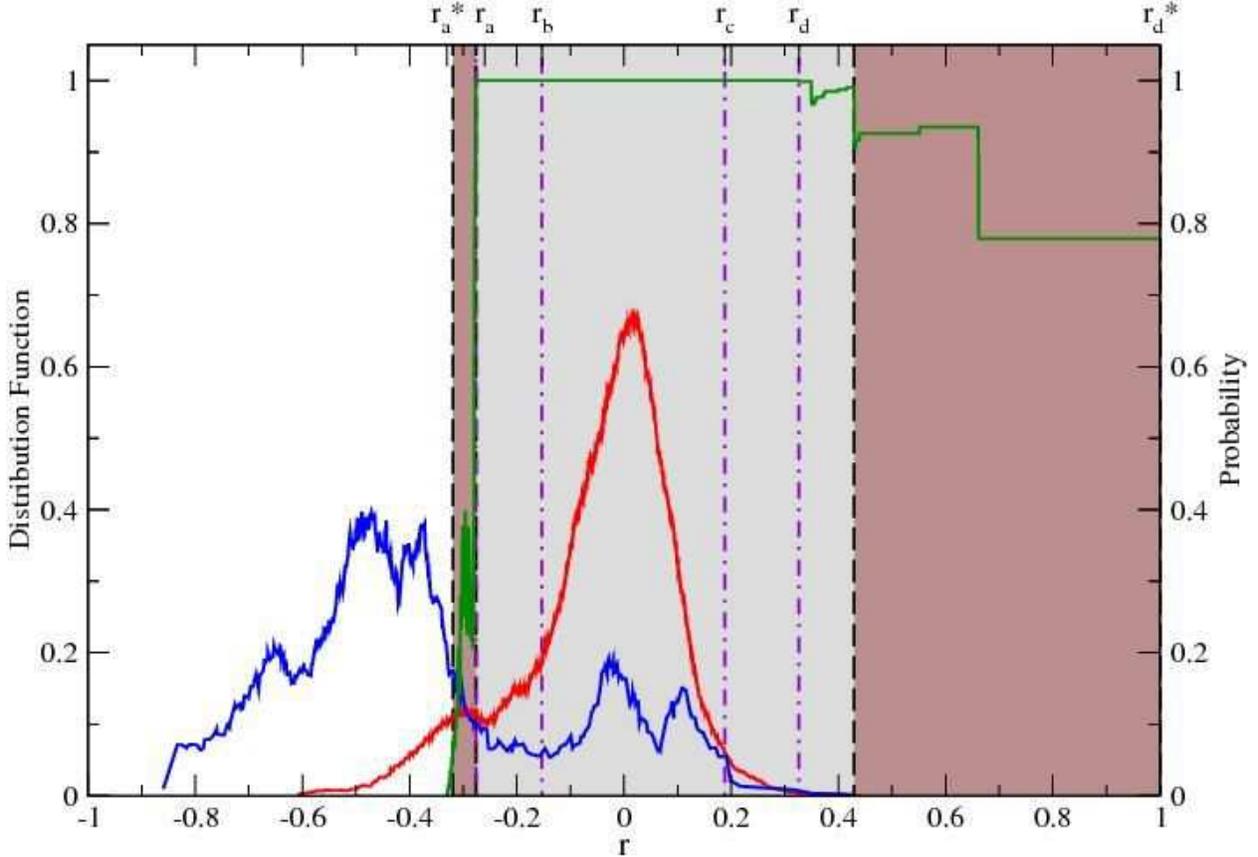}
\caption {As a function of the r-correlation statistic between J-band photometry and J-band seeing, we plot approximate distribution functions (arbitrary units) for identified variables (blue line) and non-variables (red line) in our $\left|b\right|>20\degree$ sample.  Unlike Figure 15, we group together identified extragalactic sources with the other sources.  In green, we plot the estimated probability that observed variability is intrinsic to the source, as opposed to correlated with seeing variations.  The different color regions correspond to probabilities $>$95\% (grey; same as Figure 15), between 5\% and 95\% (brown), and $<$5\% (white);  r$_{a}$* and r$_{d}$* are shown where the probability is 5\%.  The regions bounded by \{r$_{a}$,r$_{b}$\} and \{r$_{c}$,r$_{d}$\} (purple vertical dashed lines), correspond to the range of r-values for which the K-S test between the populations of variables and non-variables evaluates to a probability of 99.7\%.  See ${\S}$3.4.1 for discussion.}
\end{figure}
\clearpage
\begin{figure}
\plotone{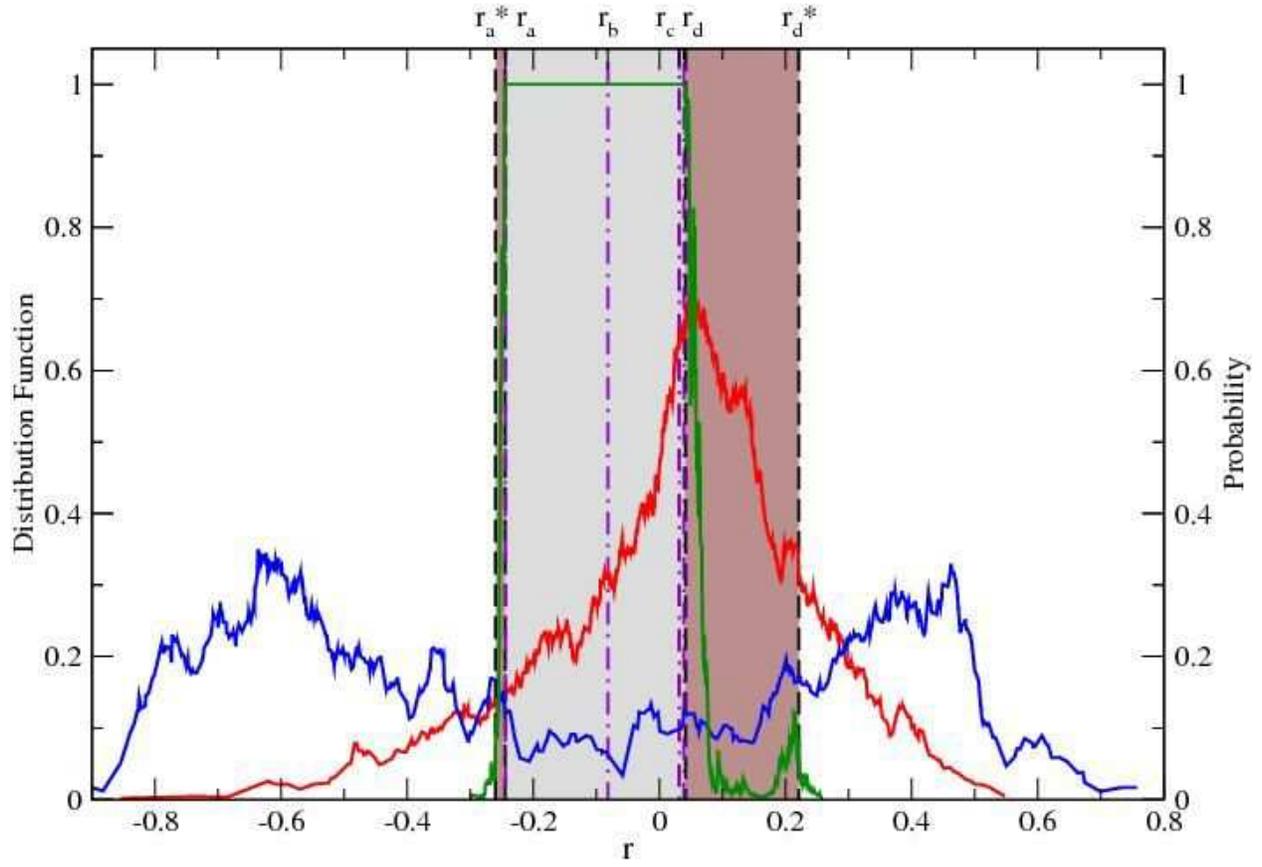}
\caption {Identical plot to Figure 17 for our $\left|b\right|<20\degree$ sample.}
\end{figure}
\clearpage
\begin{figure}
\plotone{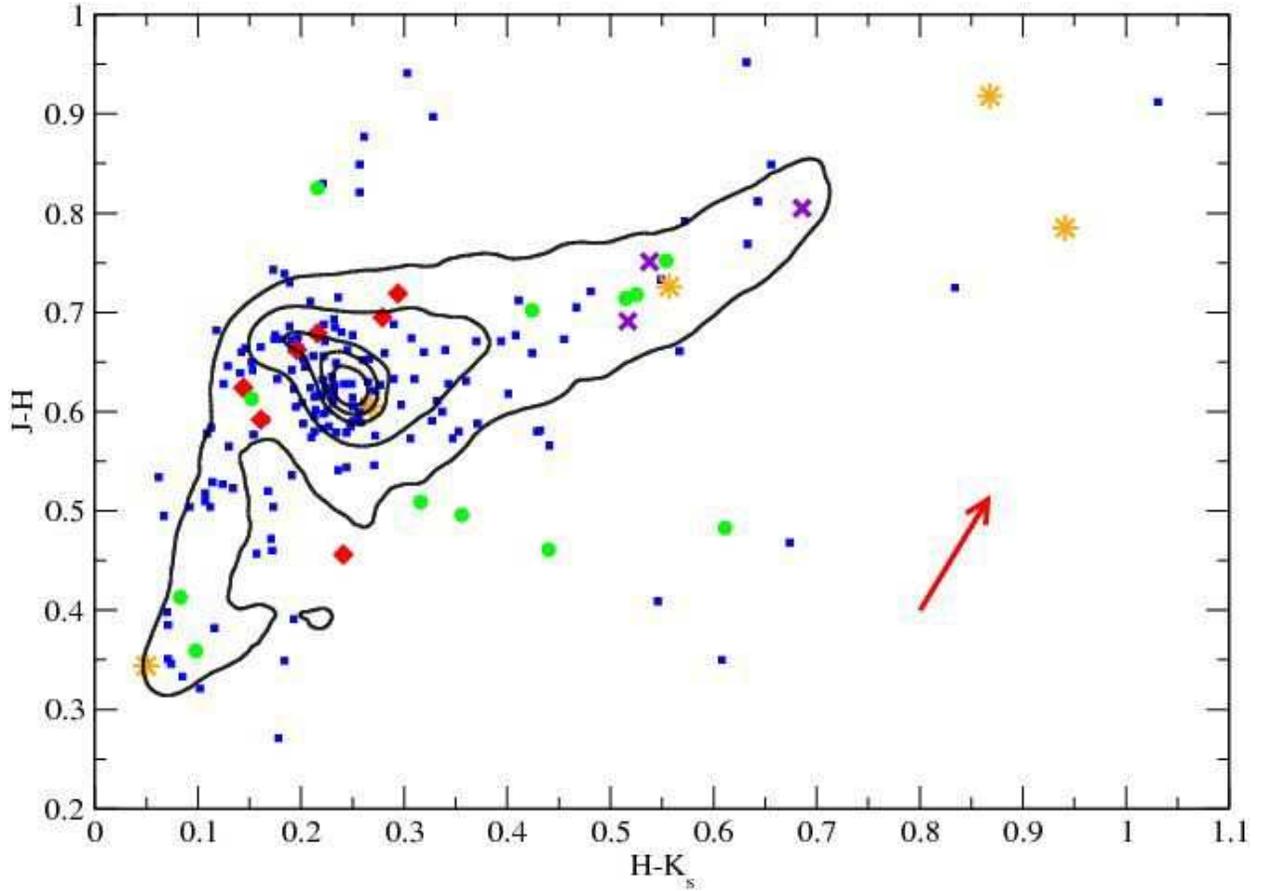}
\caption {For our $\left|b\right|>20\degree$ sample, we plot the average Cal-PSWDB colors for variables with blue squares, periodic variables with red diamonds, and identified extragalactic variables with orange stars.  Candidate variables are shown with green circles, and extragalactic candidate variables with maroon `X' symbols.  For reference, we plot the contours from Figure 4 for the entire  $\left|b\right|>20\degree$ sample.  An A$_{v}$=1 reddening vector is shown in red.}
\end{figure}
\clearpage
\begin{figure}
\plotone{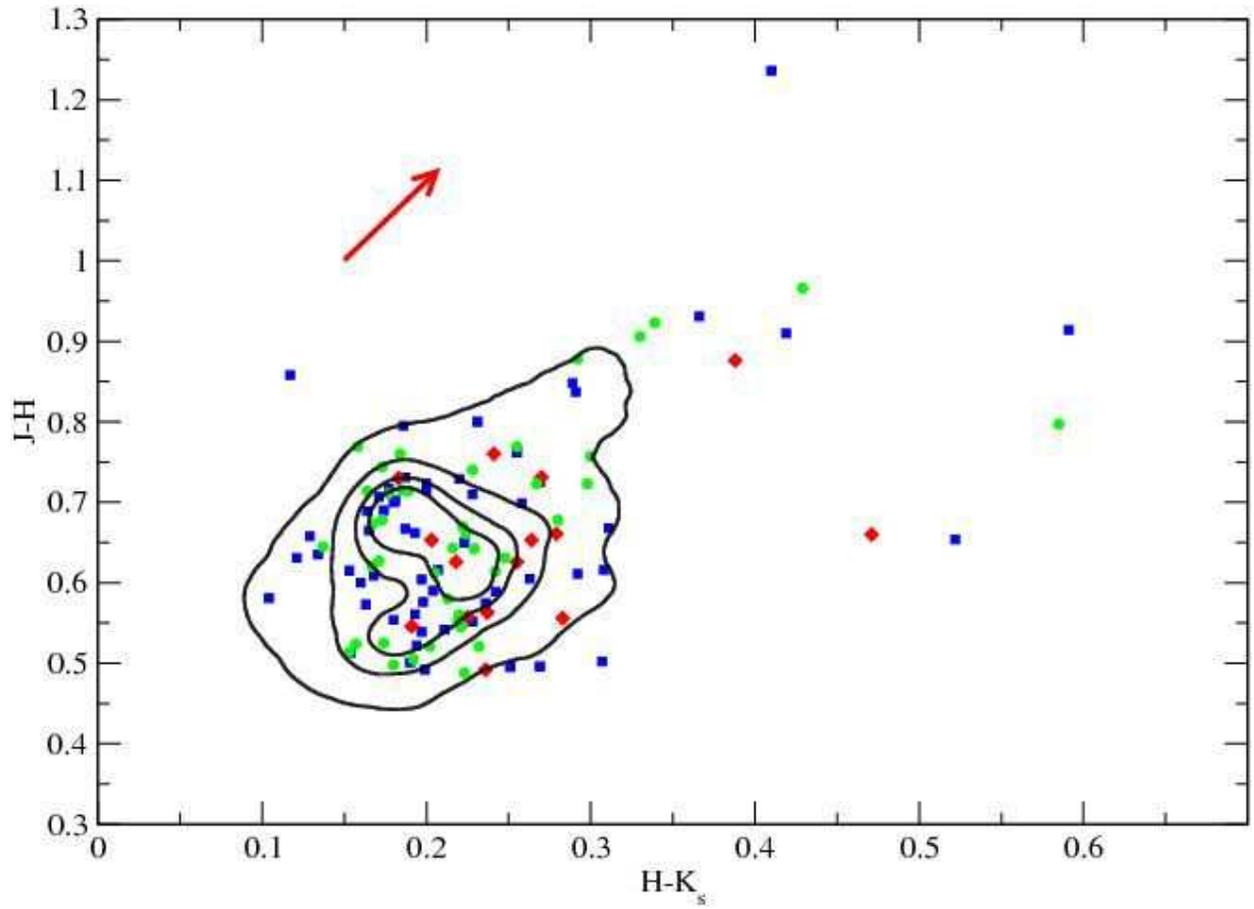}
\caption {Identical plot to Figure 19, corresponding to the variables in our $\left|b\right|<20\degree$ sample and analogous to Figure 5.}
\end{figure}
\clearpage
\begin{figure}
\plotone{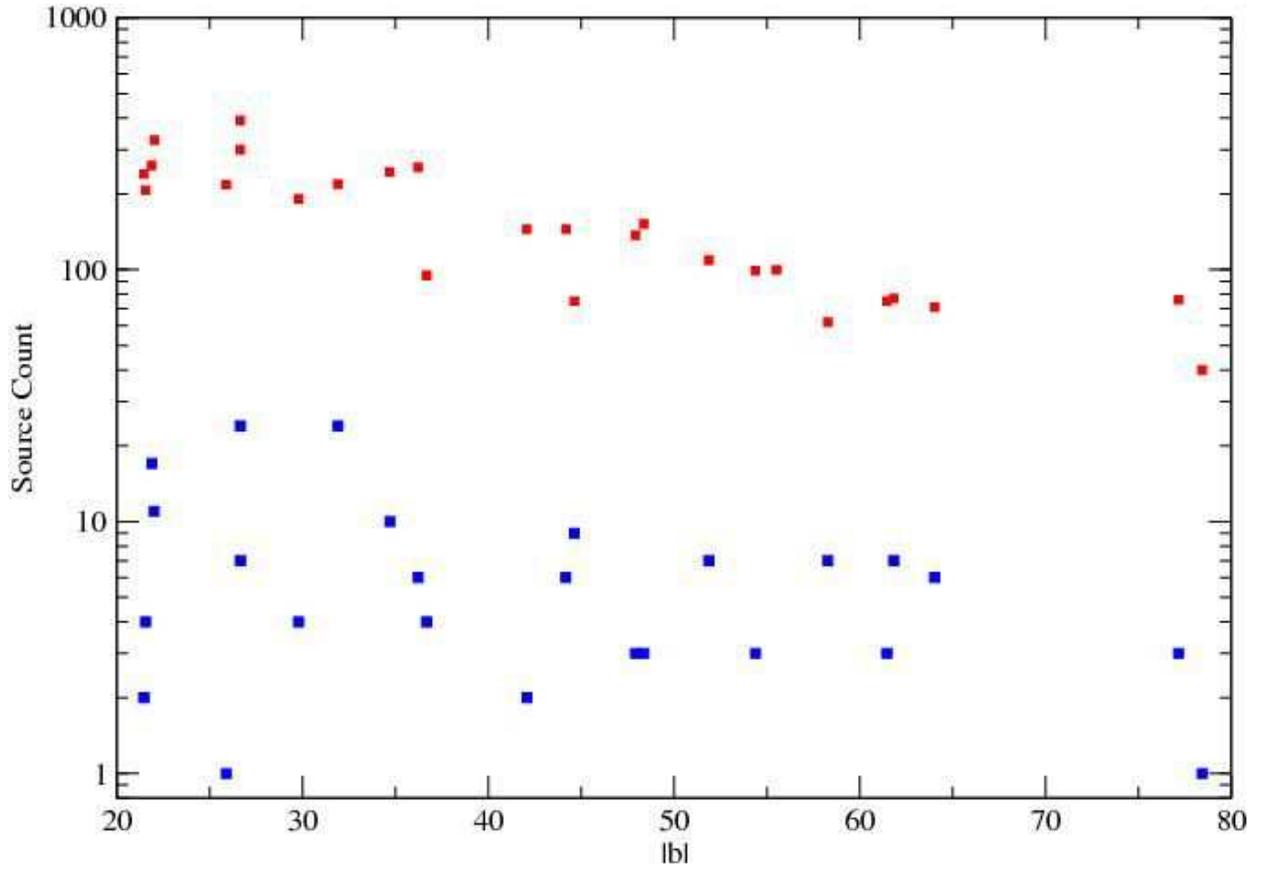}
\caption {In red we plot the number of sources in each field, as a function of the mean galactic latitude for each $\left|b\right|>20\degree$ field.  In blue, we plot the number of variables in each field. }
\end{figure}
\clearpage
\begin{figure}                             
\plotone{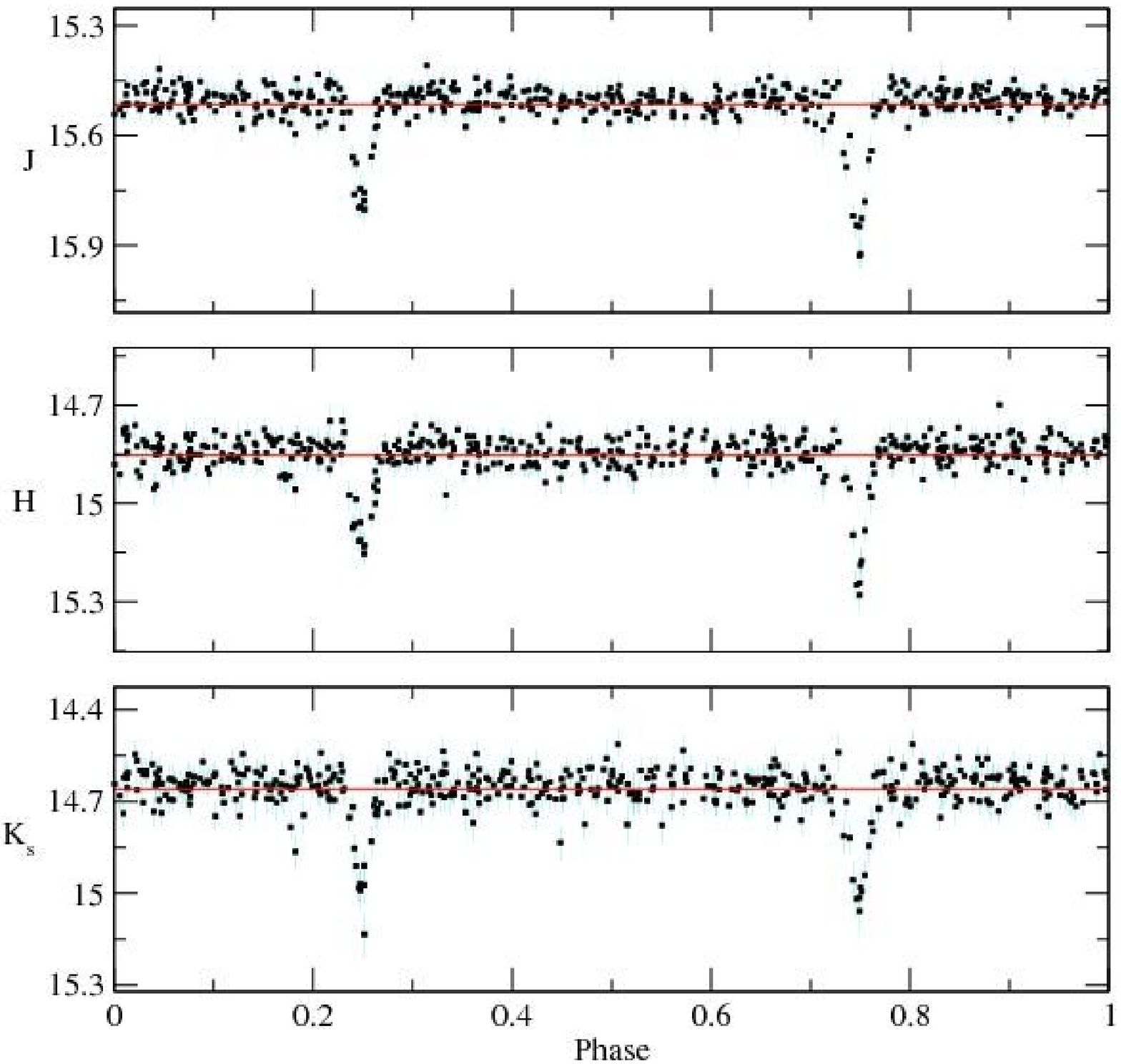}
\caption{JHK$_{s}$ data for late-type eclipsing binary 2MASS J01542930+0053266, folded to a period of 2.639 days.  Data are shown in black with error bars shown in teal.  Each data point corresponds to the unweighted average of one group of six scans ($m_{n,t_6}$).  Mean apparent magnitudes are shown with red horizontal lines.}\end{figure}
\begin{figure}%
\plotone{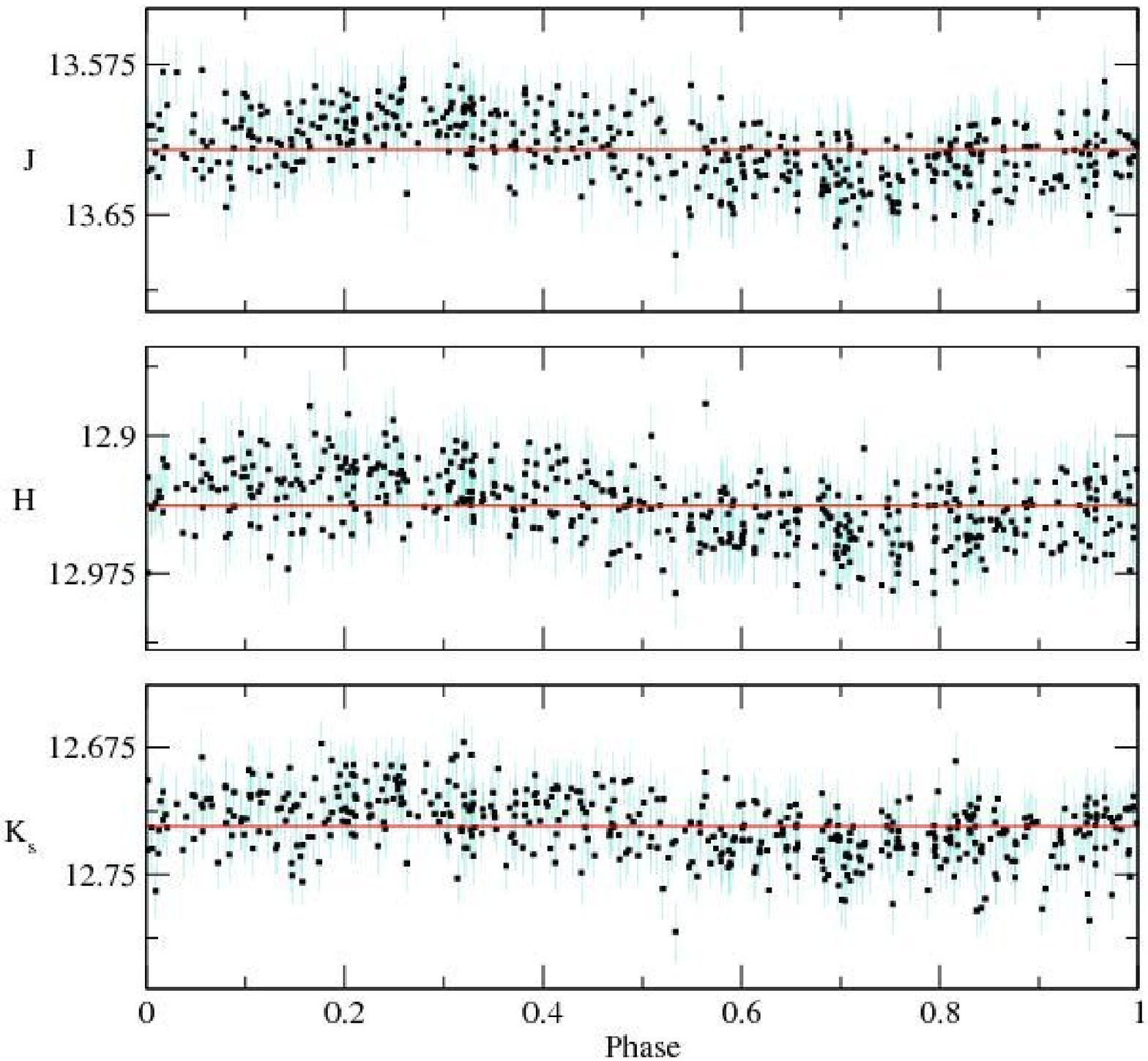}
\caption{JHK$_{s}$ data for suspected CV 2MASS J01545296+0110529, folded to a period of 0.18603 days.  }\end{figure}\clearpage
\begin{figure}
\plotone{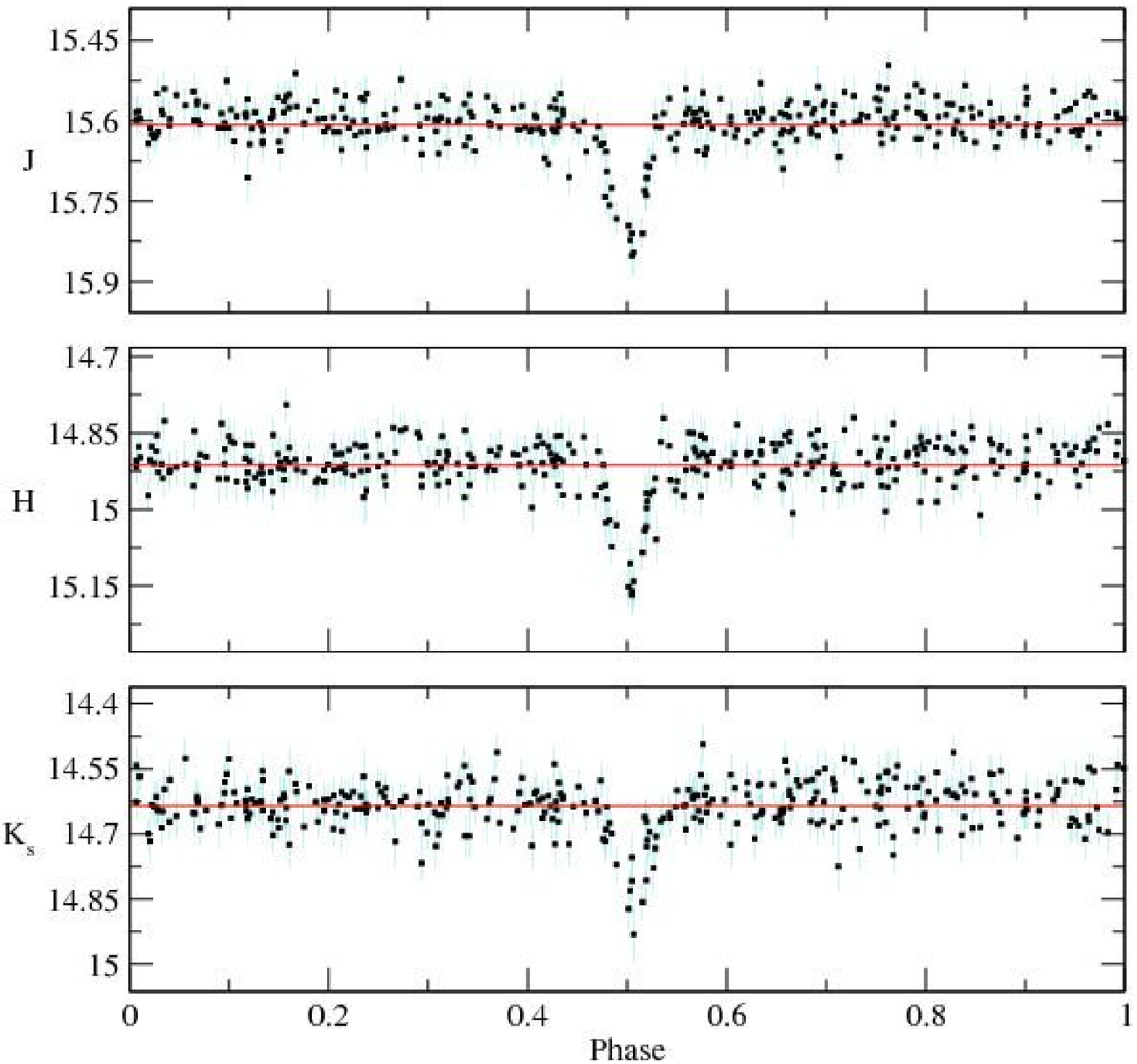}
\caption{JHK$_{s}$ data for late-type eclipsing binary 2MASS J04261603+0323578, folded to a period of 0.88320 days. }\end{figure}
\begin{figure}
\plotone{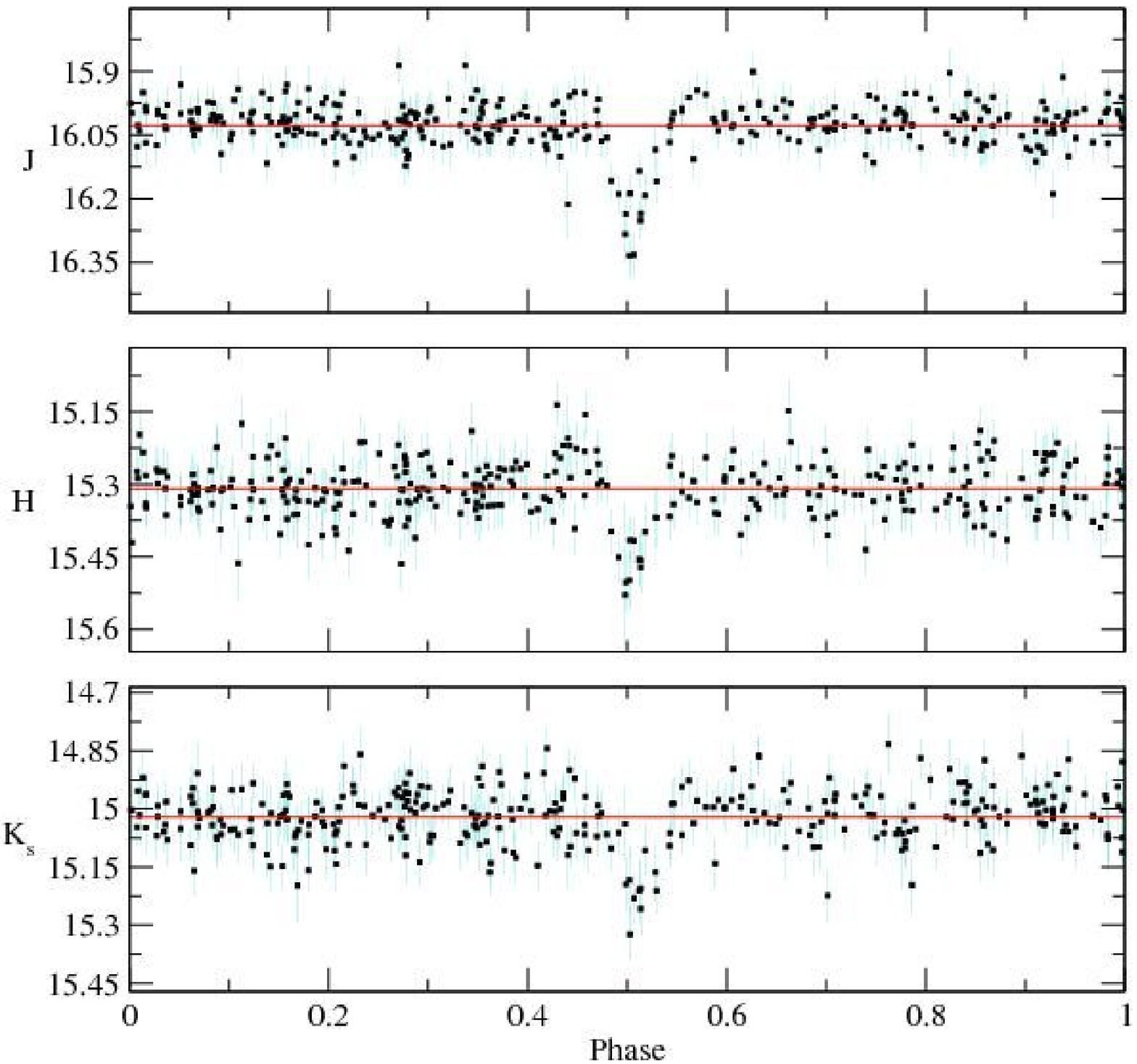}
\caption{JHK$_{s}$ data for late-type eclipsing binary 2MASS J04261900+0314008, folded to a period of 1.07631 days. }\end{figure}\clearpage
\begin{figure}
\plotone{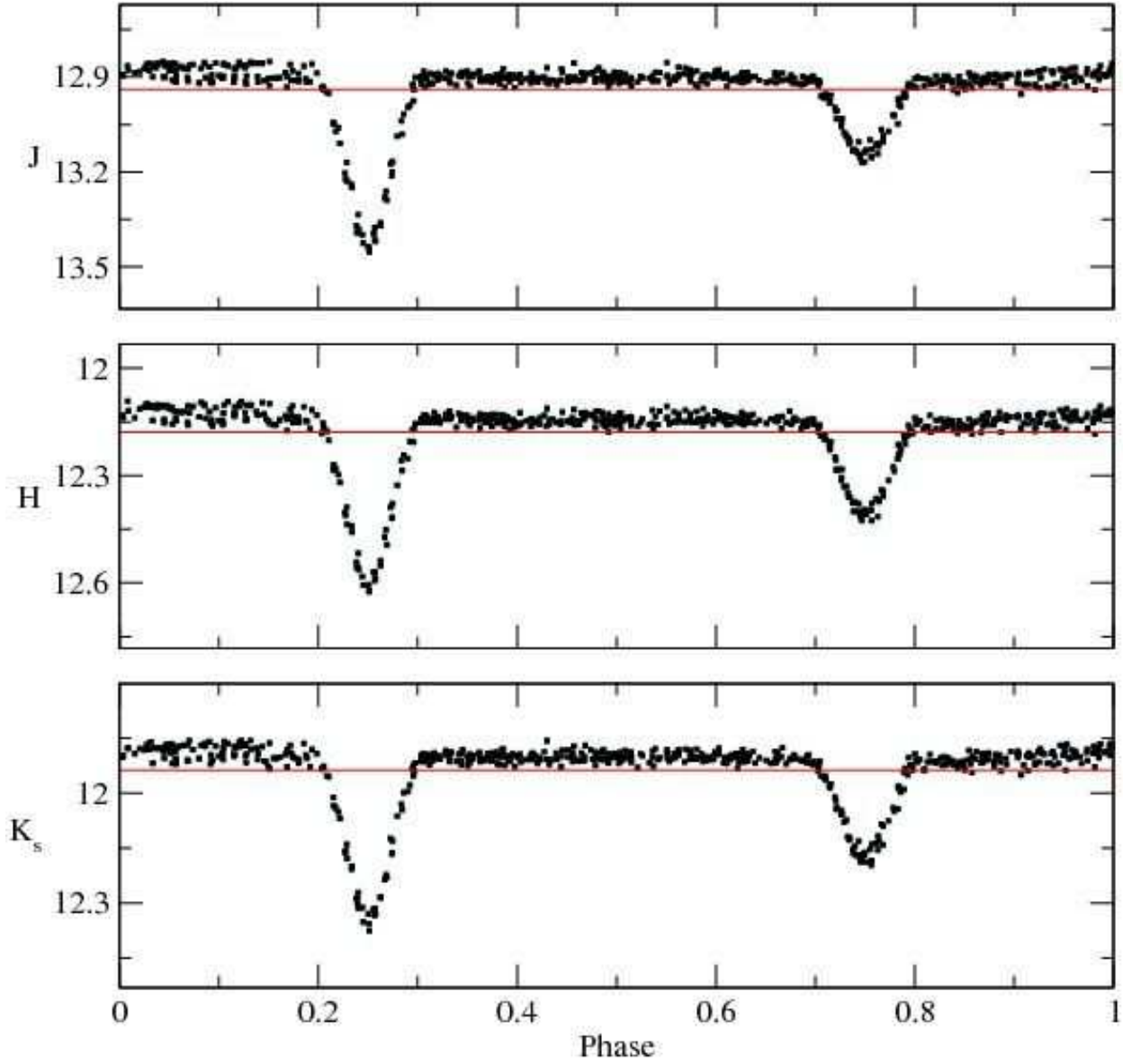}
\caption{JHK$_{s}$ data for eclipsing binary 2MASS J08255405-3908441, folded to a period of 8.08986 days.  }\end{figure}
\begin{figure}
\plotone{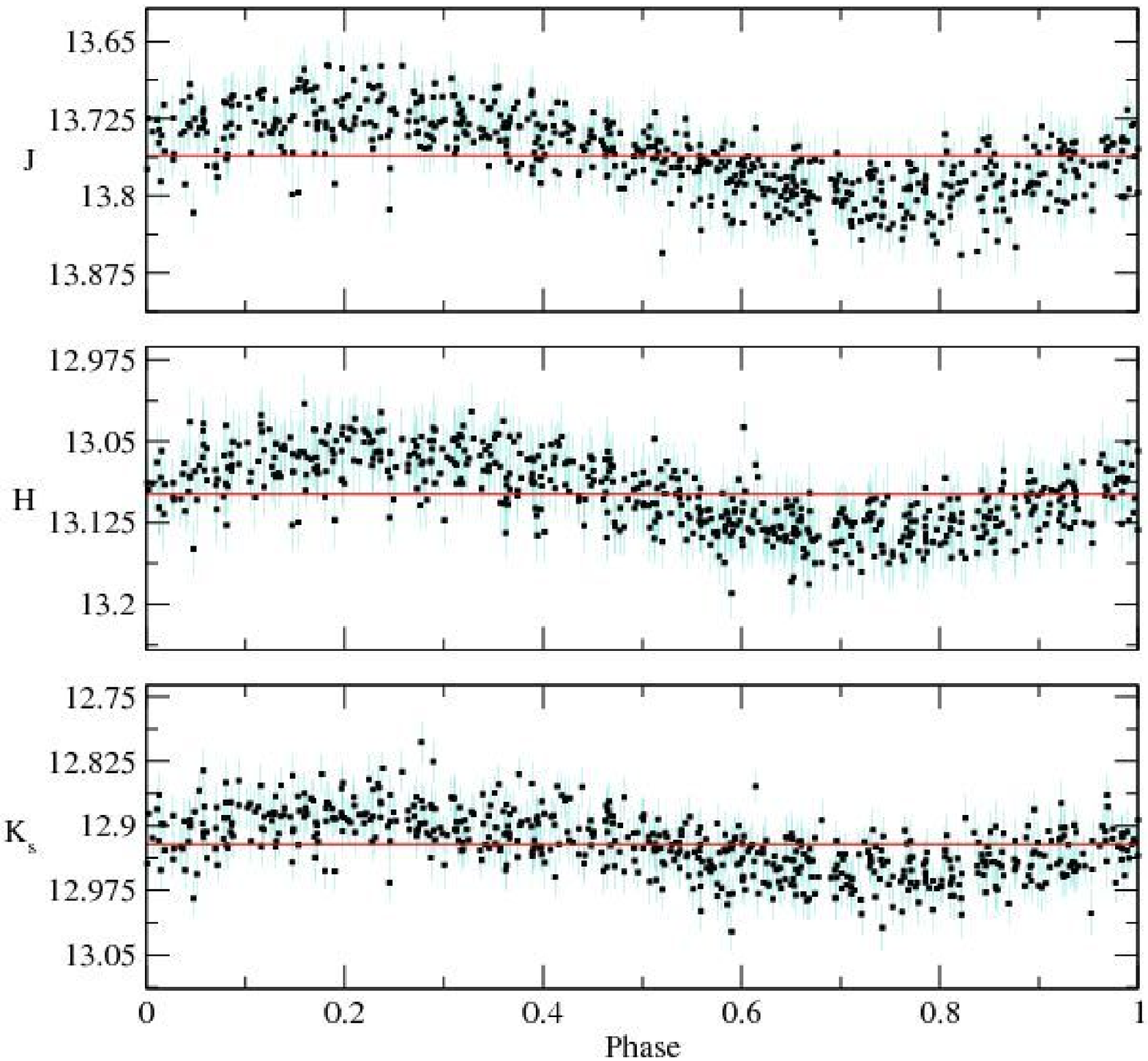}
\caption{JHK$_{s}$ data for sinusoidal variable 2MASS J08512729+1211484, folded to a period of 1.23725 days.  This object has a persistent artifact overlaid with the source every other scan, but still exhibits intrinsic periodic variability. We eliminate scans with the persistence for our analysis.}\end{figure}\clearpage
\begin{figure}
\plotone{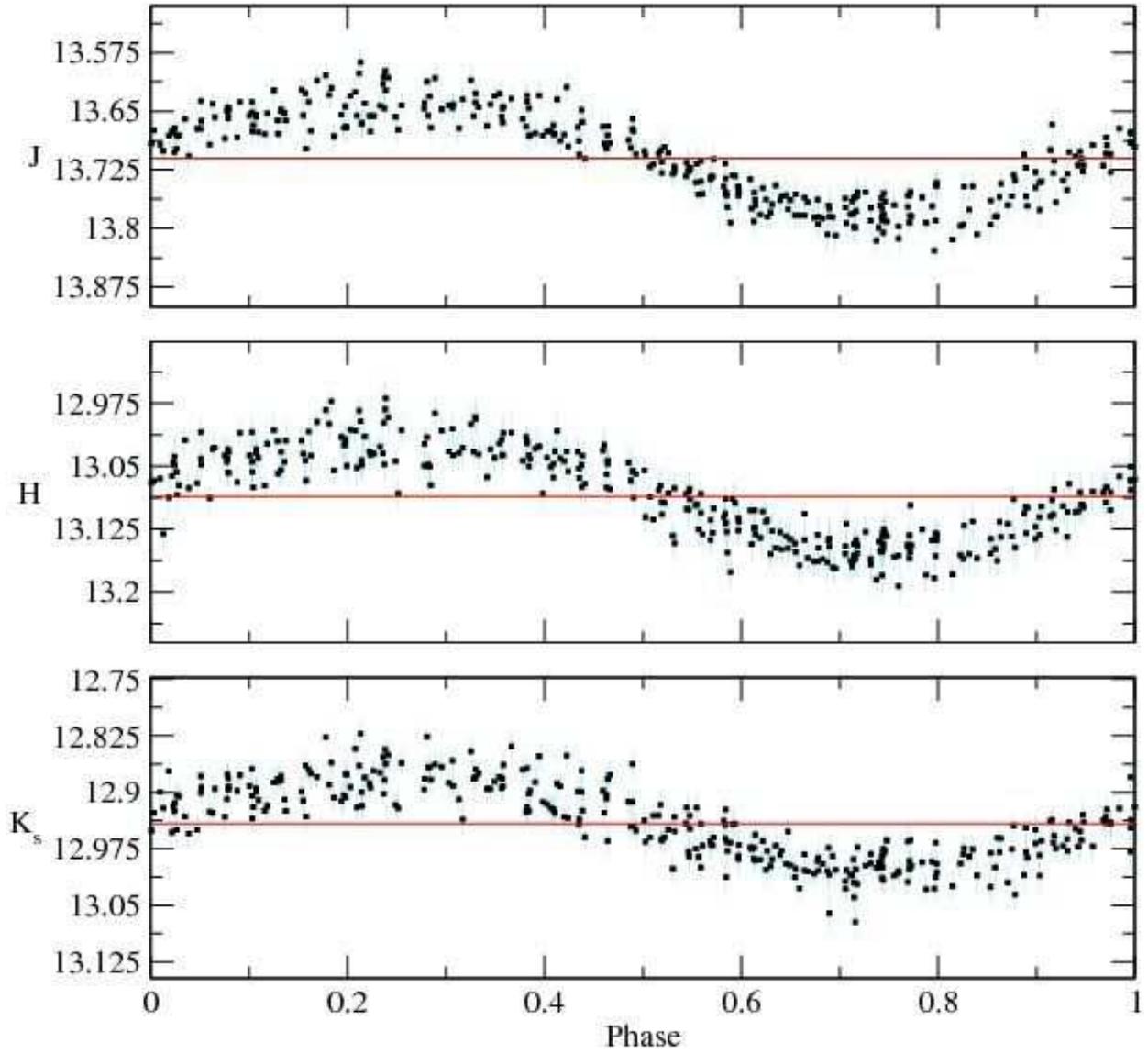}
\caption{JHK$_{s}$ data for sinusoidal variable 2MASS J15001192-0103090, folded to a period of 3.262 days.  }\end{figure}
\begin{figure}
\plotone{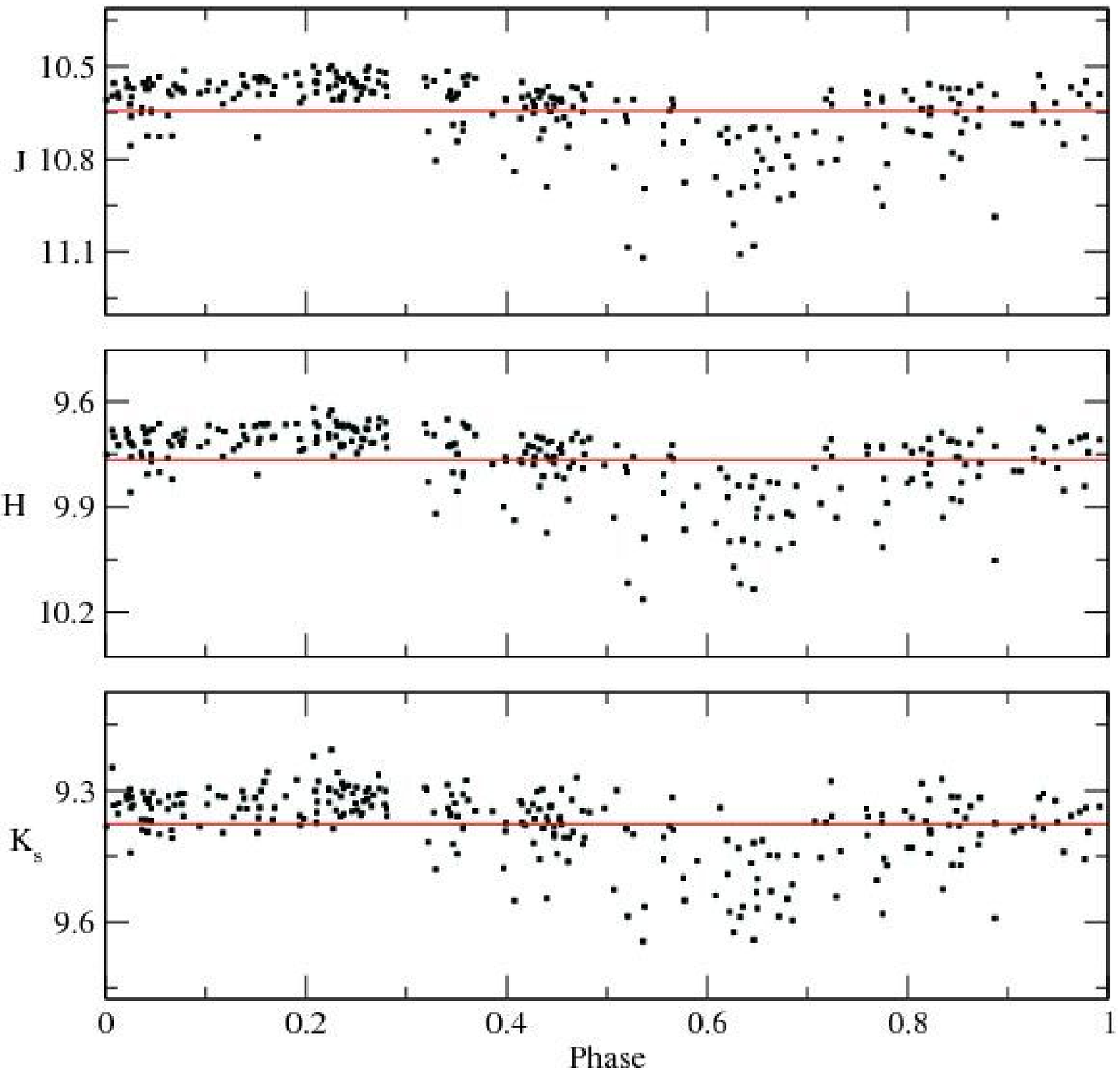}
\caption{JHK$_{s}$ data for Rho Oph YSO 2MASS J16271273-2504017, folded to a period of 0.831445 days.  }\end{figure}\clearpage
\begin{figure}
\plotone{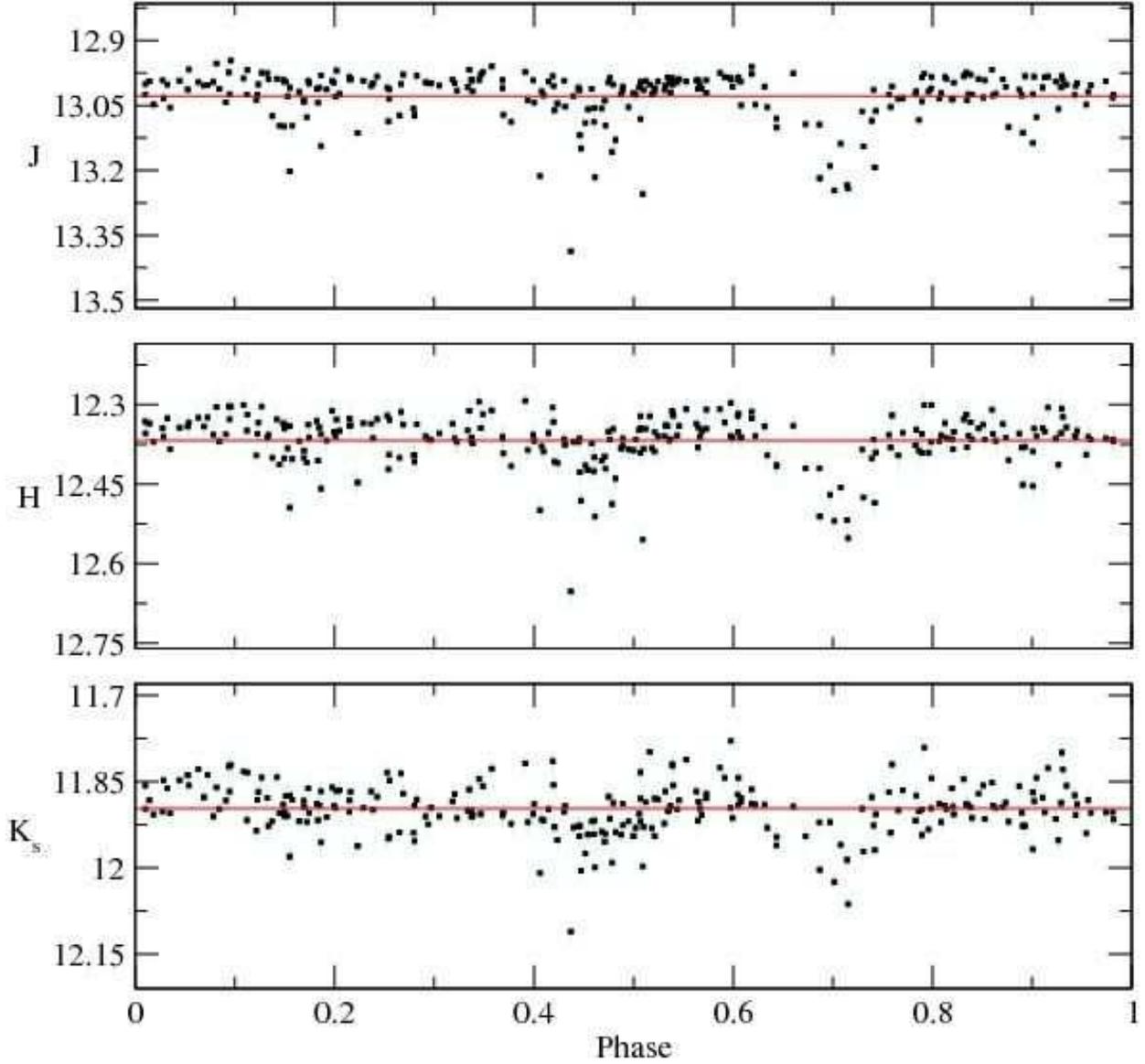}
\caption{JHK$_{s}$ data for Rho Oph YSO 2MASS J16272658-2425543, folded to a period of 2.9603 days.  The identification of this period is uncertain. The object exhibits variability on a 3-day timescale, but the observed variability is not entirely consistent with this period derived from our analysis.  Possible solutions include spotting, and semi-periodic veiling from circumstellar material.}\end{figure}
\begin{figure}
\plotone{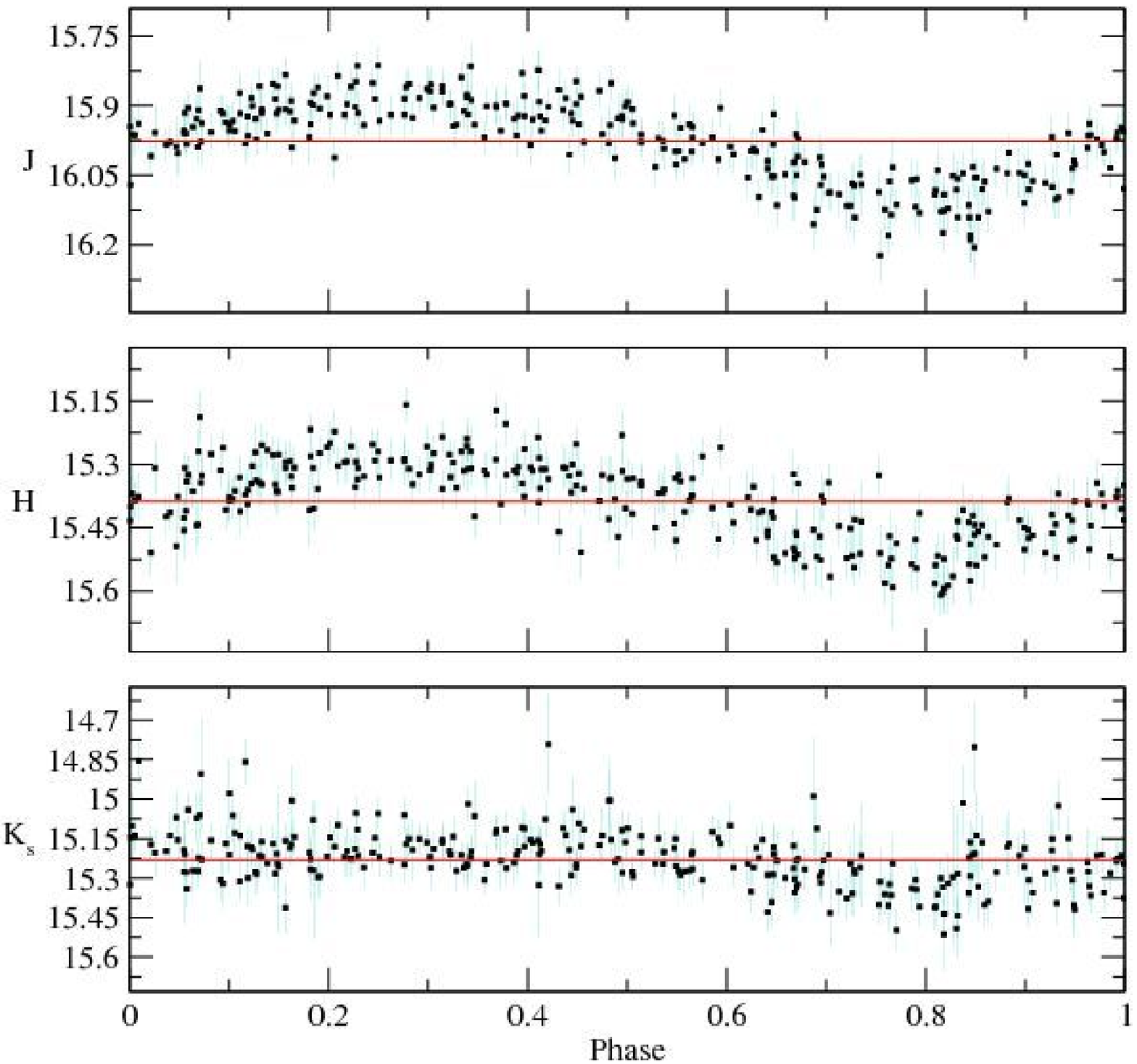}
\caption{JHK$_{s}$ data for suspected CV 2MASS J18391777+4854001, folded to a period of 0.1257 days.  }\end{figure}\clearpage
\begin{figure}
\plotone{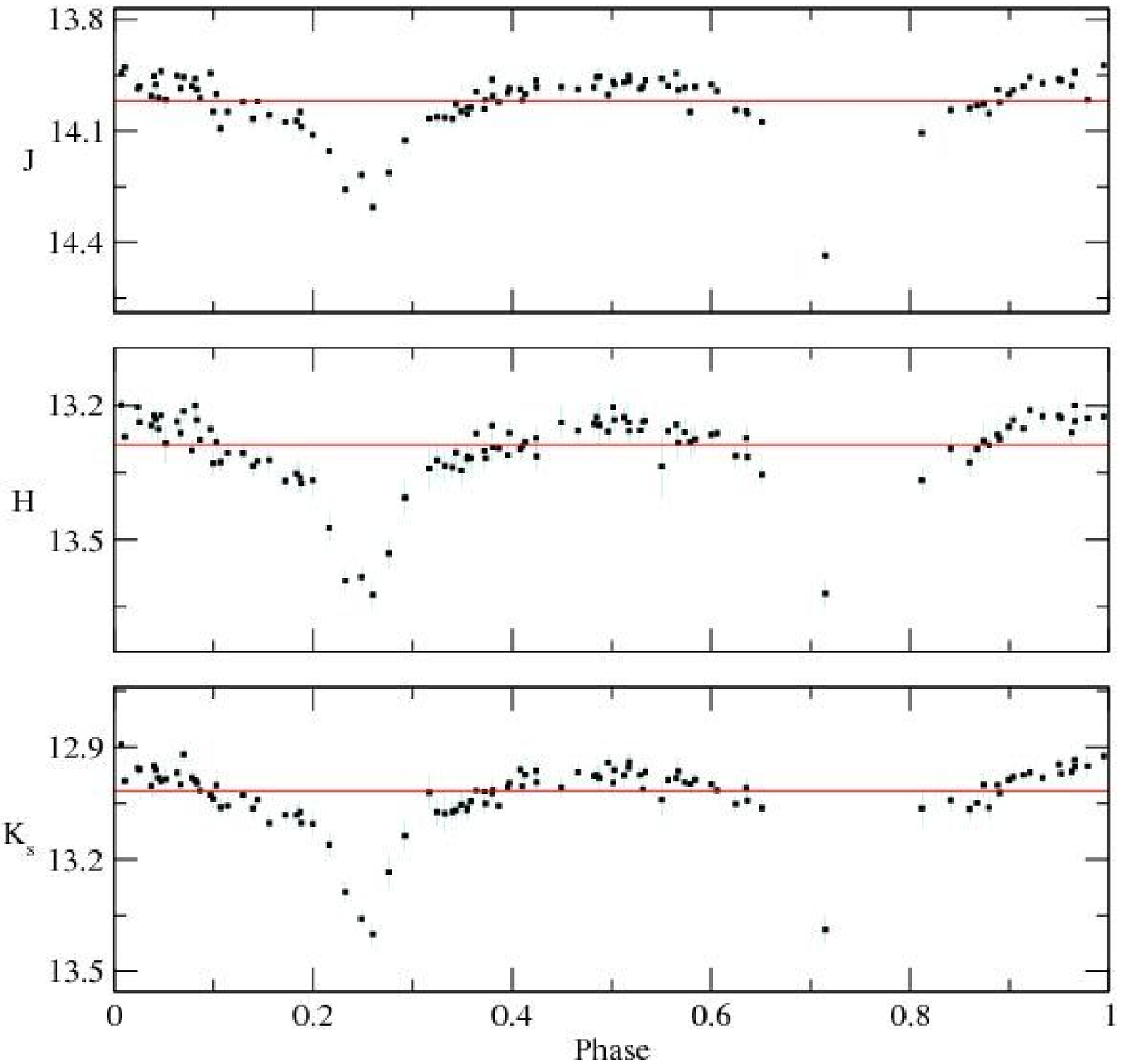}
\caption{JHK$_{s}$ data for eclipsing binary 2MASS J18510479-0442005, folded to a period of 5.657 days.   At this period, the secondary eclipse is not well sampled by the available observations. }\end{figure}
\begin{figure}
\plotone{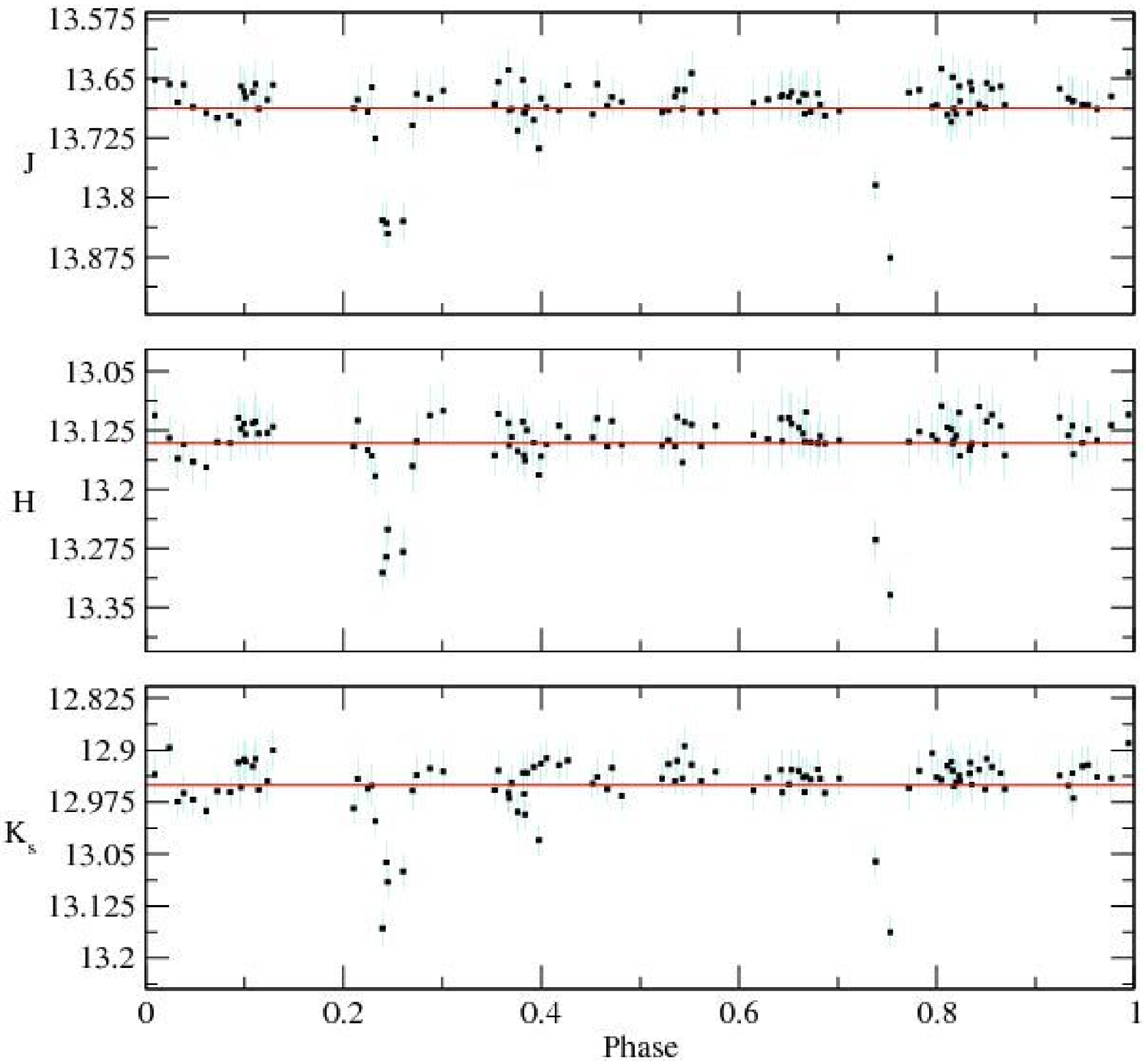}
\caption{JHK$_{s}$ data for eclipsing binary 2MASS J18510526-0437311, folded to a period of 6.2898 days.  }\end{figure}\clearpage
\begin{figure}
\plotone{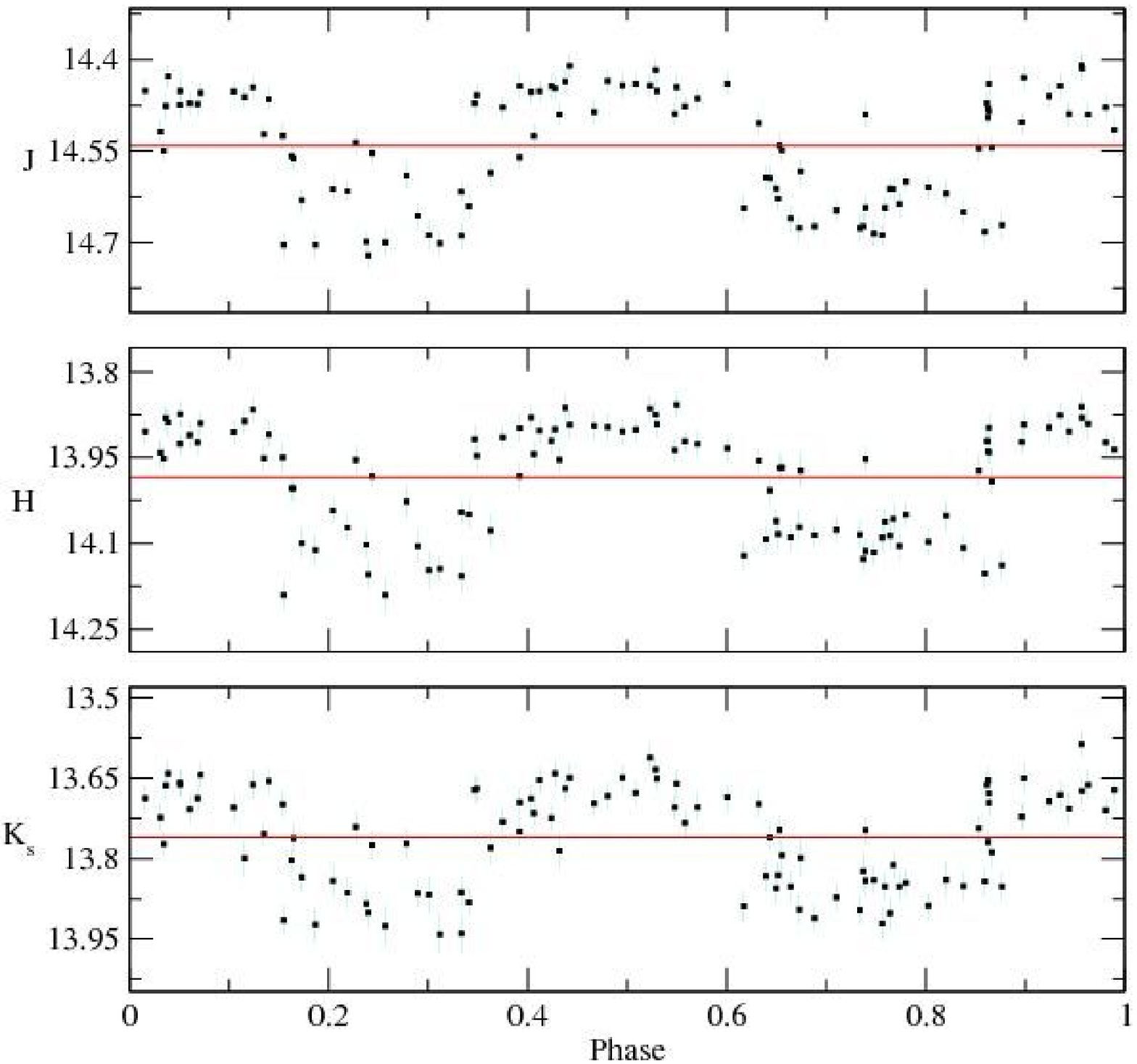}
\caption{JHK$_{s}$ data for 2MASS J18510882-0436123, folded to a period of 0.88822 days.  We do not identify an object type for this variable.}\end{figure}
\begin{figure}
\plotone{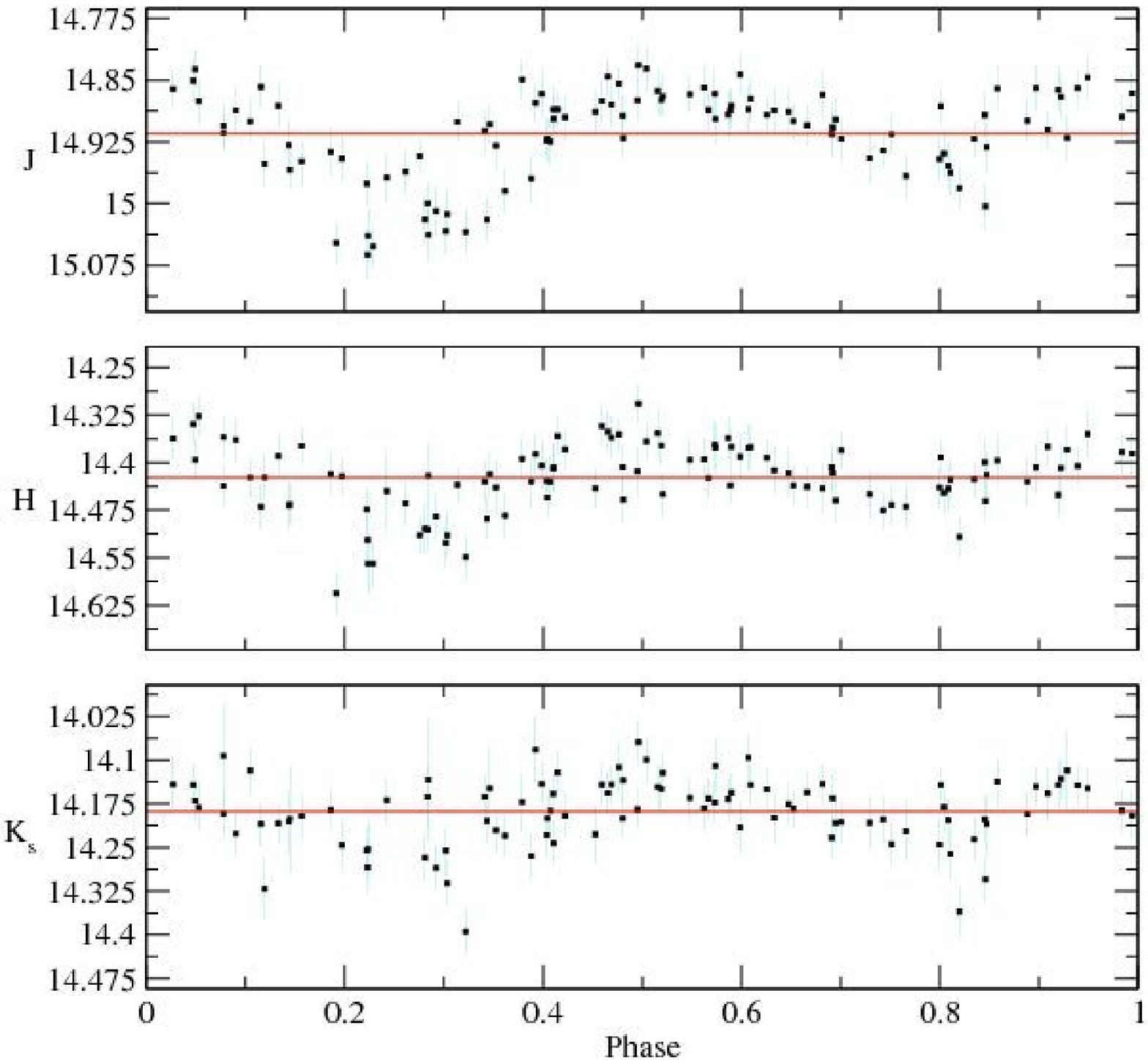}
\caption{JHK$_{s}$ data for suspected eclipsing binary 2MASS J18511786-0355311, folded to a period of 0.7804 days.  }\end{figure}\clearpage
\begin{figure}
\plotone{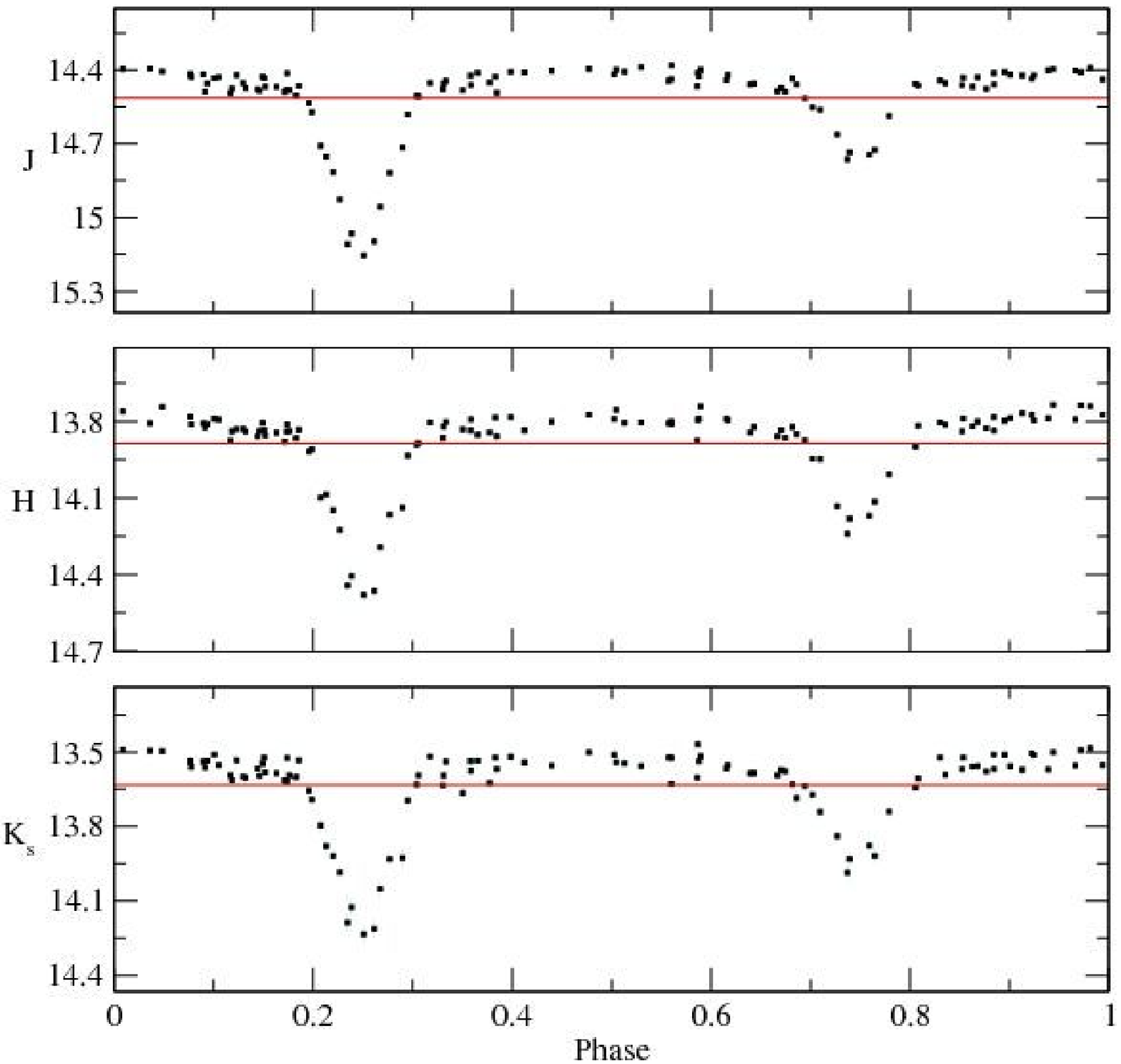}
\caption{JHK$_{s}$ data for eclipsing binary 2MASS J18512034-0426311, folded to a period of 3.3549 days.  }\end{figure}
\begin{figure}
\plotone{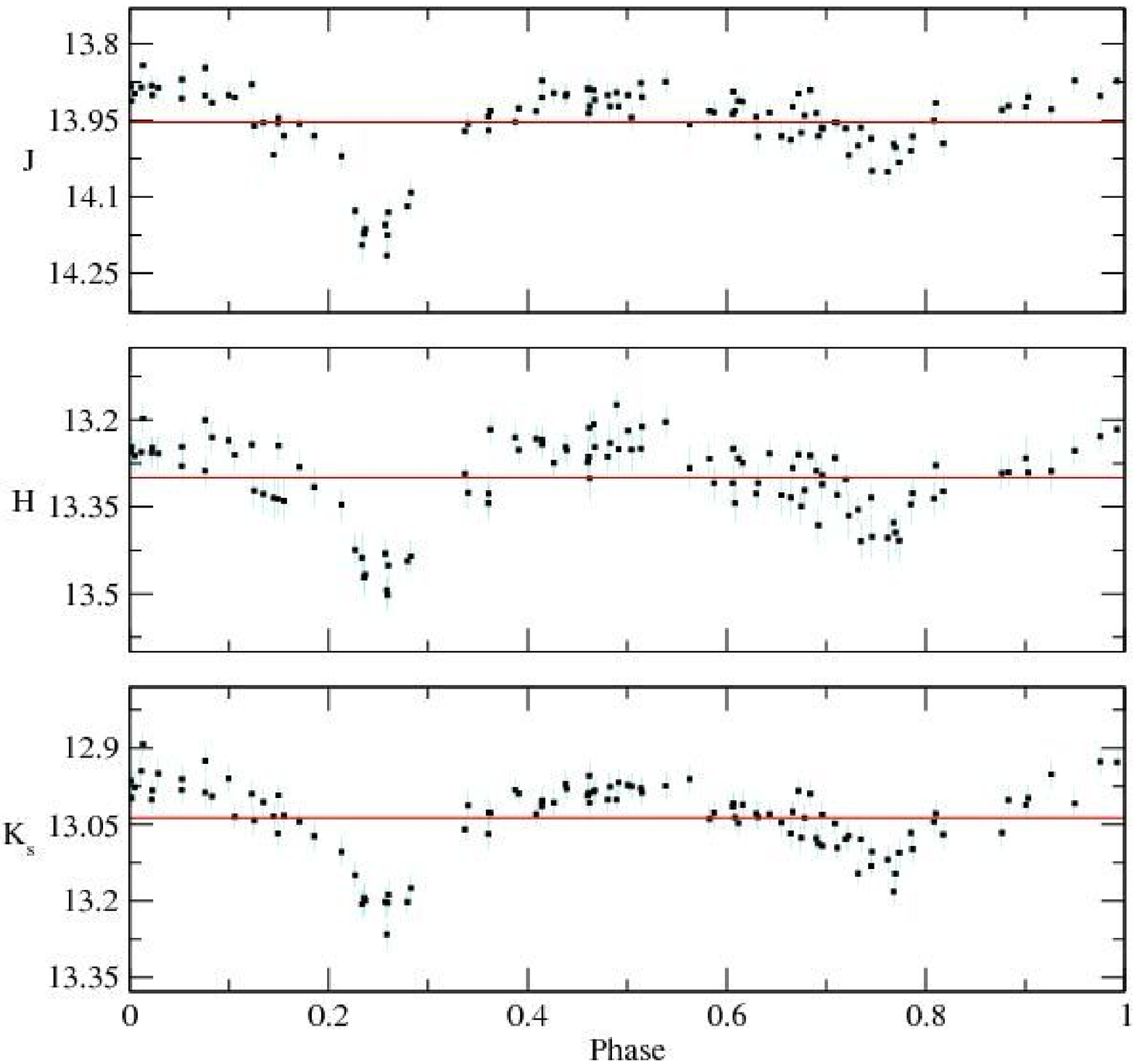}
\caption{JHK$_{s}$ data for eclipsing binary 2MASS J18512261-0409084, folded to a period of 3.916 days.  }\end{figure}\clearpage
\begin{figure}
\plotone{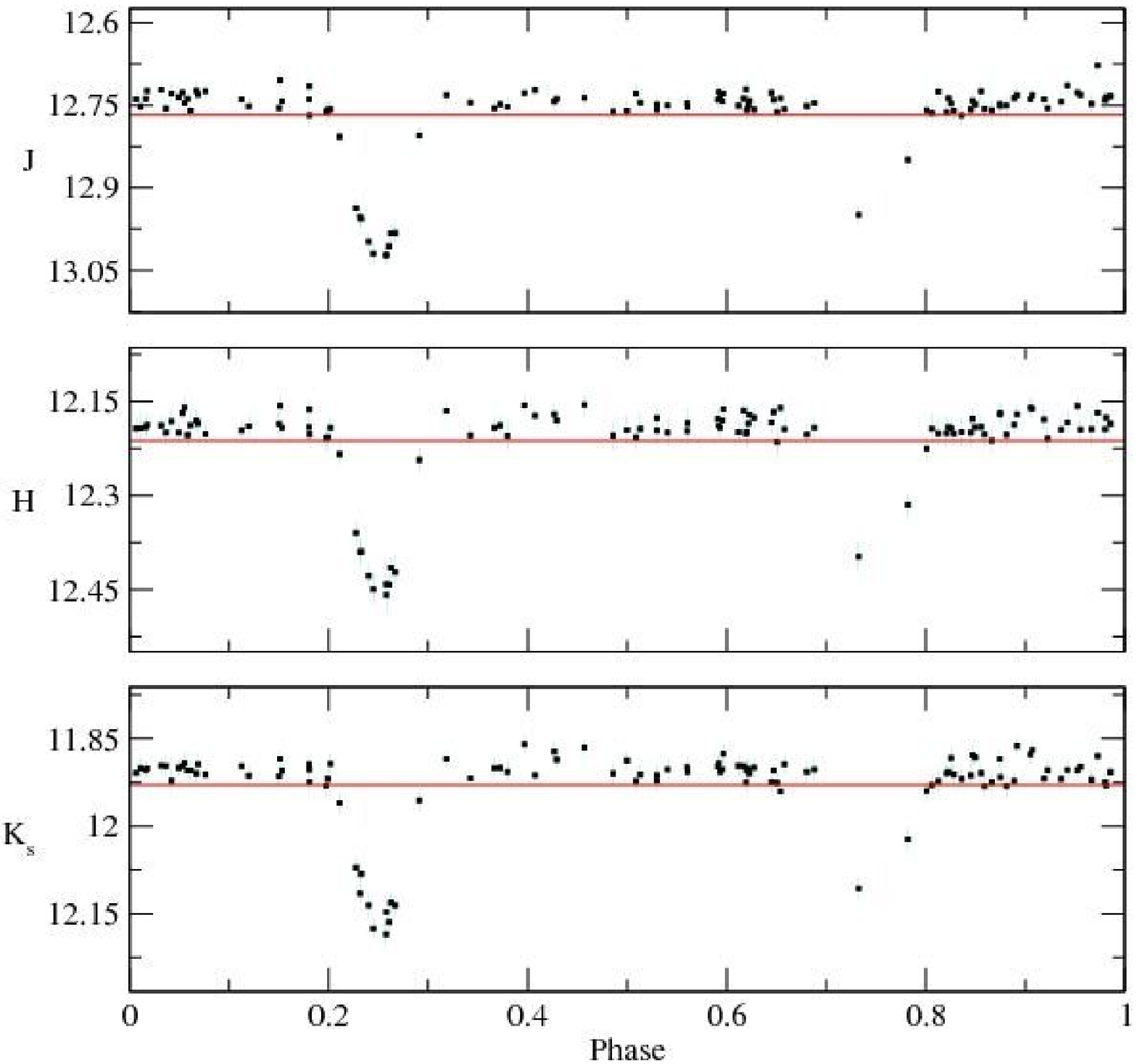}
\caption{JHK$_{s}$ data for eclipsing binary 2MASS J18512929-0412407, folded to a period of 3.0396 days.  }\end{figure}
\begin{figure}
\plotone{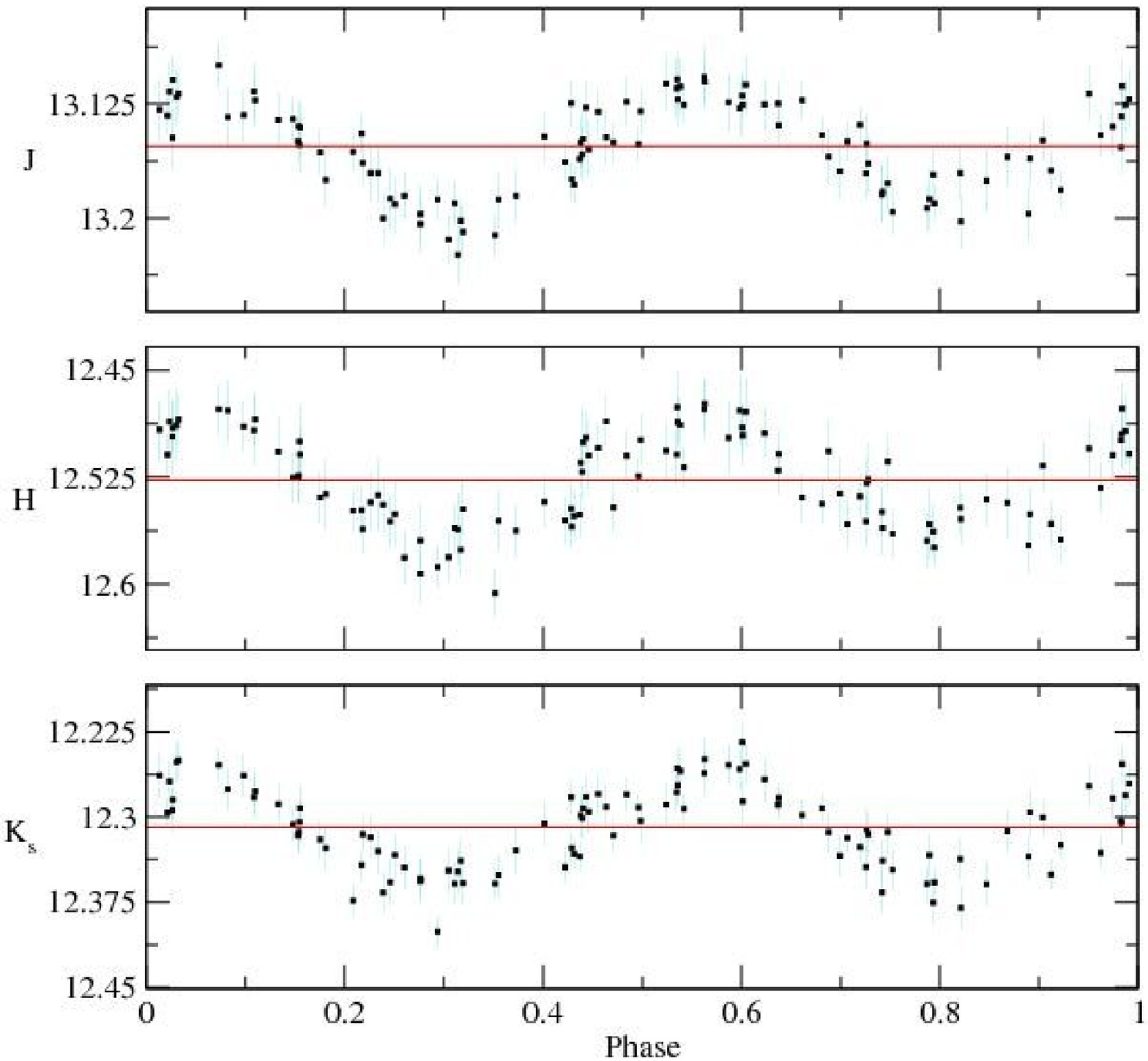}
\caption{JHK$_{s}$ data for suspected CV 2MASS J18513076-0432148, folded to a period of 0.32251 days.  }\end{figure}\clearpage
\begin{figure}
\plotone{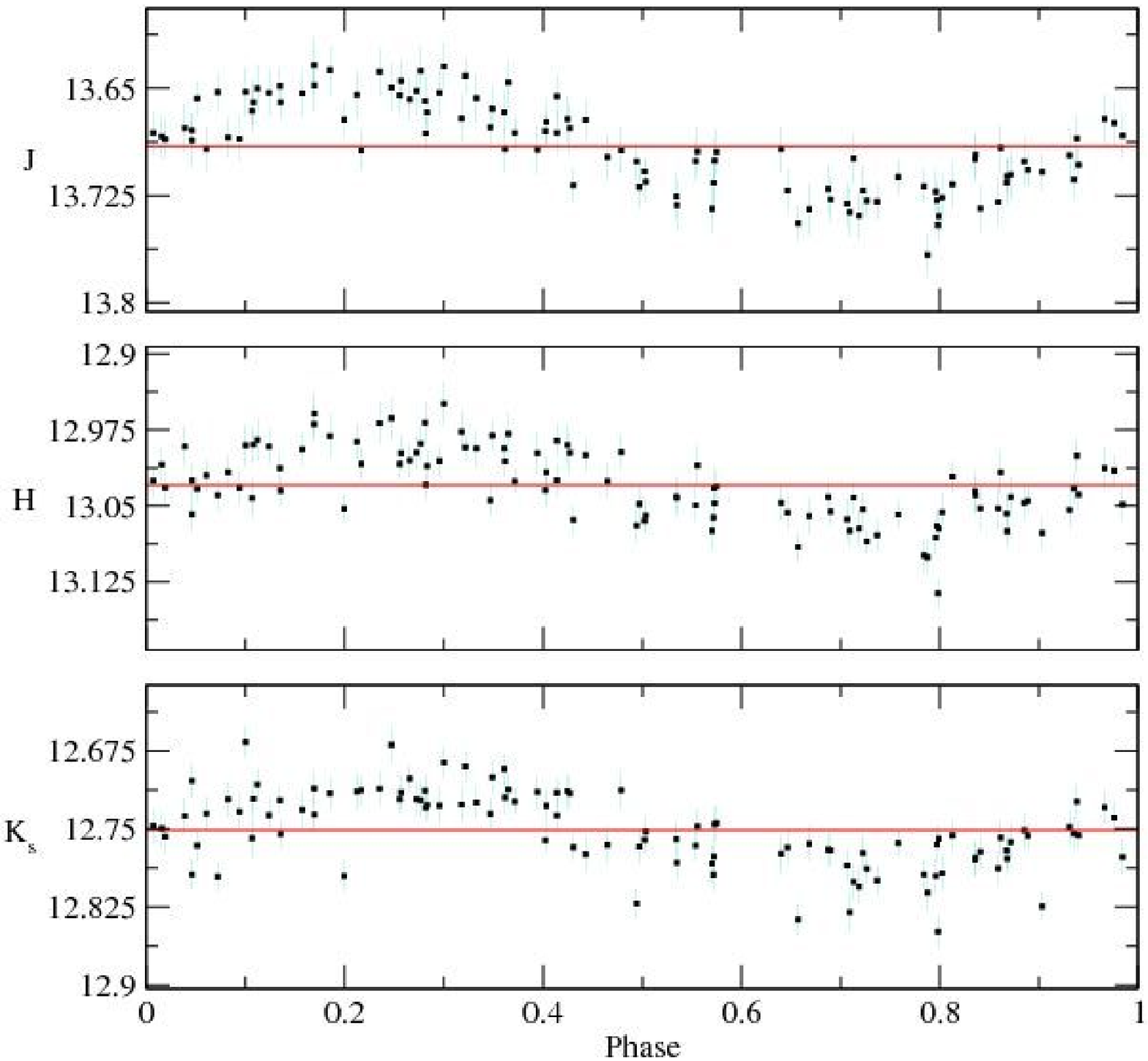}
\caption{JHK$_{s}$ data for suspected CV 2MASS J18513115-0424324, folded to a period of 0.6223 days.  }\end{figure}
\begin{figure}
\plotone{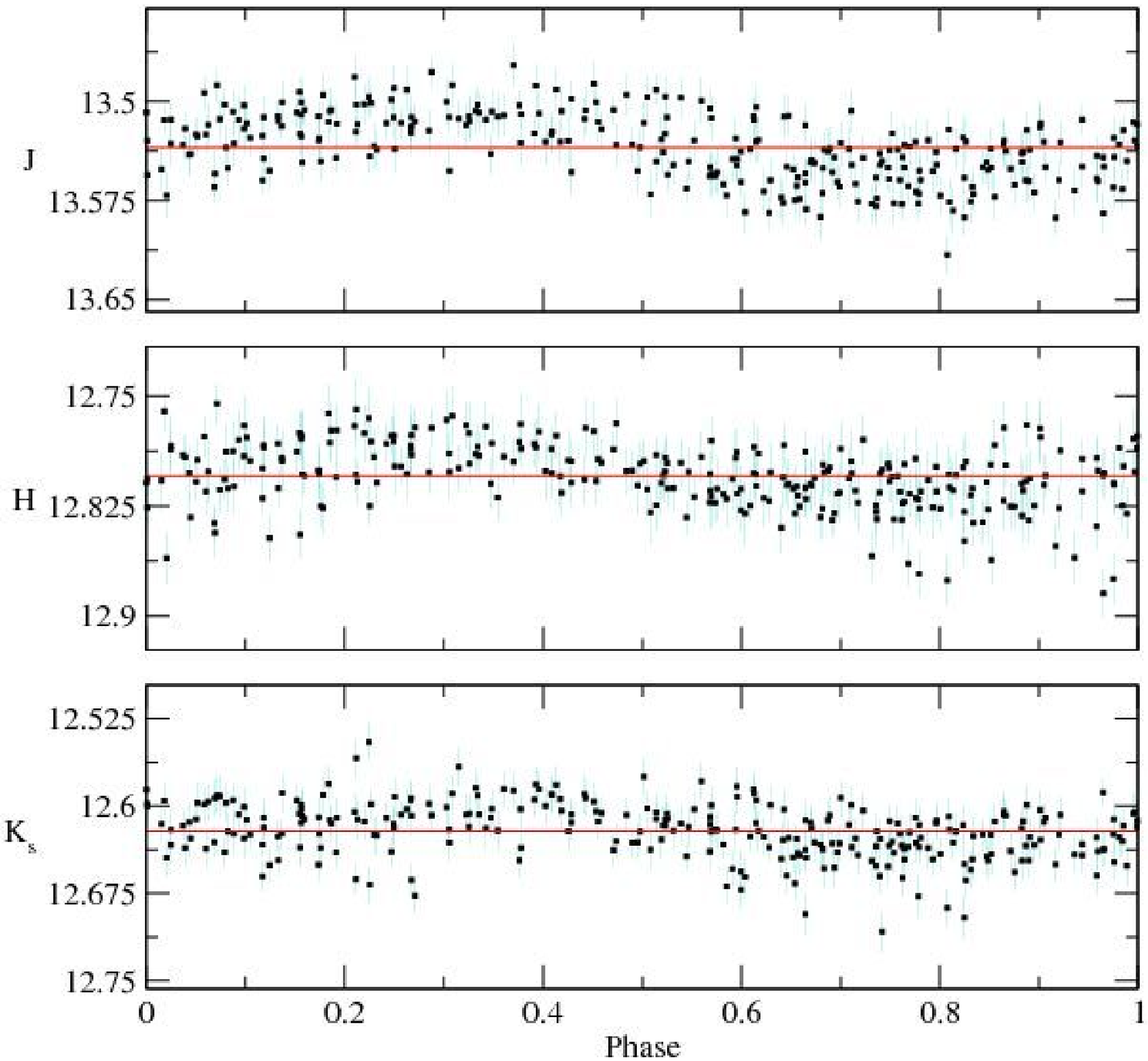}
\caption{JHK$_{s}$ data for sinusoidal variable 2MASS J19014393-0447412, folded to a period of 0.9822 days.  }\end{figure}\clearpage
\begin{figure}
\plotone{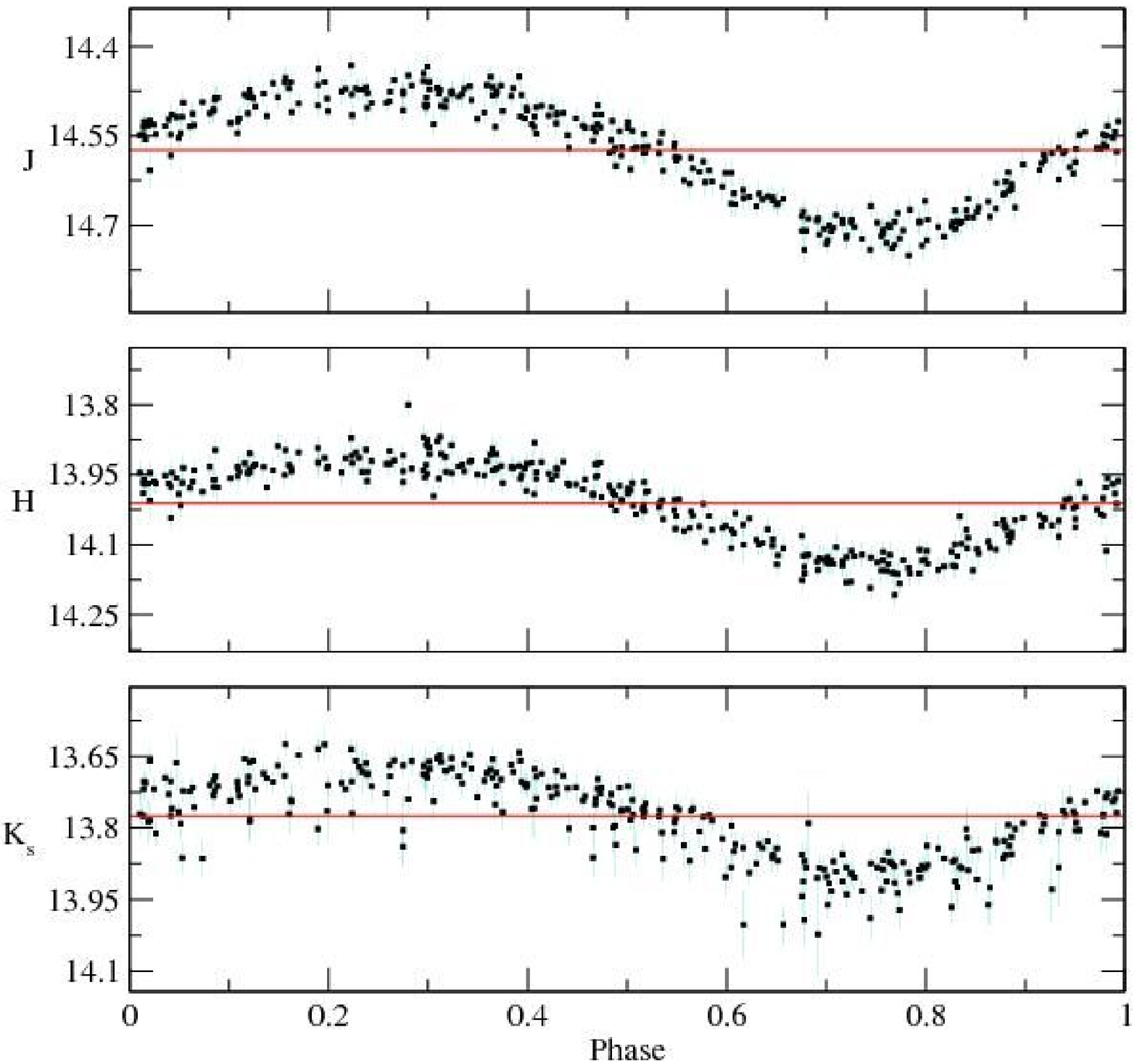}
\caption{JHK$_{s}$ data for suspected CV 2MASS J19014985-0432493, folded to a period of 0.3297 days.  }\end{figure}
\begin{figure}
\plotone{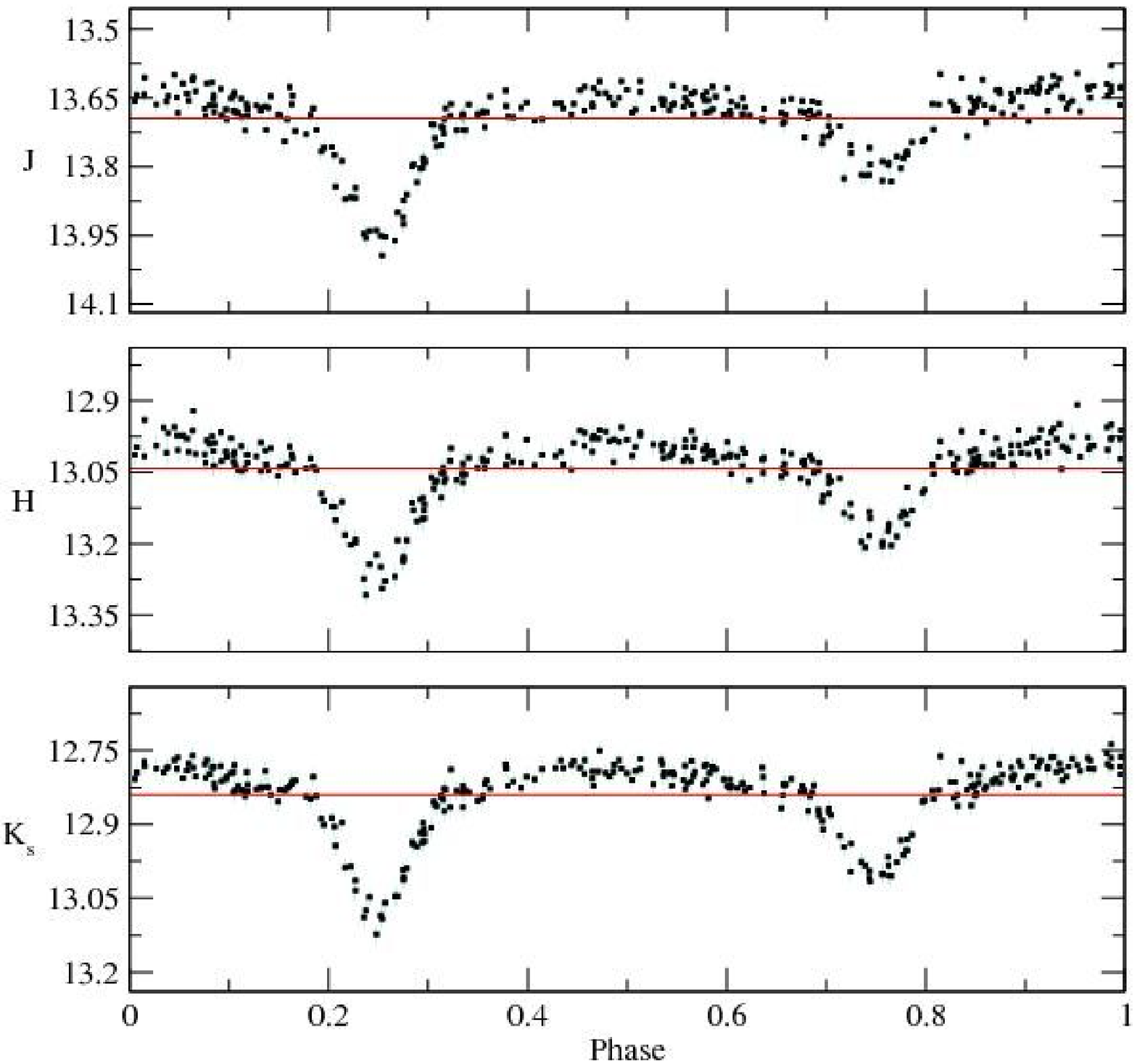}
\caption{JHK$_{s}$ data for eclipsing binary 2MASS J19020989-0439440, folded to a period of 4.3473 days.  }\end{figure}\clearpage
\begin{figure}
\plotone{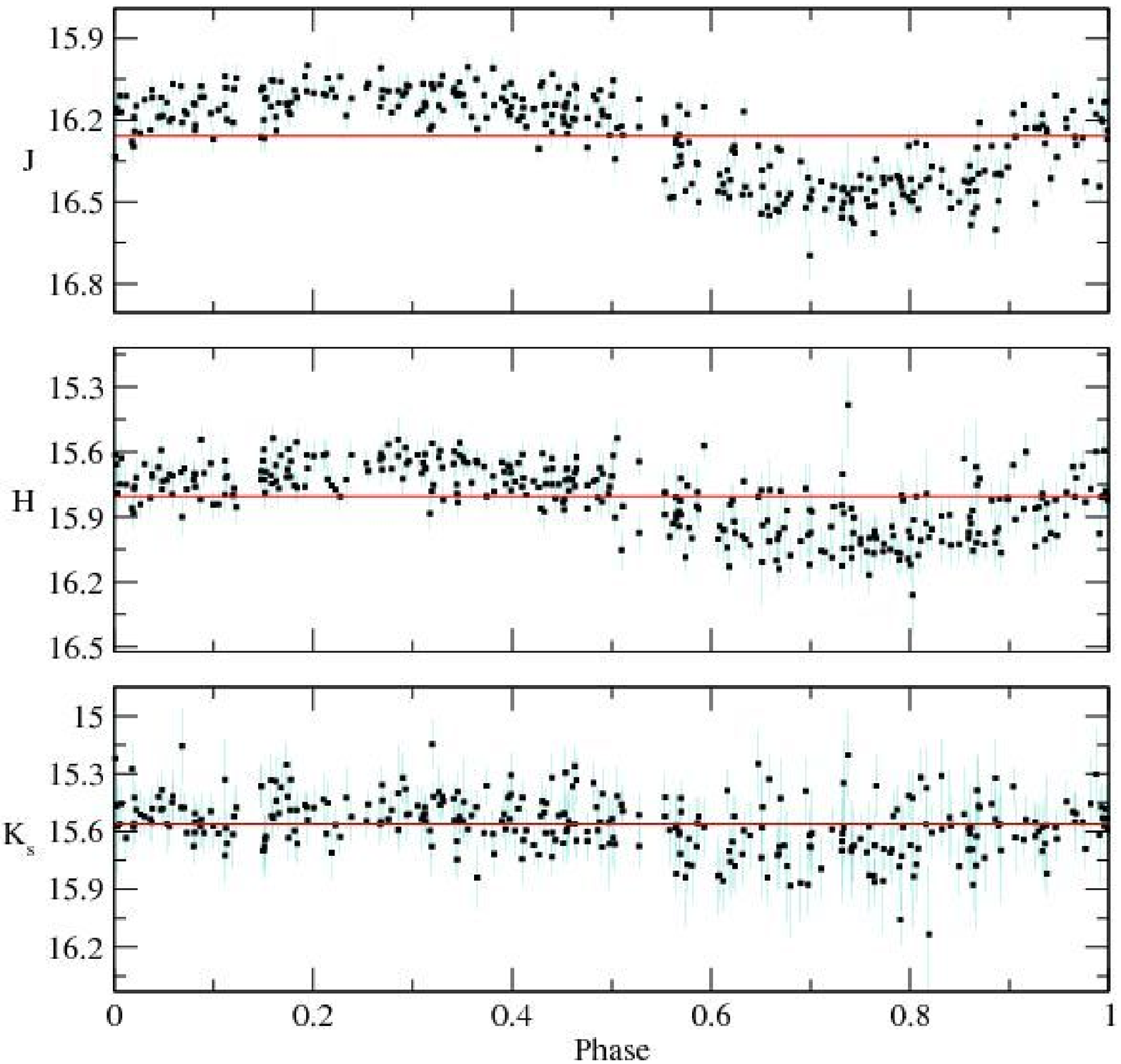}
\caption{JHK$_{s}$ data for suspected CV 2MASS J20310630-4914562, folded to a period of 0.12047 days.  }\end{figure}\clearpage
\begin{figure}                             
\plotone{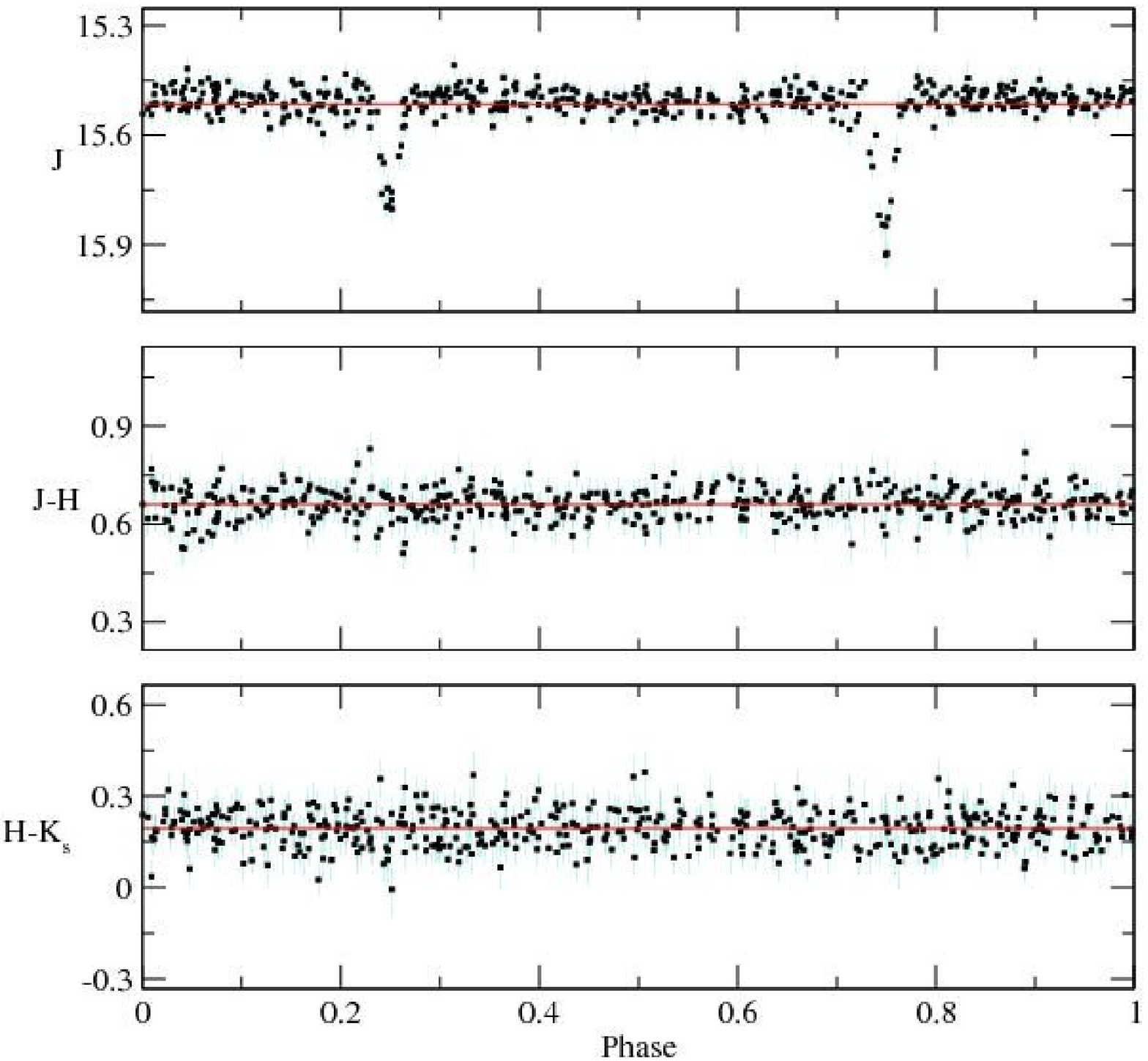}
\caption{J,J-H,H-K$_{s}$ data for late-type eclipsing binary 2MASS J01542930+0053266, folded to a period of 2.639 days.  Data are shown in black with error bars shown in teal.  Each data point corresponds to the unweighted average of one group of six scans ($m_{n,t_6}$).  Mean apparent magnitudes and colors are shown with red horizontal lines.}\end{figure}
\begin{figure}
\plotone{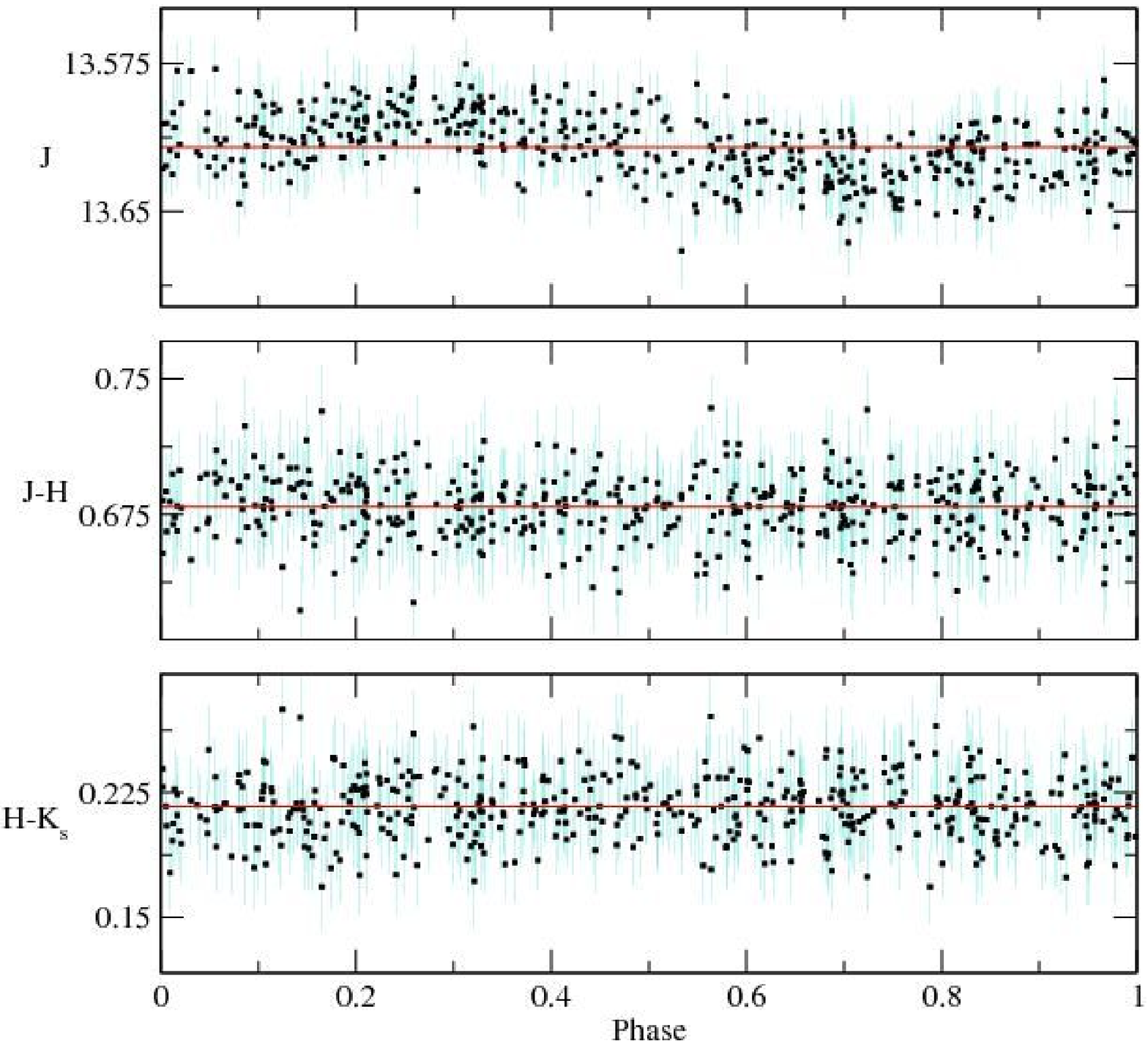}
\caption{J,J-H,H-K$_{s}$ data for suspected CV 2MASS J01545296+0110529, folded to a period of 0.18603 days.  }\end{figure}\clearpage
\begin{figure}
\plotone{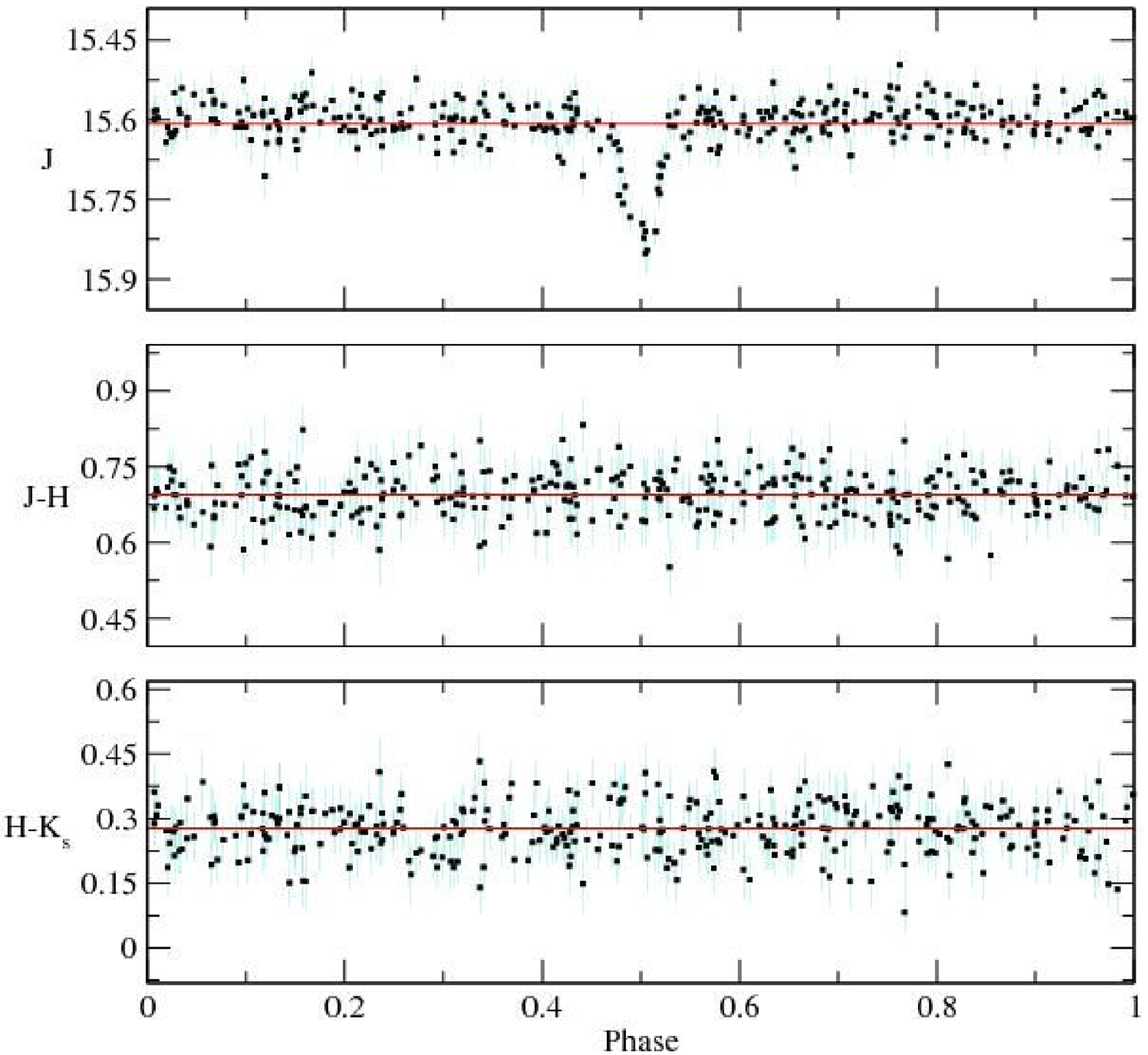}
\caption{J,J-H,H-K$_{s}$ data for late-type eclipsing binary 2MASS J04261603+0323578, folded to a period of 0.88320 days. }\end{figure}
\begin{figure}
\plotone{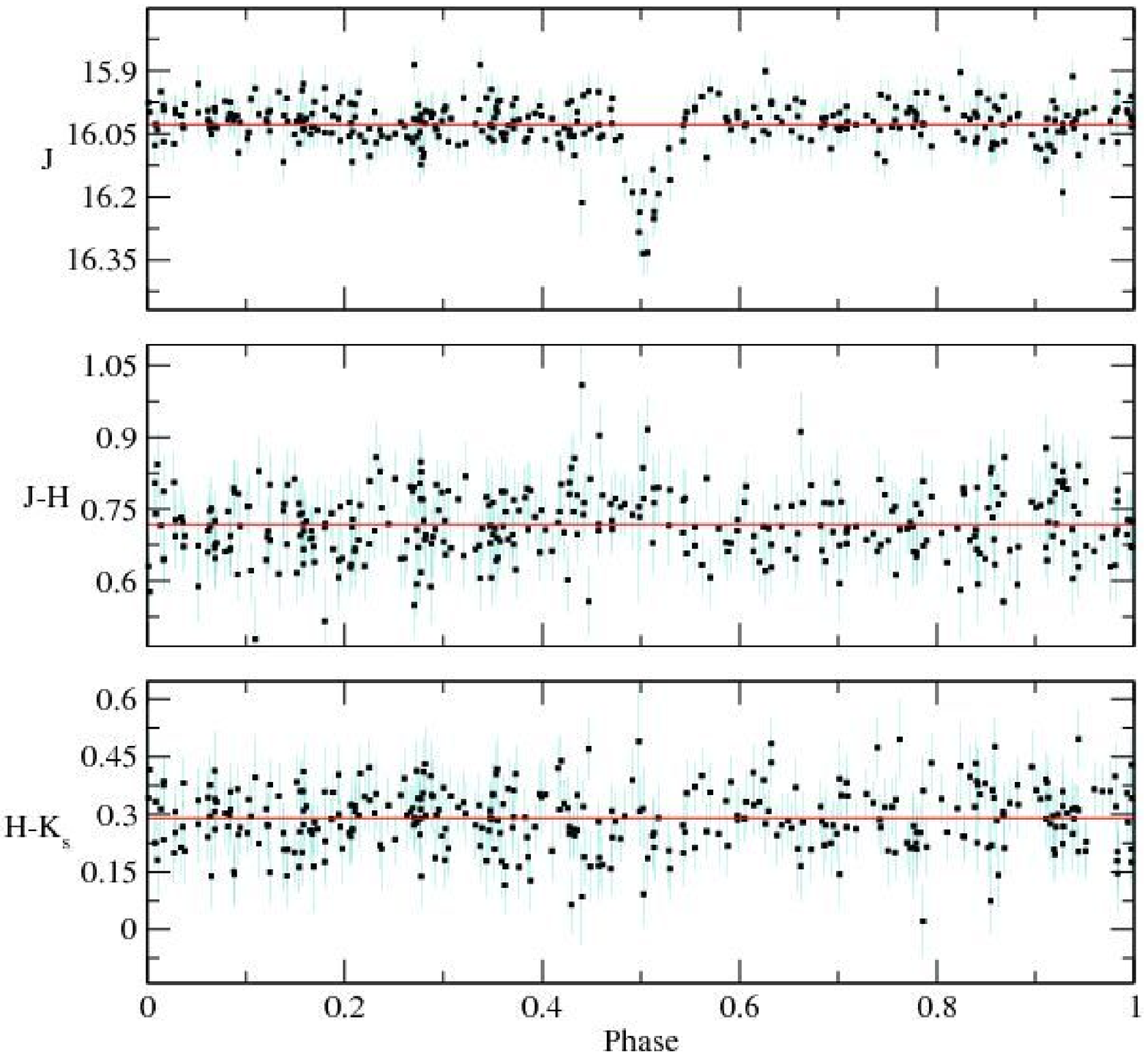}
\caption{J,J-H,H-K$_{s}$ data for late-type eclipsing binary 2MASS J04261900+0314008, folded to a period of 1.07631 days. }\end{figure}\clearpage
\begin{figure}
\plotone{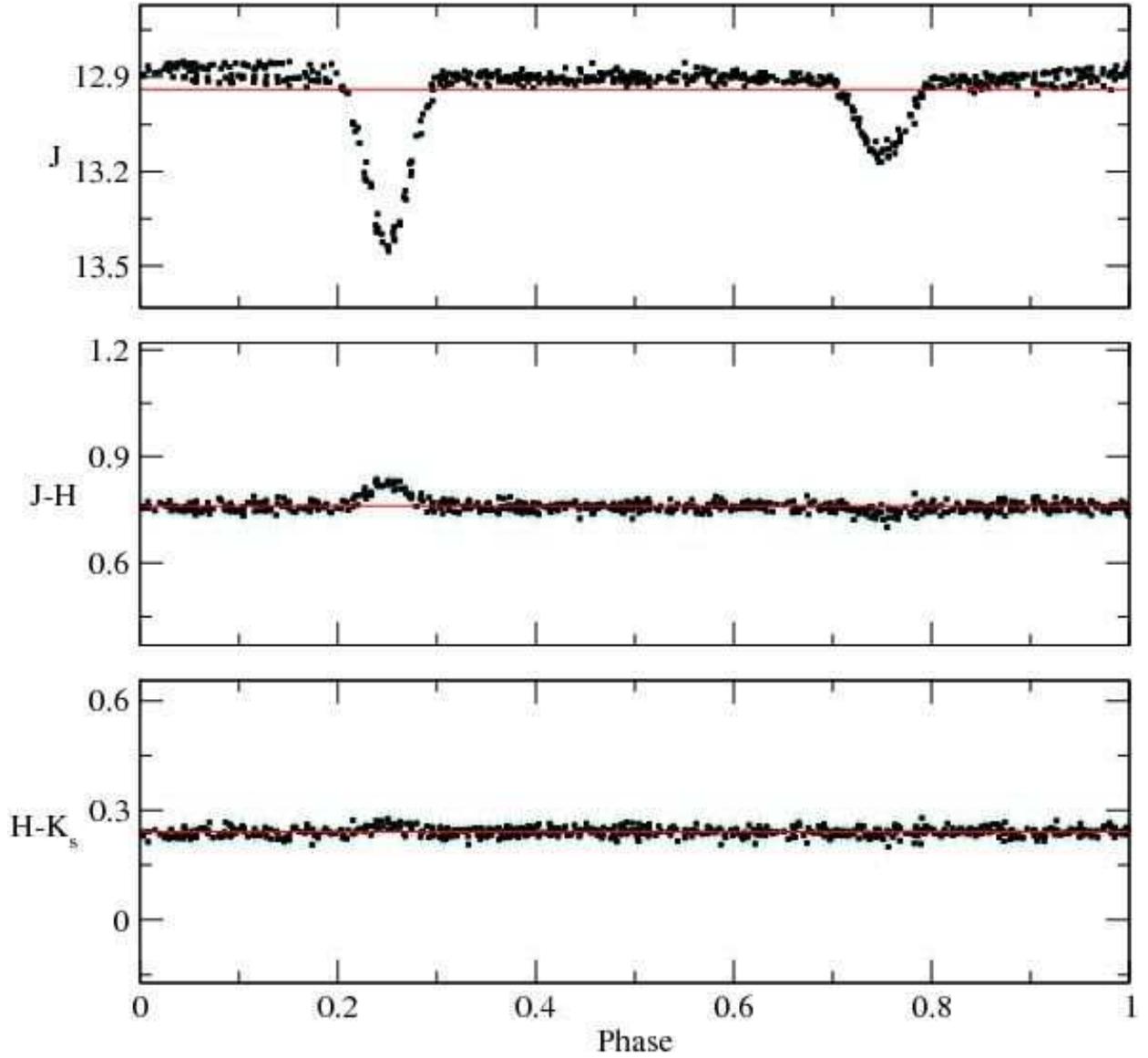}
\caption{J,J-H,H-K$_{s}$ data for eclipsing binary 2MASS J08255405-3908441, folded to a period of 8.08986 days.  The observable changes in color during primary and secondary eclipse for this system indicate the different effective temperatures of the components.}\end{figure}
\begin{figure}
\plotone{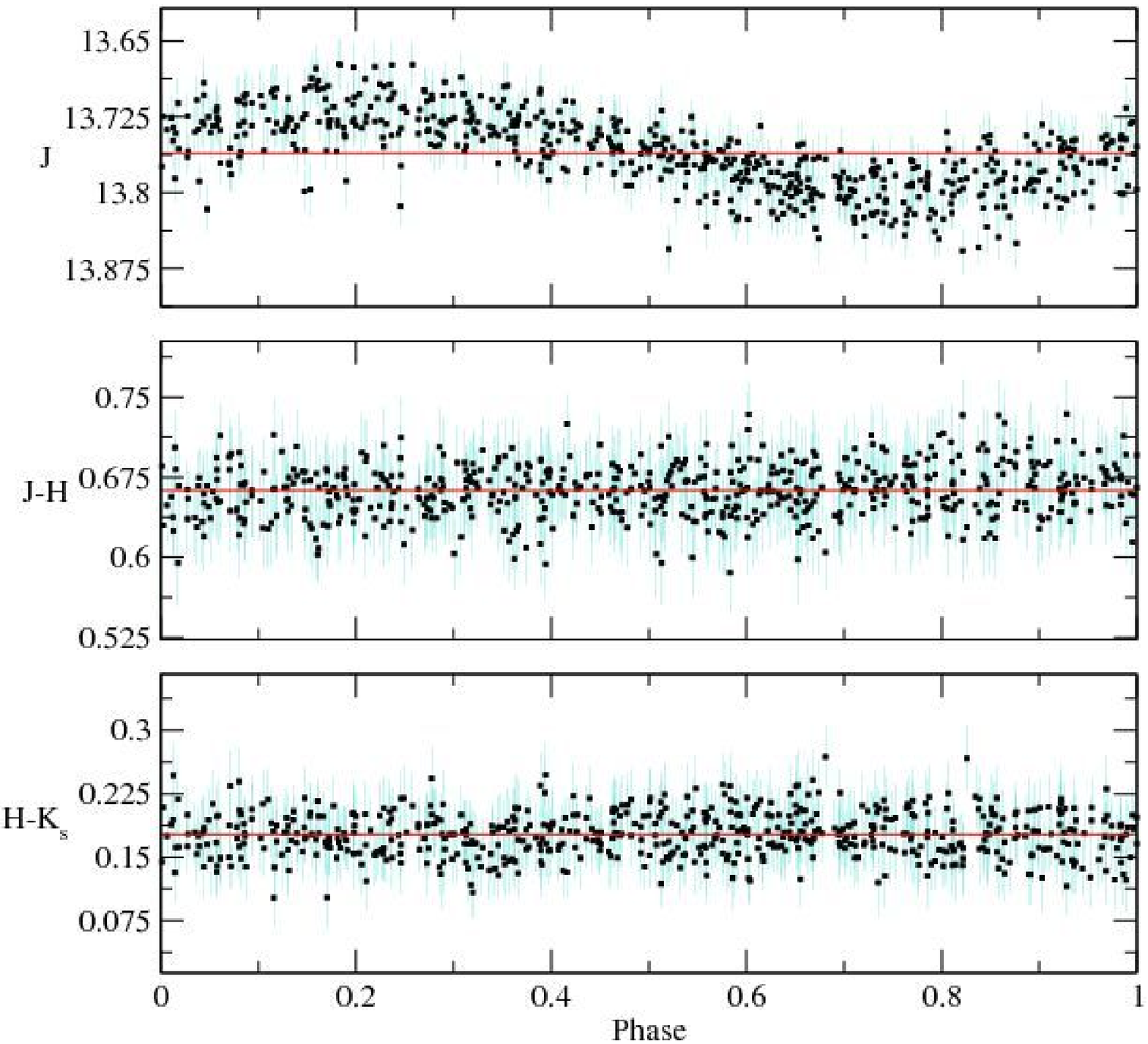}
\caption{J,J-H,H-K$_{s}$ data for the sinusoidal variable 2MASS J08512729+1211484, folded to a period of 1.23725 days.  This object has a persistent artifact overlaid with the source every other scan, but still exhibits intrinsic periodic variability. We eliminate scans with the persistence for our analysis.}\end{figure}\clearpage
\begin{figure}
\plotone{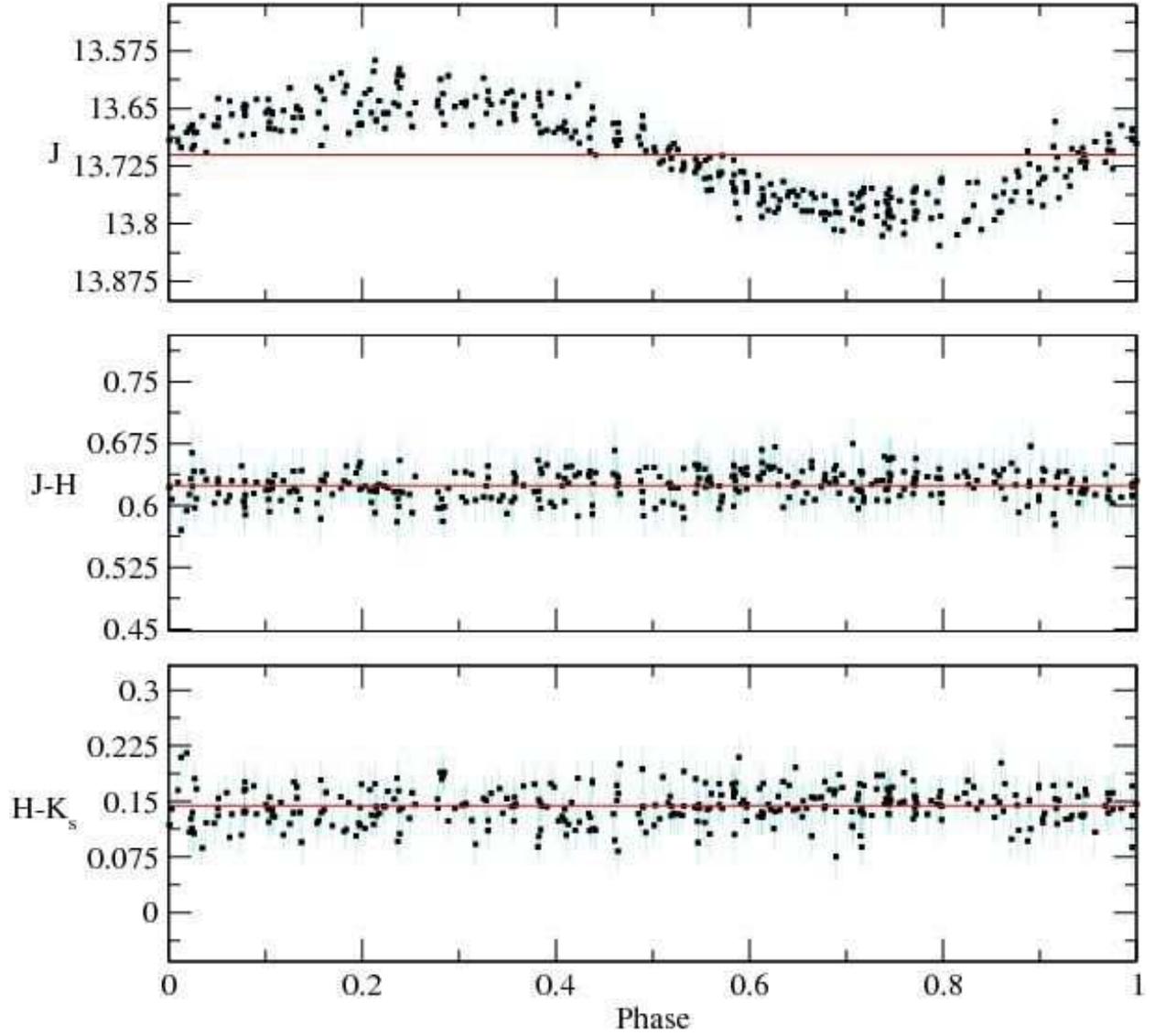}
\caption{J,J-H,H-K$_{s}$ data for the sinusoidal variable 2MASS J15001192-0103090, folded to a period of 3.262 days.  }\end{figure}
\begin{figure}
\plotone{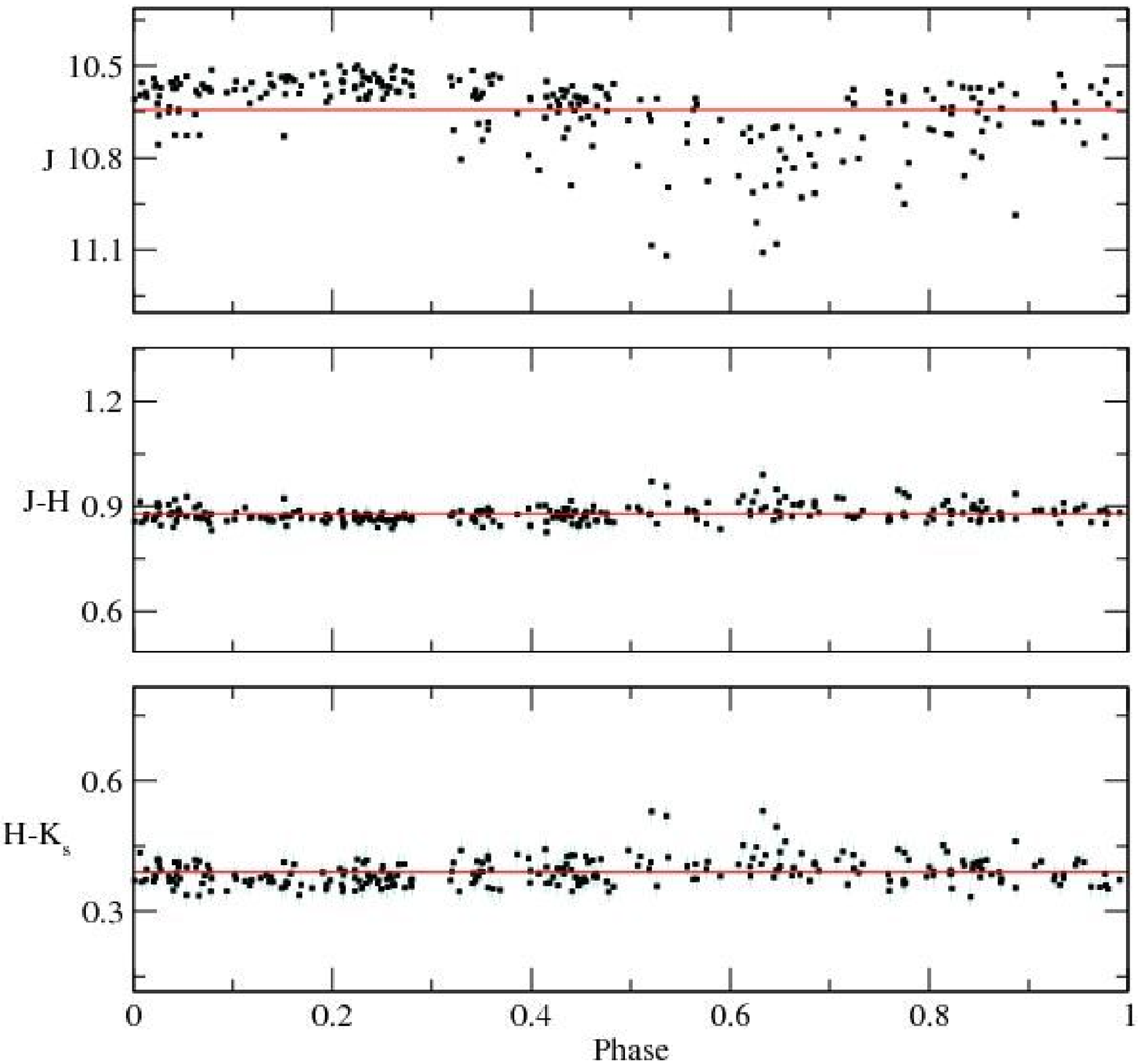}
\caption{J,J-H,H-K$_{s}$ data for Rho Oph YSO 2MASS J16271273-2504017, folded to a period of 0.831445 days.  }\end{figure}\clearpage
\begin{figure}
\plotone{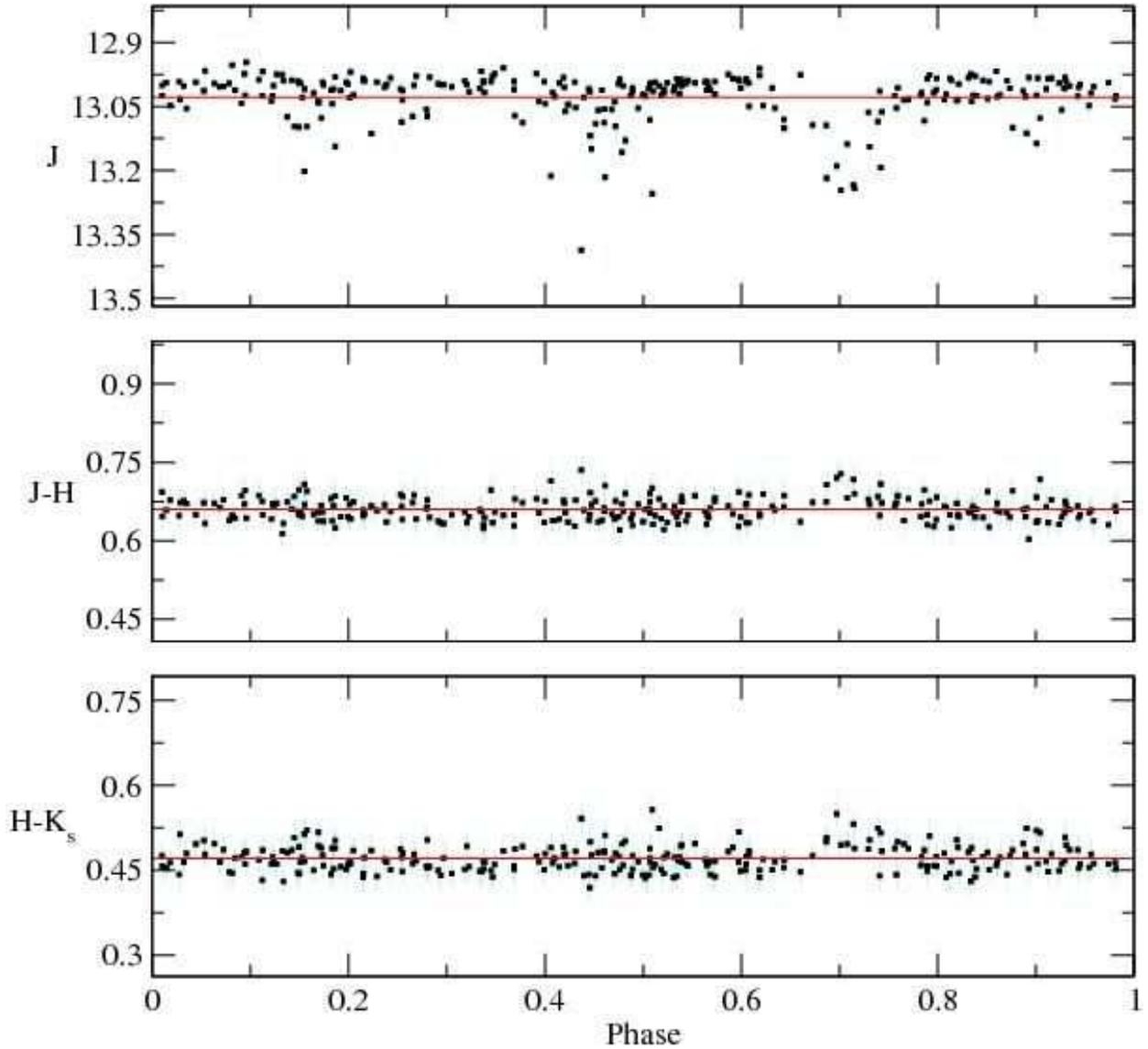}
\caption{J,J-H,H-K$_{s}$ data for Rho Oph YSO 2MASS J16272658-2425543, folded to a period of 2.9603 days. The object exhibits variability on a 3-day timescale, but the observed variability is not entirely consistent with this period derived from our analysis.  Possible solutions include spotting, and semi-periodic veiling from circumstellar material. }\end{figure}
\begin{figure}
\plotone{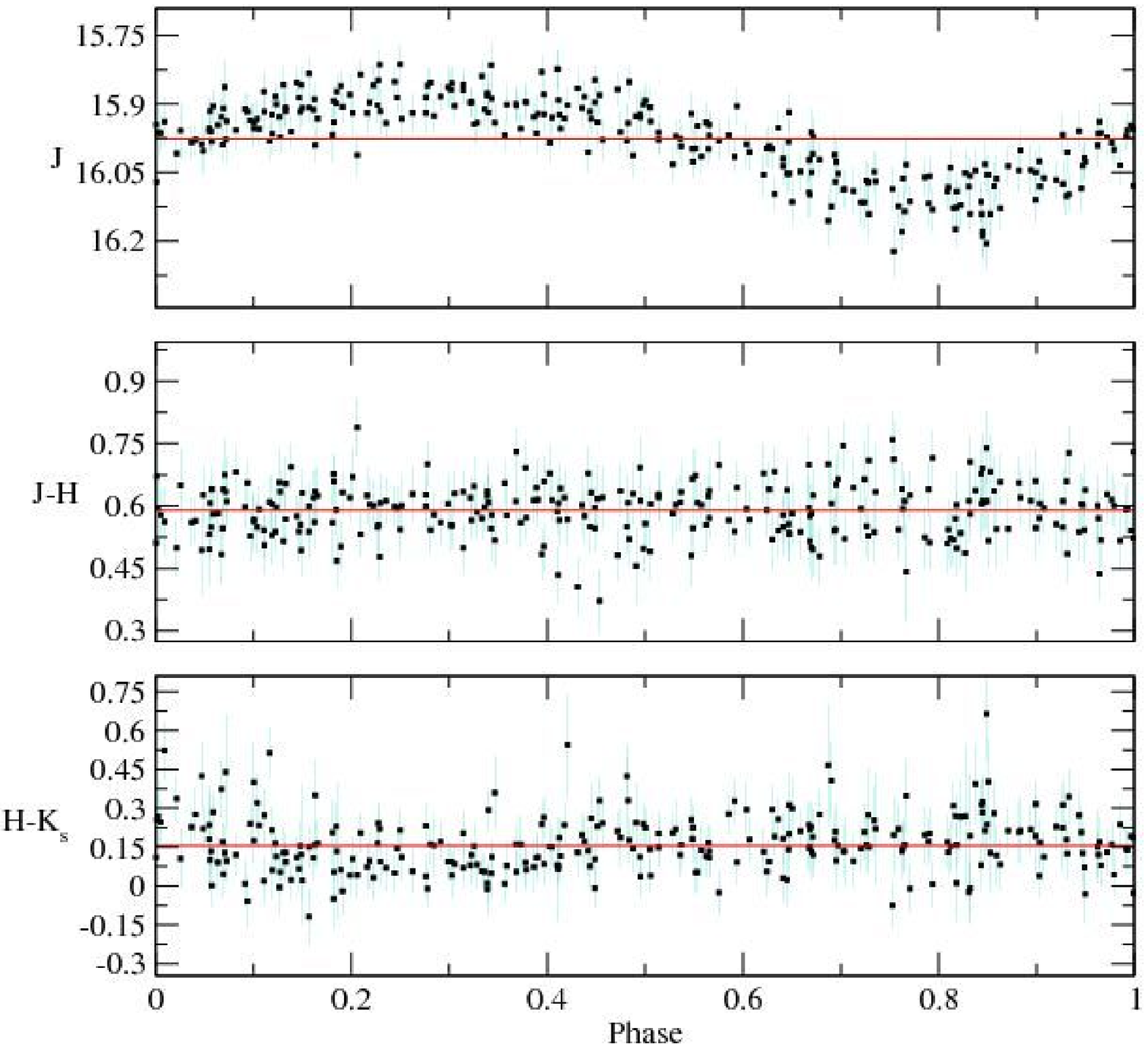}
\caption{J,J-H,H-K$_{s}$ data for suspected CV 2MASS J18391777+4854001, folded to a period of 0.1257 days.  }\end{figure}\clearpage
\begin{figure}
\plotone{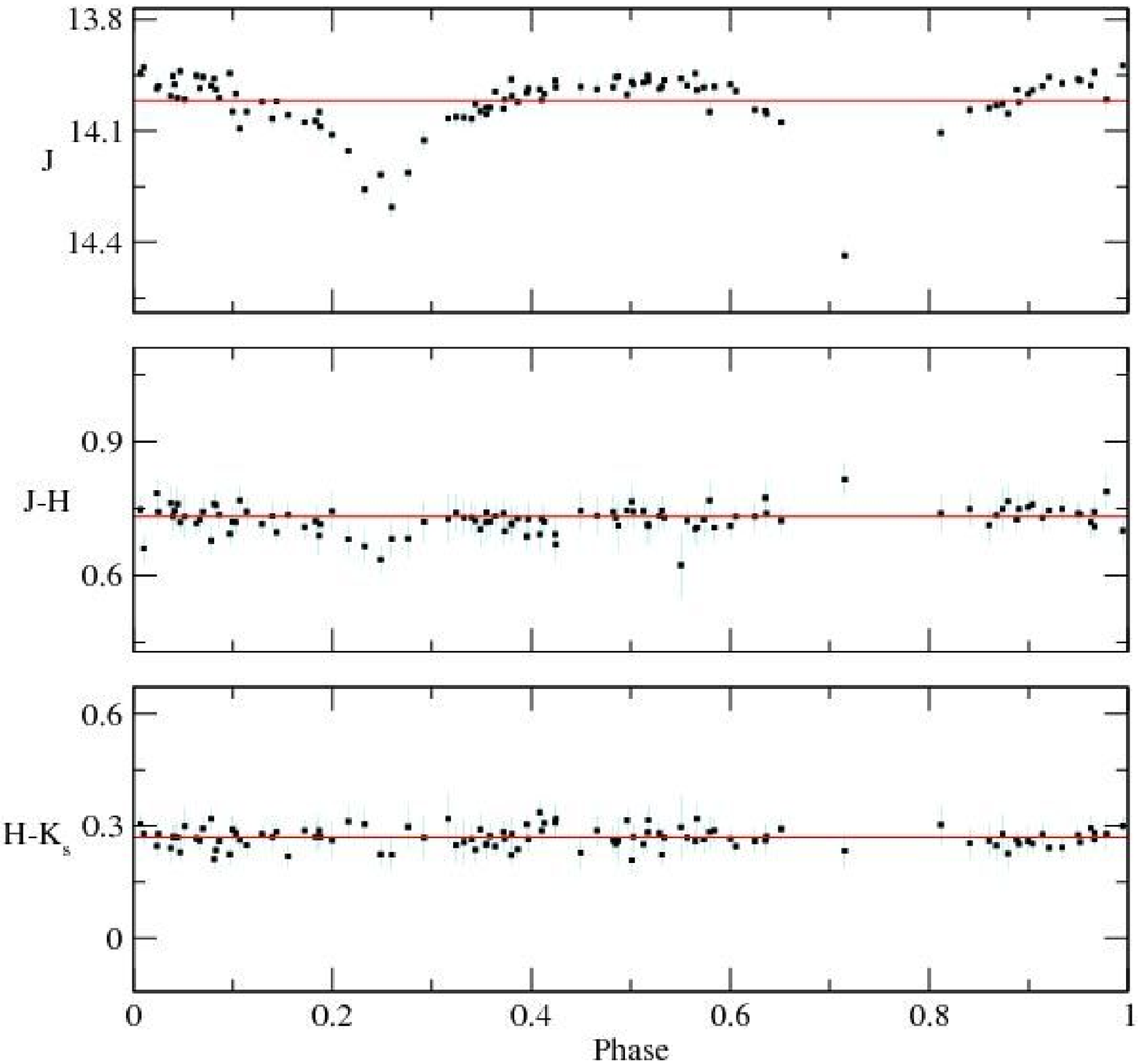}
\caption{J,J-H,H-K$_{s}$ data for eclipsing binary 2MASS J18510479-0442005, folded to a period of 5.657 days.  At this period, the secondary eclipse is not well sampled by the available observations.}\end{figure}
\begin{figure}
\plotone{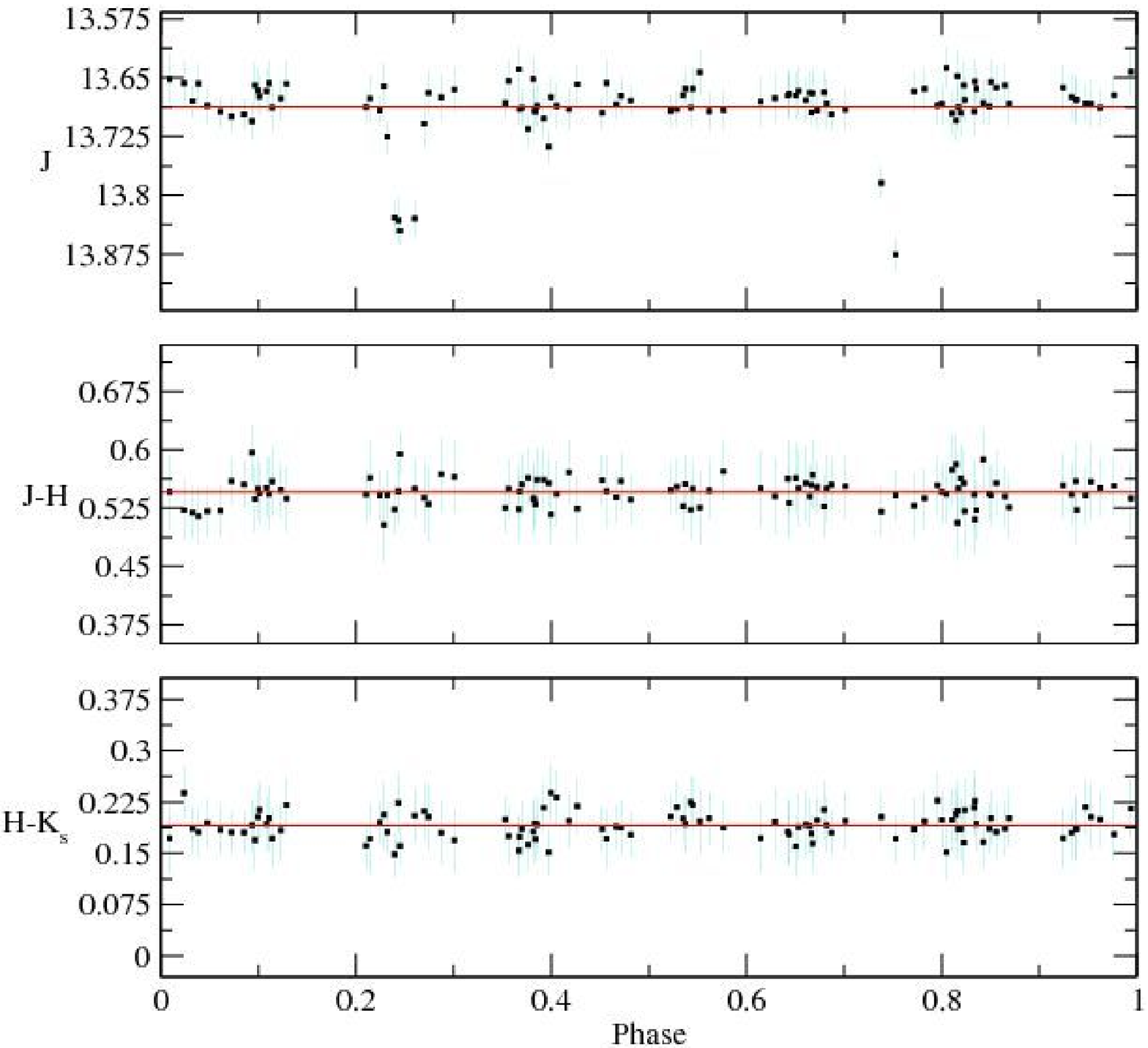}
\caption{J,J-H,H-K$_{s}$ data for eclipsing binary 2MASS J18510526-0437311, folded to a period of 6.2898 days.  }\end{figure}\clearpage
\begin{figure}
\plotone{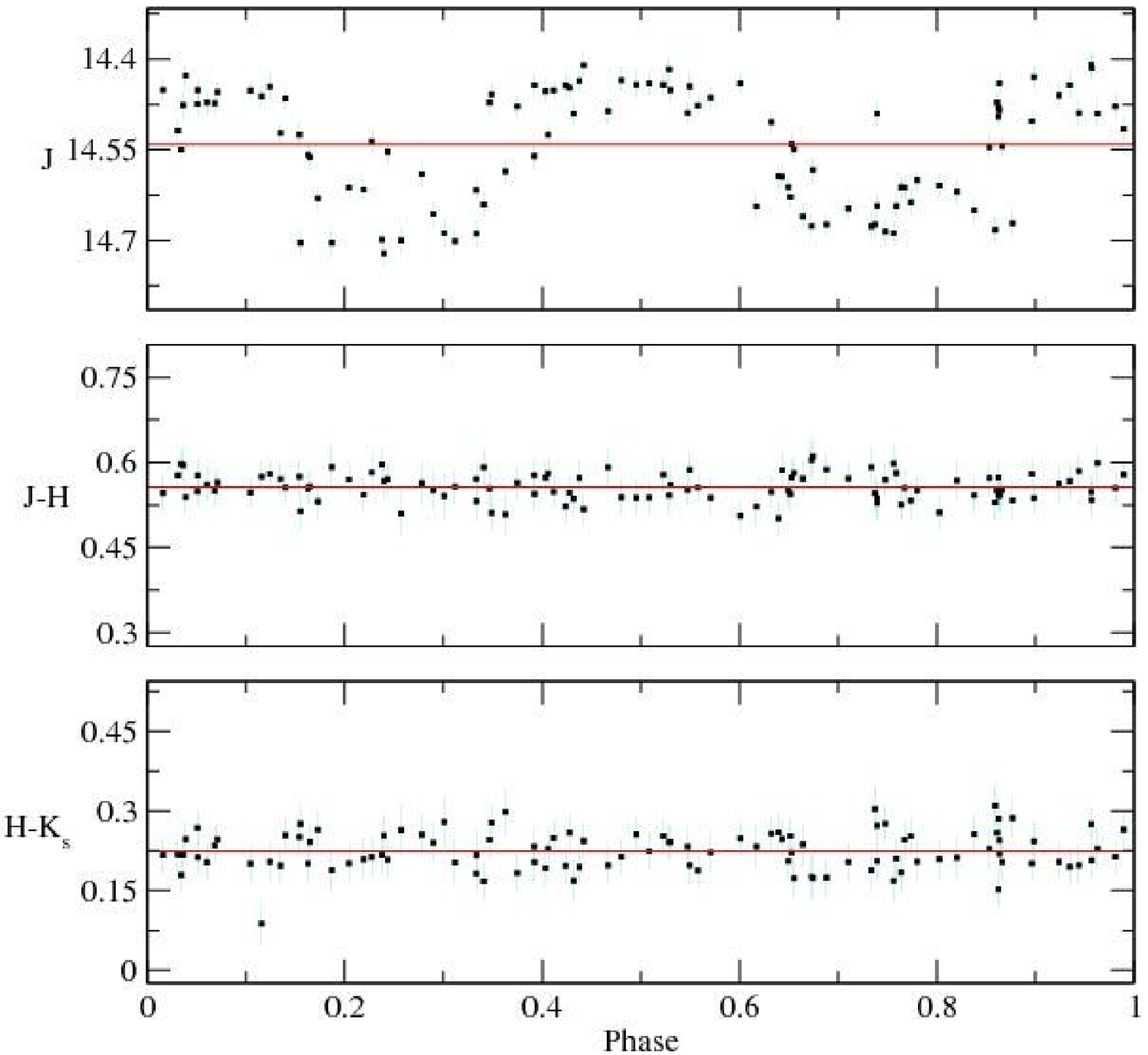}
\caption{J,J-H,H-K$_{s}$ data for 2MASS J18510882-0436123, folded to a period of 0.88822 days.  We do not identify an object type for this variable.}\end{figure}
\begin{figure}
\plotone{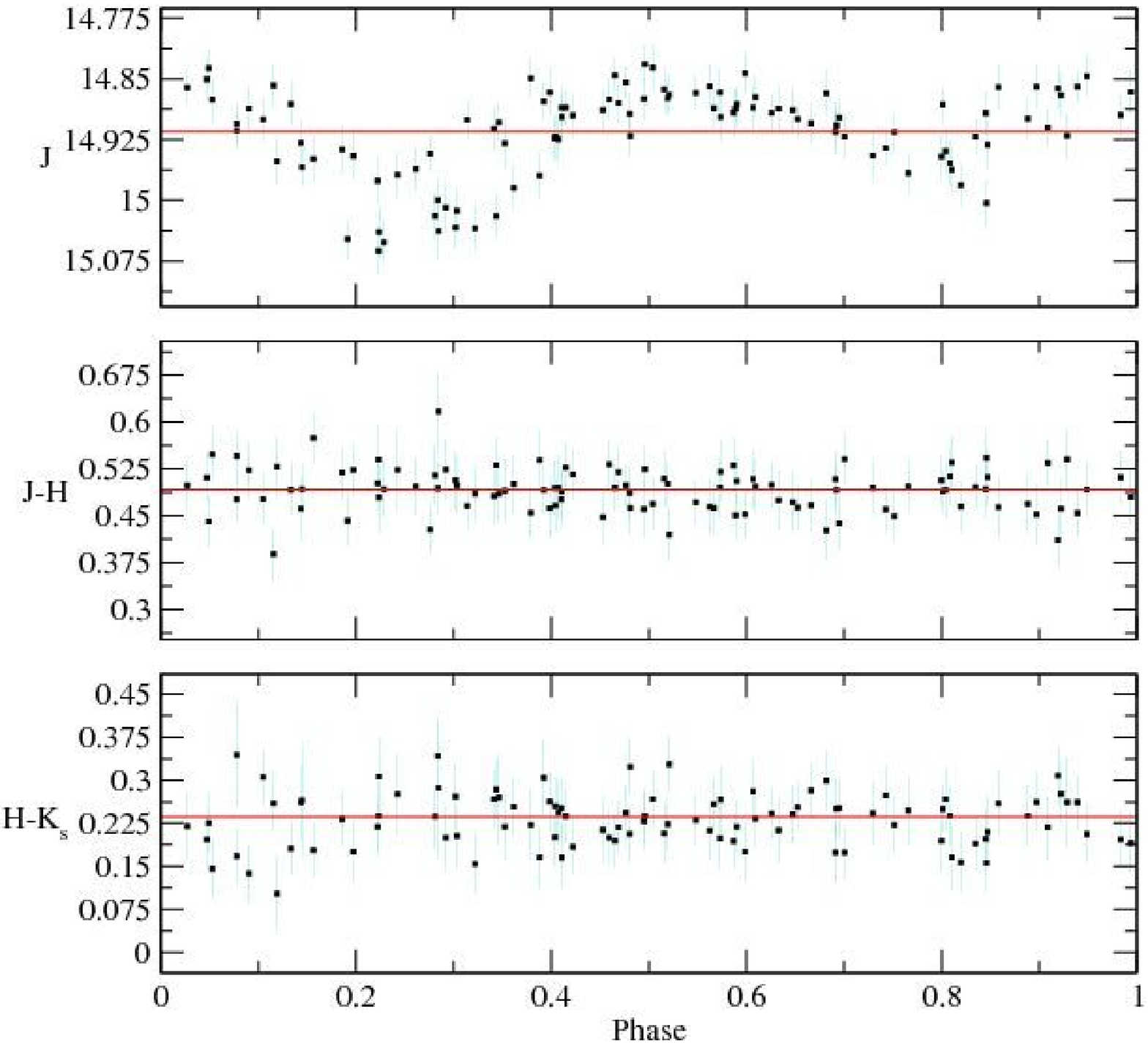}
\caption{J,J-H,H-K$_{s}$ data for suspected eclipsing binary 2MASS J18511786-0355311, folded to a period of 0.7804 days.  }\end{figure}\clearpage
\begin{figure}
\plotone{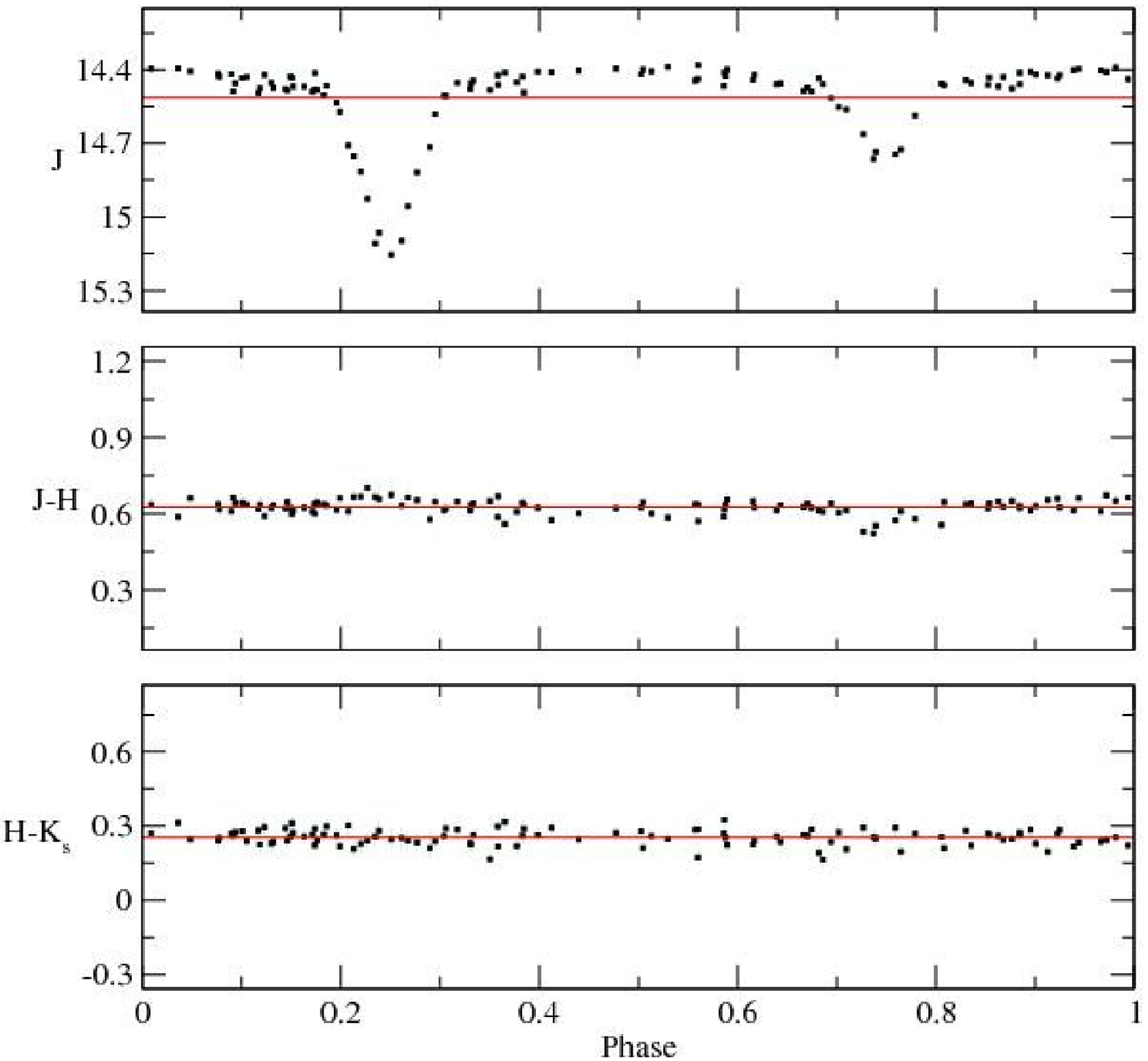}
\caption{J,J-H,H-K$_{s}$ data for eclipsing binary 2MASS J18512034-0426311, folded to a period of 3.3549 days.  }\end{figure}
\begin{figure}
\plotone{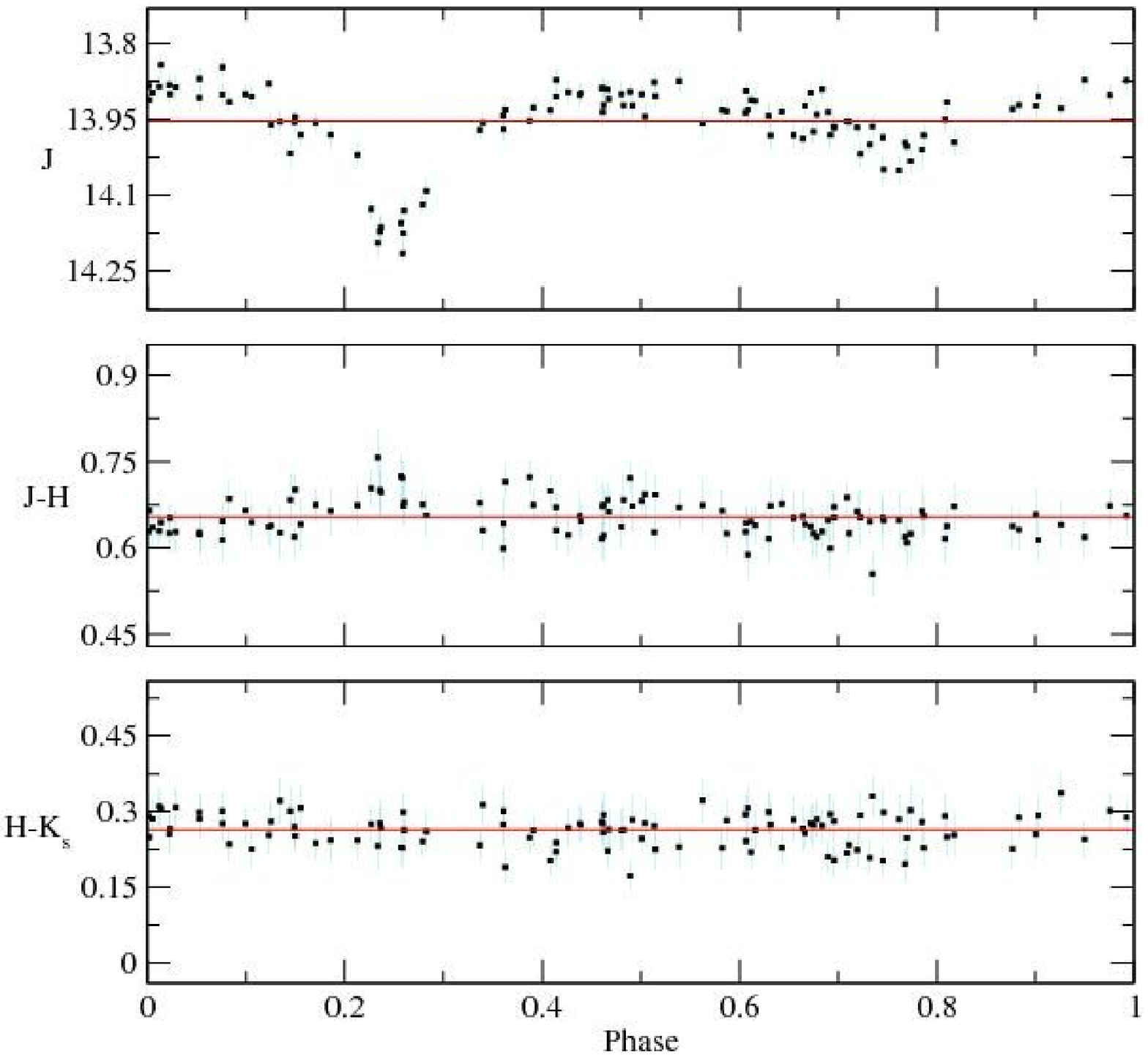}
\caption{J,J-H,H-K$_{s}$ data for eclipsing binary 2MASS J18512261-0409084, folded to a period of 3.916 days.  }\end{figure}\clearpage
\begin{figure}
\plotone{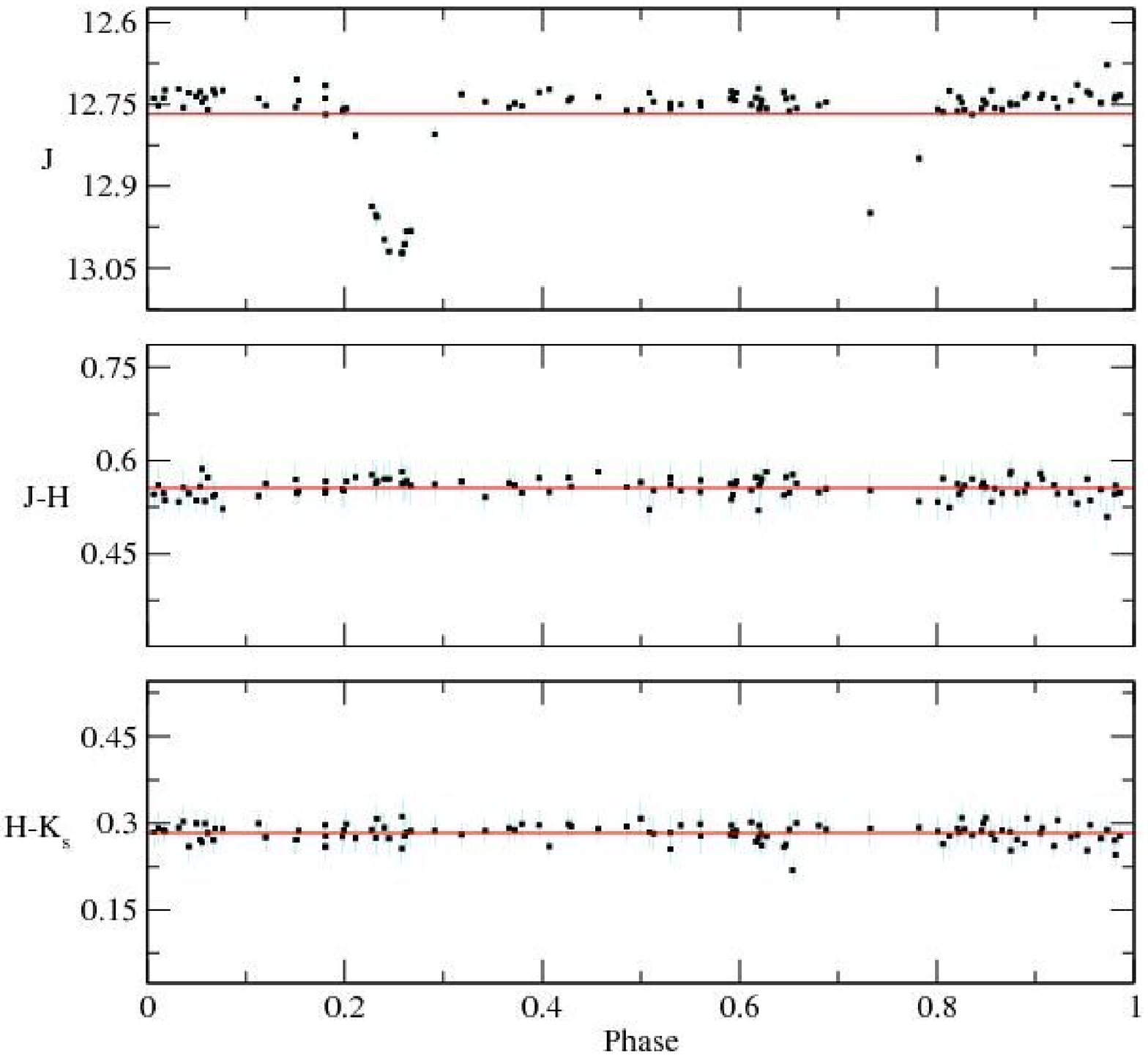}
\caption{J,J-H,H-K$_{s}$ data for eclipsing binary 2MASS J18512929-0412407, folded to a period of 3.0396 days.  }\end{figure}
\begin{figure}
\plotone{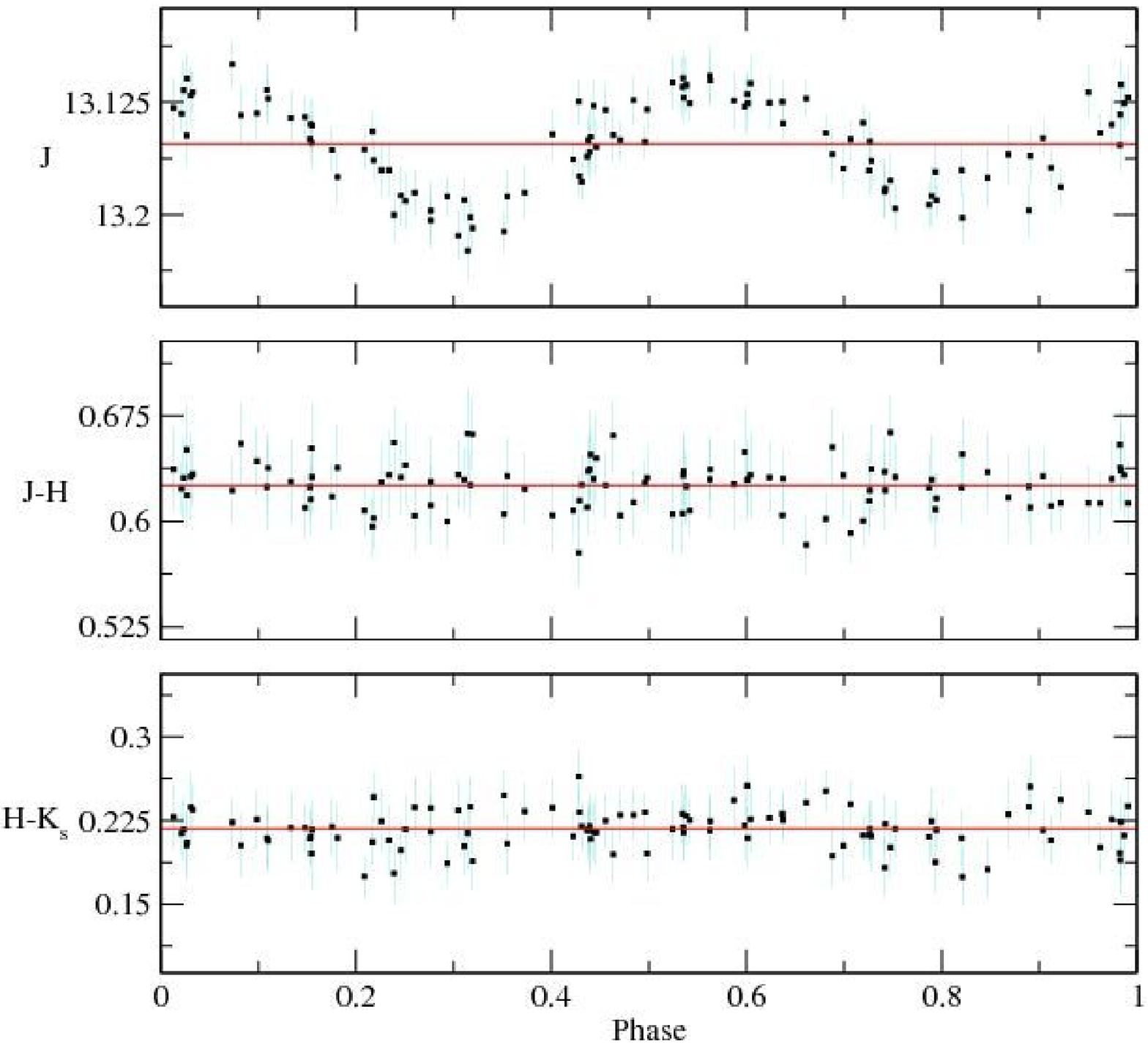}
\caption{J,J-H,H-K$_{s}$ data for suspected CV 2MASS J18513076-0432148, folded to a period of 0.32251 days.  }\end{figure}\clearpage
\begin{figure}
\plotone{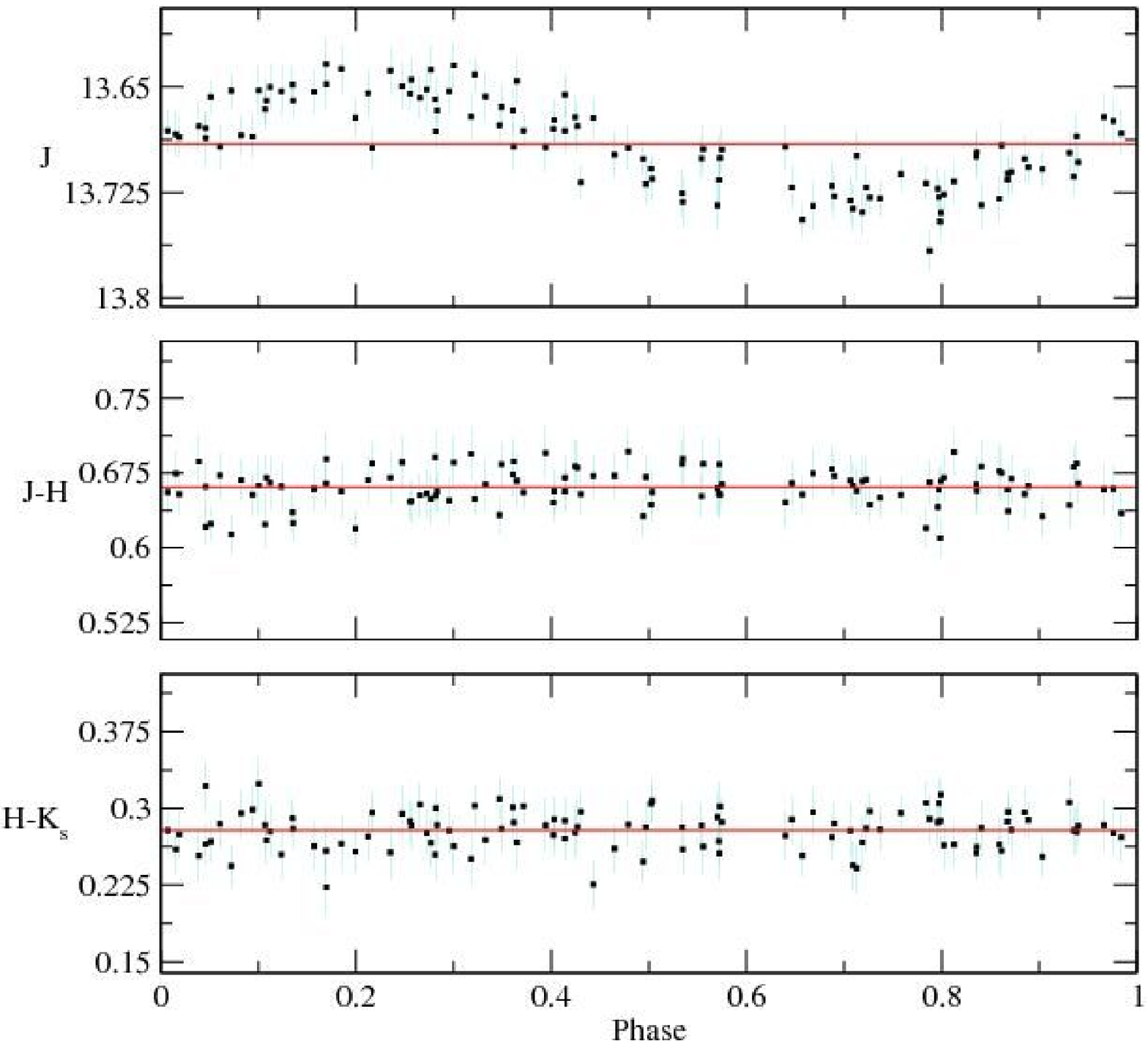}
\caption{J,J-H,H-K$_{s}$ data for suspected CV 2MASS J18513115-0424324, folded to a period of 0.6223 days.  }\end{figure}
\begin{figure}
\plotone{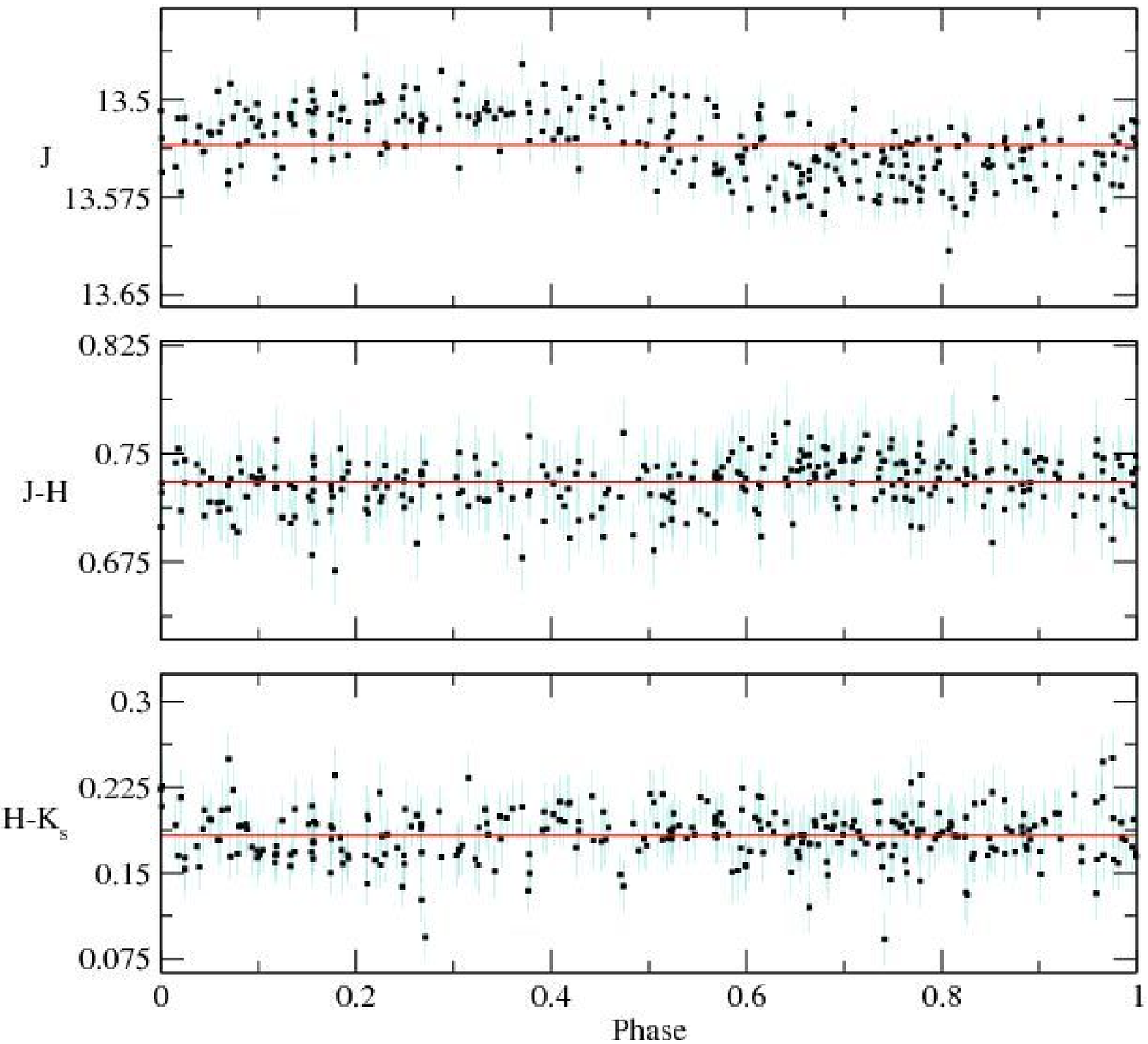}
\caption{J,J-H,H-K$_{s}$ data for sinusoidal variable 2MASS J19014393-0447412, folded to a period of 0.9822 days.  }\end{figure}\clearpage
\begin{figure}
\plotone{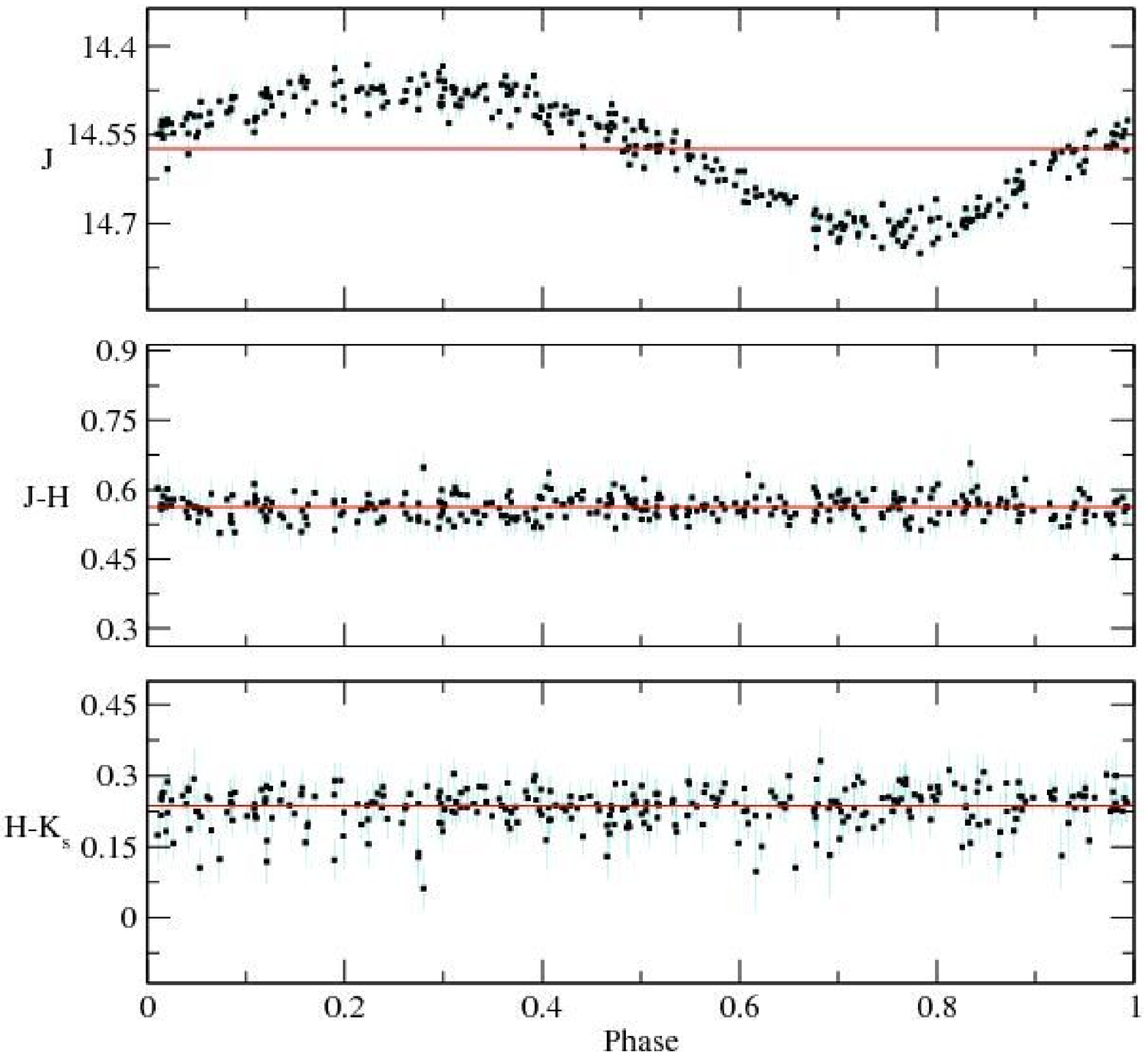}
\caption{J,J-H,H-K$_{s}$ data for suspected CV 2MASS J19014985-0432493, folded to a period of 0.3297 days.  }\end{figure}
\begin{figure}
\plotone{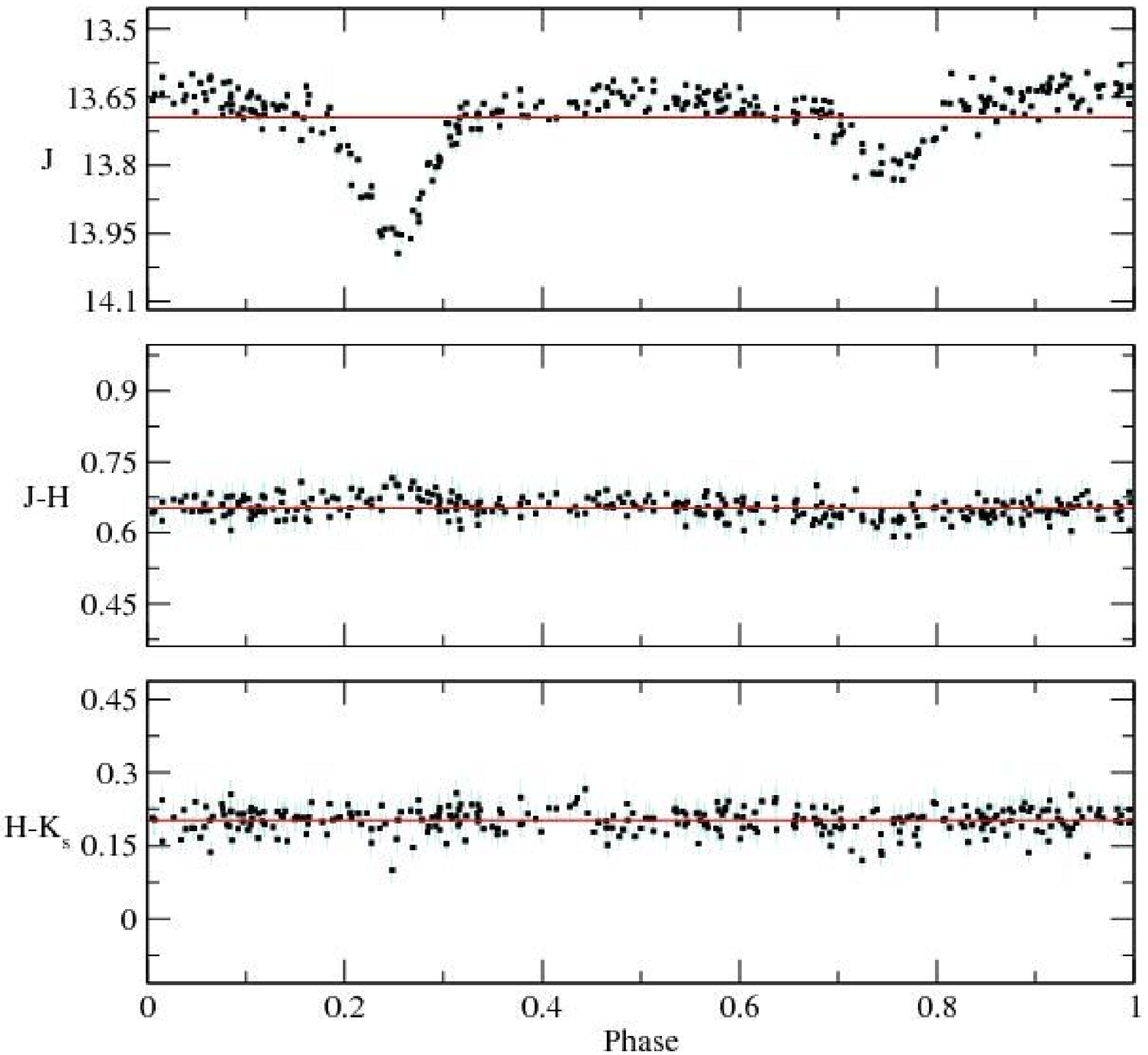}
\caption{J,J-H,H-K$_{s}$ data for eclipsing binary 2MASS J19020989-0439440, folded to a period of 4.3473 days.  }\end{figure}\clearpage
\begin{figure}%
\plotone{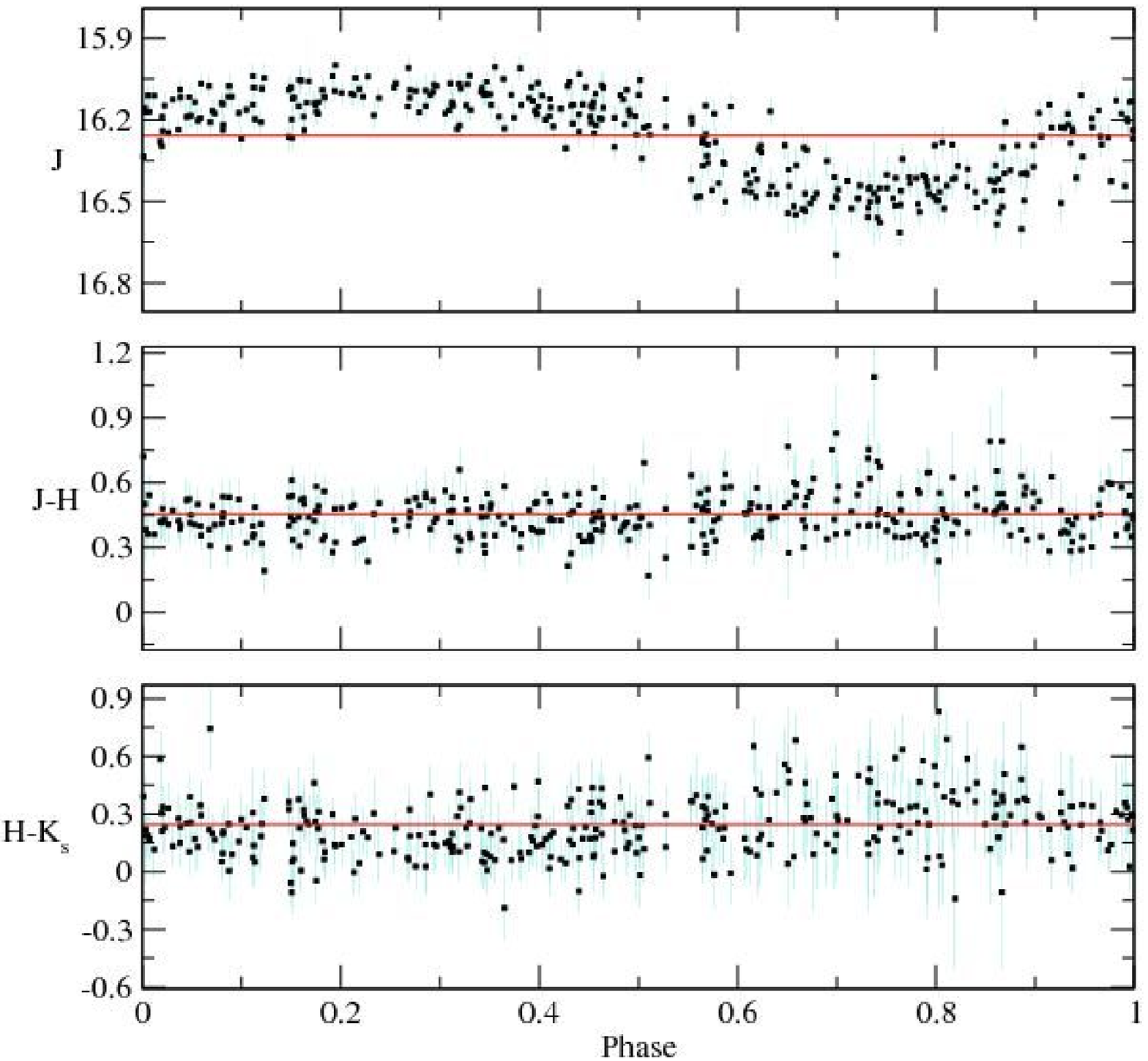}
\caption{J,J-H,H-K$_{s}$ data for suspected CV 2MASS J20310630-4914562, folded to a period of 0.12047 days.  }\end{figure}\clearpage
\begin{figure}
\plotone{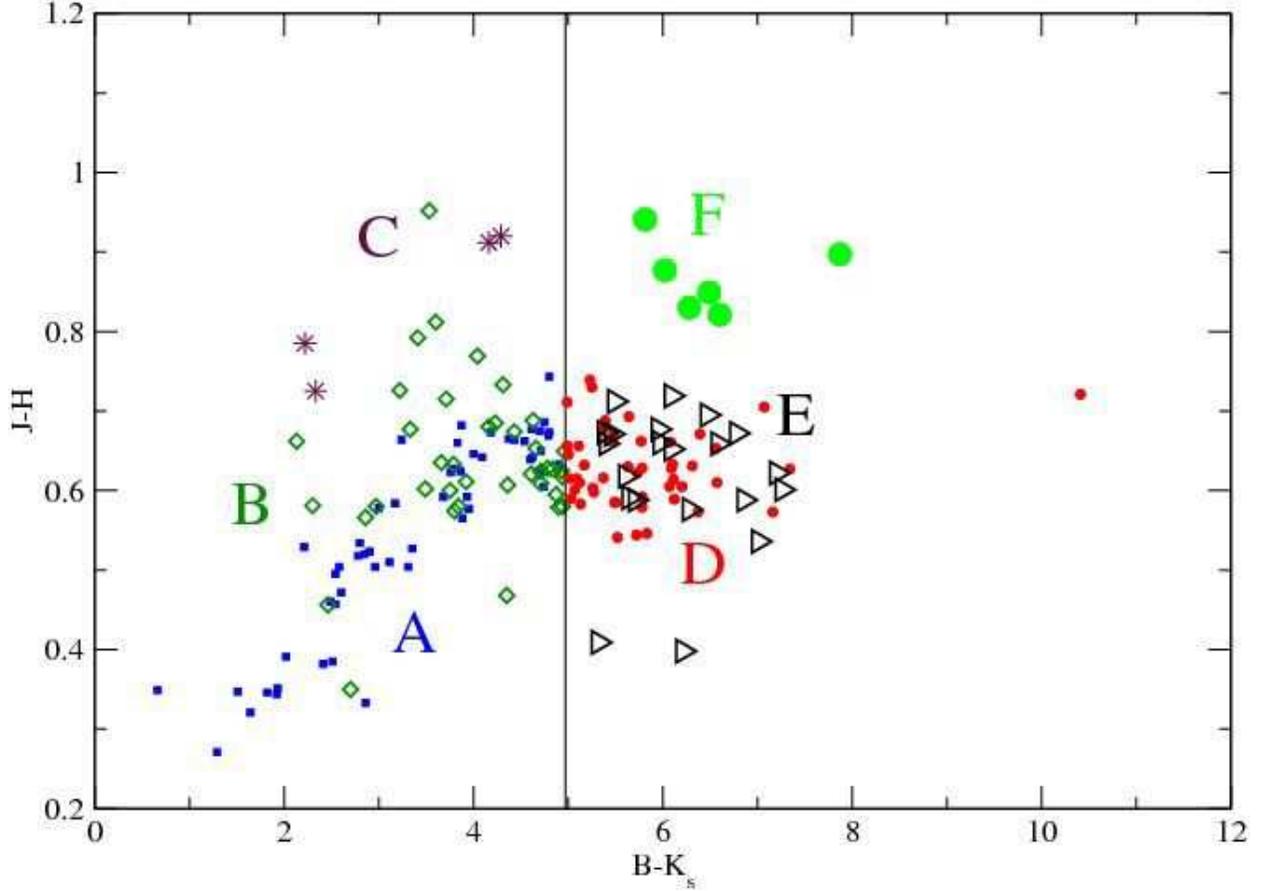}
\caption {We plot B-K$_{s}$ vs. J-H for our sample of $\left|b\right|>20\degree$ variables, using Cal-PSWDB average apparent magnitudes.  We assign different symbols to different color regimes to distinguish the different types of variables.  We identify B-magnitudes from the PSC optical counterparts taken from USNO-A2.0 and Tycho 2 \citep[Johnson B and photographic B respectively]{monet96,hog00}, SIMBAD with a 5'' search radius, and SDSS DR5.  The blue squares (A; K-type star or earlier), green open diamonds (B; Extragalactic or CV), and maroon stars (C; Extragalactic) correspond to sources with B-K$_{s}<$4.97.  The red circle (D; M dwarf) correspond to sources with B-K$_{s}>$4.97.  The open black right triangles (E; M dwarf or extragalactic) are lower limits corresponding to sources with no identified B-band counterpart; we assign a B magnitude of 21 for the points plotted.  `E' sources are consistent with the near-infrared colors of `D' sources. The green circles (F; red giant) correspond to sources with B-K$_{s}>$4.97 and J-H$>$0.79.  The letter designations A-F are the same as in Tables 12 through 14, and described in Table 12.}
\end{figure}
\begin{figure}
\plotone{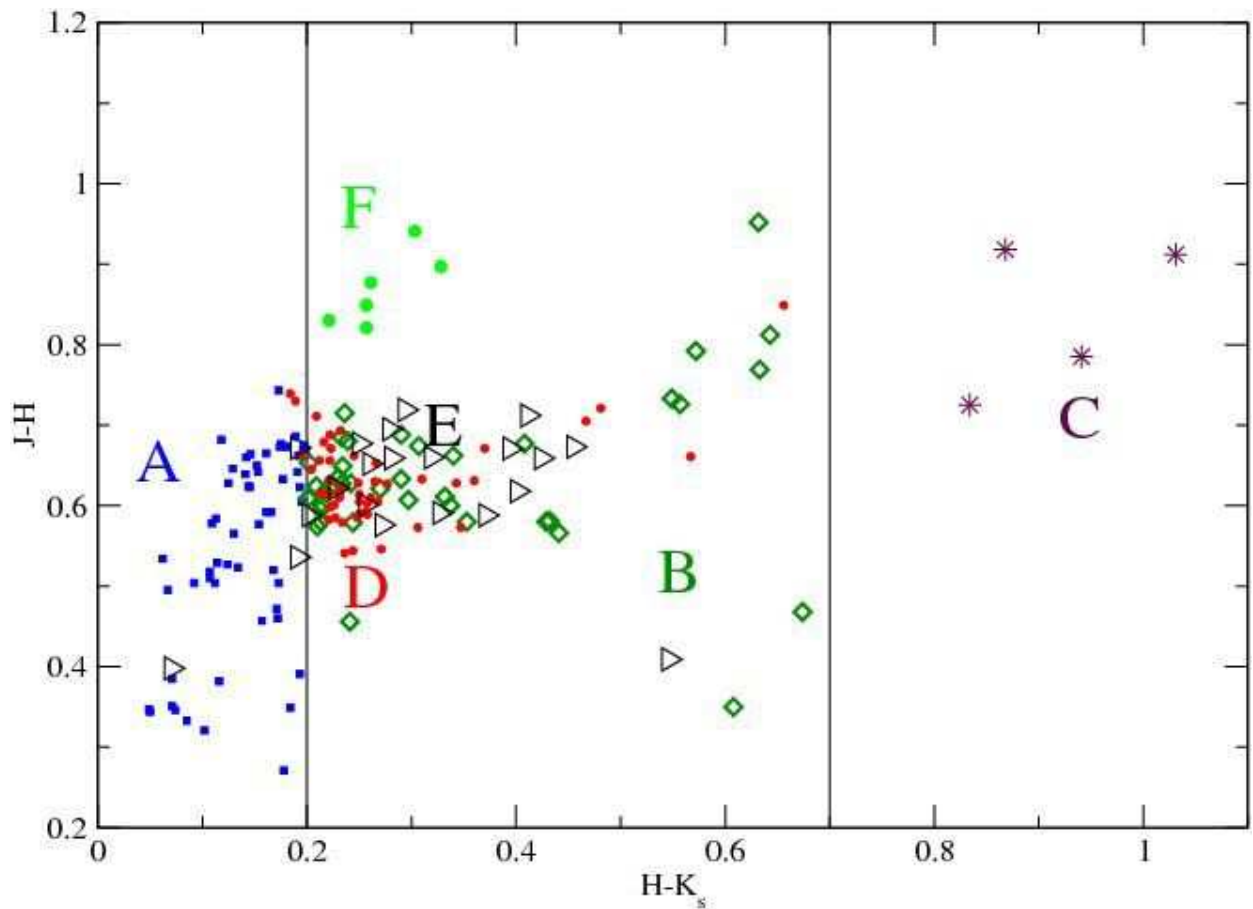}
\caption {We plot H-K$_{s}$ vs J-H for our sample of $\left|b\right|>20\degree$ variables, as in Figure 20.  Unlike Figure 19, we plot sources with the same symbols and Cal-PSWDB colors based on the B-K$_{s}$ criteria shown in Figure 68. All `A', `B' and `C' sources have B-K$_{s}<$4.97.  `A' sources have H-K$_{s}<$0.2, between 0.2 and 0.7 for `B' sources, and greater than 0.7 for `C' sources.}
\end{figure}

\end{document}